\documentclass[showpacs,twocolumn,preprintnumbers,amsmath,amssymb]{revtex4}

\usepackage{graphicx}
\usepackage{dcolumn}
\usepackage{bm}
\usepackage{hyperref}
\usepackage{array,booktabs}
\usepackage{multirow}
\usepackage{hhline}
\usepackage{amsfonts}
\usepackage{amsmath}
\usepackage{slashed}
\usepackage{verbatim}

\newcommand{\bq}{\begin{equation}}
\newcommand{\eq}{\end{equation}}
\newcommand{\bqa}{\begin{eqnarray}}
\newcommand{\eqa}{\end{eqnarray}}
\newcommand{\nn}{\nonumber \\}

\def\be     {\begin{equation}}
\def\ee     {\end{equation}}
\def\bea        {\begin{eqnarray}}
\def\eea        {\end{eqnarray}}
\def\bnn    {\begin{eqnarray*}}
\def\enn    {\end{eqnarray*}}
\def\bc      {\begin{comment}}
\def\ec      {\end{comment}}
\def\bf      {\begin{figure}}
\def\ef      {\end{figure}}
\def\bpm     {\begin{pmatrix}}
\def\epm     {\end{pmatrix}}
\def\bvm    {\begin{vmatrix}}
\def\evm    {\end{vmatrix}}

\def\ing    {\includegraphics}
\def\f      {\frac}
\def\bl      {\biggl}
\def\br     {\biggr}
\def\i         {\imath}
\def\limR    {\lim_{R\to0}}
\def\al         {\left |}
\def\ar         {\right |}
\def\tx        {\textrm}
\def\tw        {\textwidth}
\def\fs         {\footnotesize}
\def\ns         {\normalsize}
\def\no       {\nonumber}

\def\dag    {\dagger}
\def\i        {\imath}
\def\gr      {\nabla}

\def\D       {\mathcal{D}}
\def\O       {\mathcal{O}}

\def\M       {\mathcal{M}}

\def\tr        {\tx{tr}}

\def\k         {\bm{k}}
\def\x         {\bm{x}}
\def\p         {\bm{p}}
\def\q         {\bm{q}}
\def\l          {\bm{l}}
\def\c         {\bm{c}}
\def\sp        {\slashed{p}}

\def\sc        {\slashed{c}}
\def\su        {\slashed{u}}

\def\intx      {\int d^{3}\x}

\def\intt       {\int^{\beta}_{0}d\tau}

\begin{document}

\title{Two-parameter scaling theory of the longitudinal magnetoconductivity in a Weyl metal phase: Chiral anomaly, weak disorder, and finite temperature}
\author{Kyoung-Min Kim$^{1}$, Dongwoo Shin$^{1}$, M. Sasaki$^{2}$, Heon-Jung Kim$^{3}$, Jeehoon Kim$^{1,4}$, and Ki-Seok Kim$^{1}$}
\affiliation{$^{1}$Department of Physics, POSTECH, Pohang, Gyeongbuk 790-784, Korea \\ $^{2}$Department of Physics, Faculty of Science, Yamagata University, Kojirakawa, Yamagata 990-8560, Japan \\ $^{3}$Department of Physics, College of Natural Science, Daegu University, Gyeongbuk 712-714, Korea \\ $^{4}$Center for Artificial Low Dimensional Electronic Systems, Institute for Basic Science, 77 Cheongam-Ro, Nam-Gu, Pohang 790-784, Korea}

\date{\today}

\begin{abstract}
It is at the heart of modern condensed matter physics to investigate the role of a topological structure in anomalous transport phenomena. In particular, chiral anomaly turns out to be the underlying mechanism for the negative longitudinal magnetoresistivity in a Weyl metal phase. Existence of a dissipationless current channel causes enhancement of electric currents along the direction of a pair of Weyl points or applied magnetic fields ($B$). However, temperature ($T$) dependence of the negative longitudinal magnetoresistivity has not been understood yet in the presence of disorder scattering since it is not clear at all how to introduce effects of disorder scattering into the ``topological-in-origin" transport coefficient at finite temperatures. The calculation based on the Kubo formula of the current-current correlation function is simply not known for this anomalous transport coefficient. Combining the renormalization group analysis with the Boltzmann transport theory to encode the chiral anomaly, we reveal how disorder scattering renormalizes the distance between a pair of Weyl points and such a renormalization effect modifies the topological-in-origin transport coefficient at finite temperatures. As a result, we find breakdown of $B/T$ scaling, given by $B/T^{1 + \eta}$ with $0 < \eta < 1$. This breakdown may be regarded to be a fingerprint of the interplay between disorder scattering and topological structure in a Weyl metal phase.
\end{abstract}

%\pacs{71.10.Hf, 71.30.+h, 71.10.-w, 71.10.Fd}

\maketitle

\section{Introduction}

Researches on the role of topological-in-origin terms in quantum phases and their transitions have been a driving force for modern condensed matter physics, which cover quantum spin chains \cite{1D_QFT_Textbook} and deconfined quantum criticality \cite{DQCP_Senthil,DQCP_Tanaka}, quantum Hall effects and topological phases of matter \cite{Fradkin_Textbook}, Anderson localization for the classification of topological phases and their phase transitions \cite{Mirlin_Review}, and so on. In particular, renormalization effects of such topological terms are responsible for novel universality classes beyond the Landau-Ginzburg-Wilson paradigm of phase transitions with symmetry breaking. However, it is quite a nontrivial task to perform the renormalization group analysis in the presence of the topological-in-origin term, even if it can be taken into account perturbatively for the contribution of a bulk sometimes. Frequently, non-perturbative effects should be introduced into the renormalization group analysis \cite{PP_Transition_IQHE,Fu_Kane_NLsM}, uncontrolled in this situation and thus, being under debates as an open question.

In this study we investigate disorder-driven renormalization of a topological-in-origin term referred to as an inhomogeneous $\theta-$term in three spatial dimensions \cite{WM_Review_Kim}, defined by \bqa && \mathcal{F} = - \frac{1}{\beta} \int_{-\infty}^{\infty} d v(\bm{r}) P[v(\bm{r})] \ln \int D \bar{\psi}(\bm{r},\tau) D \psi(\bm{r},\tau) \nn && \exp\Big[ - \int_{0}^{\beta} d \tau \int d^{3} \bm{r} \Big\{ \bar{\psi}(\bm{r},\tau) i \gamma_{\mu} [\partial_{\mu} - i e A_{\mu}(\bm{r},\tau)] \psi(\bm{r},\tau) \nn && + v(\bm{r}) \bar{\psi}(\bm{r},\tau) \gamma_{\tau} \psi(\bm{r},\tau) - \frac{1}{4} F_{\mu\nu}(\bm{r},\tau) F_{\mu\nu}(\bm{r},\tau) \nn && + \theta(\bm{r}) \frac{e^{2}}{16 \pi^{2}} \varepsilon_{\mu\nu\gamma\delta} F_{\mu\nu}(\bm{r},\tau) F_{\gamma\delta}(\bm{r},\tau) \Big\} \Big] . \eqa $\psi(\bm{r},\tau)$ is a four-component Dirac spinor to describe an electron field of spin $1/2$ in two orbitals. Its dynamics is given by a Dirac theory, where $\gamma_{\mu}$ with $\mu = (\tau, x, y, z)$ is a Dirac matrix to satisfy the Clifford algebra. $A_{\mu}(\bm{r},\tau)$ and $F_{\mu\nu}(\bm{r},\tau) = \partial_{\mu} A_{\nu}(\bm{r},\tau) - \partial_{\nu} A_{\mu}(\bm{r},\tau)$ are an externally applied electromagnetic field and its field strength tensor, respectively. $v(\bm{r})$ is a potential configuration, given randomly and described by the Gaussian probability distribution $P[v(\bm{r})] = \mathcal{N} \exp\Big( - \int d^{3} \bm{r} \frac{[v(\bm{r})]^{2}}{2 \Gamma} \Big)$. $\Gamma$ is the variance of the disorder distribution and $\mathcal{N}$ is a normalization constant, determined by $\int_{-\infty}^{\infty} d v(\bm{r}) P[v(\bm{r})] = 1$. The last term is an inhomogeneous $\theta-$term, topological in its origin and keeping chiral anomaly that the chiral current is not conserved in the quantum level \cite{QFT_Textbook}, given by \bqa && \partial_{\mu} [\bar{\psi}(\bm{r},\tau) \gamma_{\mu} \gamma_{5} \psi(\bm{r},\tau)] = - \frac{e^{2}}{16 \pi^{2}} \varepsilon_{\mu\nu\gamma\delta} F_{\mu\nu}(\bm{r},\tau) F_{\gamma\delta}(\bm{r},\tau) . \nn \eqa $\gamma_{5}$ is chiral Dirac matrix to anticommute with $\gamma_{\mu}$. Here, the problem is how the inhomogeneous $\theta-$term becomes renormalized via the disorder scattering.

This problem can be cast into more physical terms. Introducing the chiral-anomaly equation into the effective field theory and performing the integration-by-parts for the chiral-current term with the $\theta(\bm{r})$ coefficient \cite{Kyoung}, we obtain \bqa && \mathcal{F} = - \frac{1}{\beta} \int_{-\infty}^{\infty} d v(\bm{r}) P[v(\bm{r})] \ln \int D \bar{\psi}(\bm{r},\tau) D \psi(\bm{r},\tau) \nn && \exp\Big[ - \int_{0}^{\beta} d \tau \int d^{3} \bm{r} \Big\{ \bar{\psi}(\bm{r},\tau) i \gamma_{\mu} [\partial_{\mu} - i e A_{\mu}(\bm{r},\tau)] \psi(\bm{r},\tau) \nn && + c_{\mu}(\bm{r},\tau) \bar{\psi}(\bm{r},\tau) \gamma_{\mu} \gamma_{5} \psi(\bm{r},\tau) + v(\bm{r}) \bar{\psi}(\bm{r},\tau) \gamma_{\tau} \psi(\bm{r},\tau) \nn && - \frac{1}{4} F_{\mu\nu}(\bm{r},\tau) F_{\mu\nu}(\bm{r},\tau) \Big\} \Big] . \eqa $c_{\mu}(\bm{r},\tau) = \partial_{\mu} \theta(\bm{r})$ is referred to as chiral gauge field, regarded to be a background potential given by the inhomogeneous $\theta$ coefficient. When the background chiral gauge field serves a homogeneous potential, the resulting spectrum turns out to describe dynamics of Weyl electrons. The right-handed helicity part shifts into the right hand side and the left-handed helicity part does into the left \cite{Weyl_Metal_I,Weyl_Metal_II,Weyl_Metal_III}. Physically, this homogeneous chiral-gauge-field potential is realized as $\bm{c} = \bm{\nabla} \theta(\bm{r}) = g \bm{B}$, applying a homogeneous magnetic field $\bm{B}$ into a gapless semiconductor described above. The Dirac point separates into a pair of Weyl points along the direction of the applied magnetic field and the distance of the pair of Weyl points is proportional to the strength of the applied magnetic field with a Lande$-g$ factor. See the supplementary material. As a result, the previous mathematically defined problem is actually how the background chiral gauge field, more physically, the distance between a pair of Weyl points becomes renormalized by random elastic scattering.

The renormalization effect of the distance between a pair of Weyl points is measurable experimentally since the information is encoded into the negative longitudinal magnetoresistivity. This anomalous transport phenomena in a Weyl metal phase has been well known for more than thirty years \cite{NLMR_Theory_Original} and experimentally confirmed firstly in 2013 \cite{NLMR_Experiment_Original}. The electrical resistivity measured along the direction of the applied magnetic field becomes smaller than that measured in other directions. More quantitatively, the magnetoconductivity is enhanced in the longitudinal setup, i.e., $\bm{E} ~ \| ~ \bm{B}$, as follows \bqa && \sigma_{L}(B) = \sigma_{D} (1 + \mathcal{C}_{W} B^{2}) , \eqa where $\bm{E}$ is an applied electric field \cite{Boltzmann_Spivak_Tong}. $\sigma_{D}$ is the Drude conductivity determined purely by disorder scattering. In real experiments, quantum corrections by weak anti-localization are introduced into the Drude conductivity \cite{Boltzmann_Kim_I}. $\mathcal{C}_{W}$ is a positive coefficient, discussed later in more detail. An essential point is that the enhancement of the longitudinal magnetoconductivity is given by the square of the distance between the pair of Weyl points. This longitudinal enhancement can be figured out in the following way: There exists a dissipationless current channel as a vacuum state, which connects the pair of Weyl points, responsible for the chiral anomaly. As a result, electrical currents are allowed to flow better along this direction through this vacuum channel although the measured longitudinal magnetoconductivity does not result from such dissipationless electrical currents \cite{NLMR_Theory_Original}. When the distance between the pair of Weyl points is renormalized by random elastic scattering, the positive coefficient $\mathcal{C}_{W}$ would evolve as a function of an energy scale, here, temperature. It is natural to expect finding a scaling theory for the chiral-anomaly-driven enhanced longitudinal magnetoconductivity.

The above discussion reminds us of a two-parameter scaling theory for the Anderson localization in topological phases of matter \cite{Altland_Two_Parameter_Scaling}, including the plateau-plateau transition in the integer quantum Hall effect \cite{PP_Transition_IQHE}. There, the transport phenomenon of the Anderson localization transition is determined by the ``transverse" conductivity $\sigma_{xx}$ and the Hall conductivity $\sigma_{xy}$, where the latter encodes the topological information of the integer quantum Hall effect. The present situation is quite analogous to that of the integer quantum Hall effect. $\sigma_{xx}$ in the quantum Hall effect is identified with the Drude conductivity $\sigma_{D}$, determined by disorder scattering directly. On the other hand, $\sigma_{xy}$ in the quantum Hall effect is analogous to the distance between the pair of Weyl points, where the renormalization effect is introduced into the temperature dependence of $\mathcal{C}_{W}$.

In this study we investigate the longitudinal magnetoconductivity at finite temperatures and find a two-parameter scaling theory, where renormalization effects result from random elastic scattering. There is one difficult point in the calculation of the longitudinal magnetoconductivity in a Weyl metal phase. It turns out that a naive Kubo-formula calculation does not incorporate the role of the chiral anomaly in the longitudinal magnetoconductivity \cite{TFLT_Kim}. As a result, we fail to find the $B^{2}$ enhancement of the longitudinal magnetoconductivity within the Kubo-formula calculation. In this respect our strategy consists of a two-fold way: First, we perform the renormalization group analysis and find how the distance between a pair of Weyl points evolves as a function of an energy scale or temperature. Second, introducing this information into the Boltzmann transport theory with chiral anomaly, we reveal the longitudinal negative magnetoconductivity as a function of both the applied magnetic field and temperature, given by \bqa && \sigma_{L}(B,T) \approx \sigma_{D}(T) [1 + \mathcal{C}_{W}(T) B^{2}] . \eqa In particular, we find breakdown of $B/T$ scaling \bqa && \Delta \sigma_{L}(B,T) \equiv \frac{\sigma_{L}(B,T) - \sigma_{D}(T)}{\sigma_{D}(T)} = \mathcal{C}_{W} T_{0}^{2 (1 + \eta)} \Big( \frac{B}{T^{1 + \eta}} \Big)^{2} , \nn \eqa where $\eta$ is a scaling exponent with $0 < \eta < 1$ and $T_{0}$ is an energy scale. We claim that this breakdown may be regarded to be a fingerprint of the interplay between disorder scattering and topological structure in a Weyl metal phase.

\section{Renormalization for the distance between a pair of Weyl points via disorder-driven inter-valley scattering}

\subsection{Effective field theory for a Weyl metal phase with disorder: Replica theory}

We start from an effective Hamiltonian density for a Weyl metal phase with time reversal symmetry breaking
\bqa
\mathcal{H}_{B} = \psi^{\dag}_{B}(\x) \Big( v_{B} \bm{\alpha} \cdot (-\i\gr) + g_{B} \bm{B}\cdot\bm{\sigma}\otimes I_{2\times2} \Big) \psi_{B}(\x) .
\eqa
$\psi_{B}(\x) = (\psi_{B R}(\x),\psi_{B L}(\x))^{T}$ is a four-component Dirac-spinor field in a two-component Weyl-spinor field with right(R)-left(L) chirality, and $v_{B}$ is the velocity of such fermions. $\bm{B}$ is an externally applied magnetic field with a Lande$-g$ factor $g_{B}$, splitting the Dirac band into a pair of Weyl bands (Fig. \ref{f.bandstructure}). $\bm{\alpha}$ is a four-by-four matrix, given by $\bm{\alpha} = \bm{\sigma} \otimes \sigma_{z}$, where $\bm{\sigma}$ is a Pauli matrix. The subscript $B$ denotes ``bare", meaning that this effective Hamiltonian density is defined at an ultraviolet (UV) scale.

\bf[b]
\ing[width=0.35\tw]{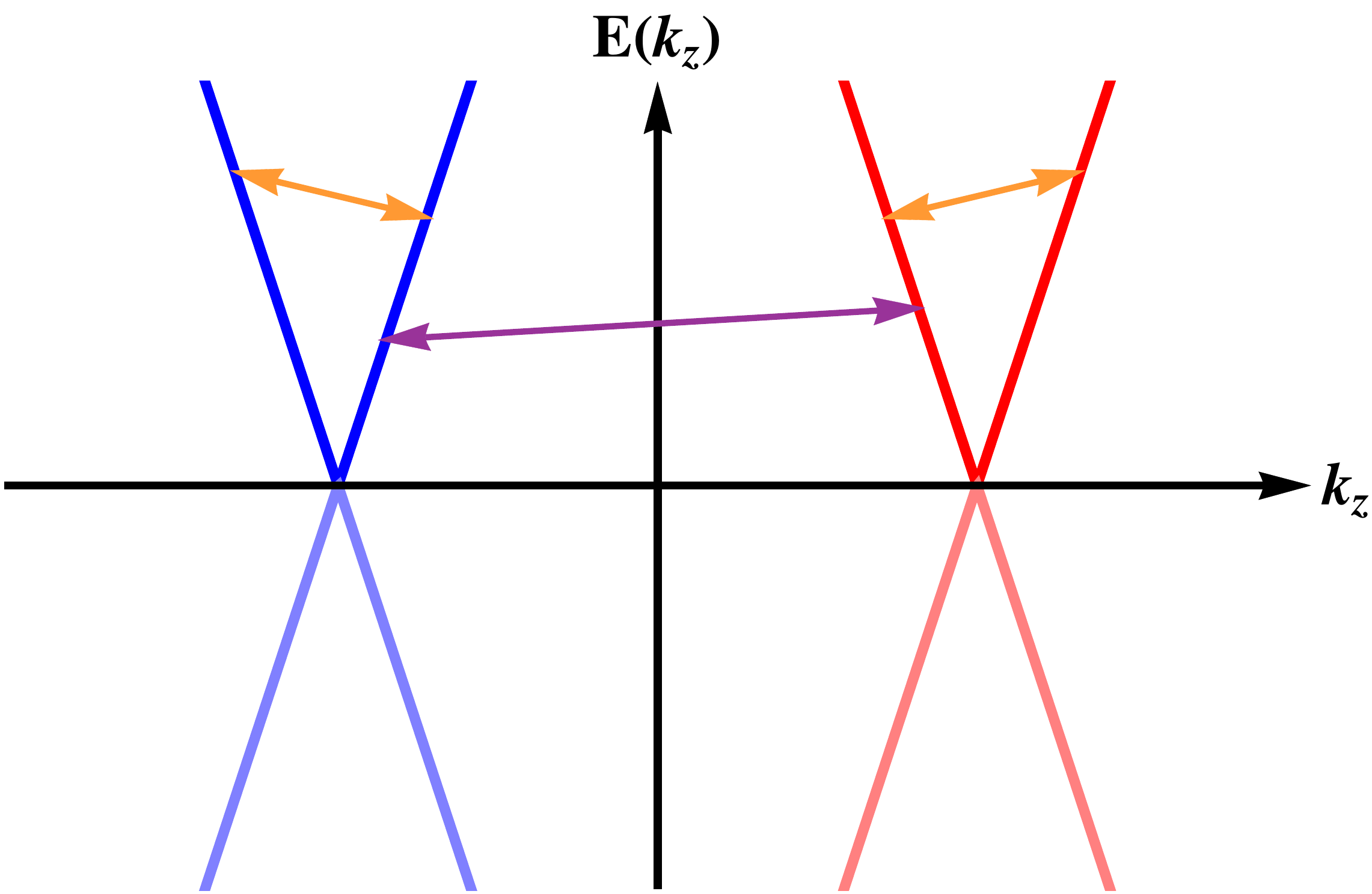} \caption{A band structure of a Weyl metal phase, projected on the plane of $k_{x} = k_{y} = 0$. Here, the direction of an applied magnetic field is the $z$-axis. Each band has definite chirality: $-1$ for the blue cone and $+1$ for the red cone. The orange arrows represent intra-valley scattering while the purple arrow stands for inter-valley scattering.} \label{f.bandstructure}
\ef

We consider two types of random potentials, introducing ``intra-valley scattering" $\psi_{B}^{\dag}(\x)V_{B}(\x)\psi_{B}(\x)$ and ``inter-valley scattering" $\psi_{B}^{\dag}(\x)U_{B}(\x)\big(I_{2\times2}\otimes\sigma_{x}\big)\psi_{B}(\x)$ into the effective Hamiltonian. Then, we obtain the following effective action
\bqa
&& S_{B}[\bar{\psi}_{B}(x),\psi_{B}(x);V_{B}(\x),U_{B}(\x)] \nn && = \int d^{4} x \big\{\bar{\psi}_{B}(x)\big(\gamma^{0}\partial_{0}+\i v_{B}\gamma^{k}\partial_{k}+c_{{B}\mu}\gamma^{\mu}\gamma^{5}\big)\psi_{B}(x) \nn && + \bar{\psi}_{B}(x) \gamma^{0} V_{B}(\x) \psi_{B}(x) + \bar{\psi}_{B}(x) U_{B}(\x) \psi_{B}(x)\big\}
\eqa
with $\bar{\psi}_{B}(x) \equiv \psi_{B}^{\dag}(x) \gamma^{0}$. Here, gamma matrices are given in the Weyl representation, for example, $\gamma^{0} = I_{2\times2} \otimes \sigma_{x}$. A magnetic field is generalized to be a chiral gauge field $c_{{B}\mu}=(c_{B0},c_{Bk} \equiv g_{B} B_{k})$. $x$ means ``space-time", given by $x^{\mu}=(\tau,\x)$. See the supplementary material.

A physical observable in this system is measured as follows
\begin{widetext}
\bqa
&& \left \langle \O[\bar{\psi}_{B}(x),\psi_{B}(x)] \right \rangle = \int\D V_{B}(\x) \D U_{B}(\x) P_{B}[V_{B}(\x),U_{B}(\x)] \nn && \f{ \int\D \bar{\psi}_{B}(x) \D \psi_{B}(x) \O[\bar{\psi}_{B}(x),\psi_{B}(x)]e^{-S_{B 0}[\bar{\psi}_{B}(x),\psi_{B}(x)]}e^{-\int d^{4} x \bar{\psi}_{B}(x) [\gamma^{0} V_{B}(\x) + U_{B}(\x) ] \psi_{B}(x)}}{\int\D \bar{\psi}_{B}(x) \D \psi_{B}(x) e^{-S_{B 0}[\bar{\psi}_{B}(x),\psi_{B}(x)]}e^{-\int d^{4} x \bar{\psi}_{B}(x)[\gamma^{0}V_{B}(\x)+U_{B}(\x)]\psi_{B}(x)}} ,
\eqa
\end{widetext}
where the free part of the effective action is $S_{B 0}[\bar{\psi}_{B}(x),\psi_{B}(x)] = \int d^{4} x \bar{\psi}_{B}(x) \big( \gamma^{0} \partial_{0} + \i v_{B} \gamma^{k} \partial_{k} + c_{B \mu} \gamma^{\mu} \gamma^{5} \big) \psi_{B}(x)$. Resorting to the replica trick and performing the average for disorder with the Gaussian distribution function of $P_{B}[V_{B}(\x),U_{B}(\x)] = N_{B} \exp{\Big[- \f{\intx V_{B}^{2}(\x)}{2 \Gamma_{B V}} - \f{\intx U_{B}^{2}(\x)}{2\Gamma_{B U}}\Big]}$, the above expression is reformulated as follows
\bqa
&& \left \langle \O[\bar{\psi}_{B}(x),\psi_{B}(x)] \right \rangle = \limR\f{1}{R} \sum_{a=1}^{R} \int\D \bar{\psi}_{B}^{a}(x) \D\psi_{B}^{a}(x) \nn && \O[\bar{\psi}_{B}^{a}(x),\psi_{B}^{a}(x)] \exp \Big\{- \sum_{a=1}^{R} S_{B 0}[\bar{\psi}_{B}^{a}(x),\psi_{B}^{a}(x)] \nn && - \sum_{b,c=1}^{R} S_{B \tx{dis}}[\bar{\psi}_{B}^{b}(x),\psi_{B}^{b}(x),\bar{\psi}_{B}^{c}(x),\psi_{B}^{c}(x)] \Big\} . \label{e.observable}
\eqa
Here, $N_{B}$ is a normalization constant and $\Gamma_{B V(U)}$ is a variance for the disorder distribution. As a result, the effective interaction term induced by disorder scattering is
\begin{widetext}
\bea
S_{B \tx{dis}}[\bar{\psi}_{B}^{b}(x),\psi_{B}^{b}(x),\bar{\psi}_{B}^{c}(x),\psi_{B}^{c}(x)] &=& - \intt\intt'\int d^{3}\x\f{\Gamma_{B V}}{2} \bar{\psi}_{B}^{b}(\tau,\x) \gamma^{0} \psi_{B}^{b}(\tau,\x) \bar{\psi}_{B}^{c}(\tau',\x) \gamma^{0} \psi_{B}^{c}(\tau',\x) \nn && - \intt\intt'\int d^{3}\x\f{\Gamma_{B U}}{2} \bar{\psi}_{B}^{b}(\tau,\x) \psi_{B}^{b}(\tau,\x) \bar{\psi}_{B}^{c}(\tau',\x) \psi_{B}^{c}(\tau',\x) . \label{e.effectivetheory}
\eea
\end{widetext}
The effective field theory is given by $S_{B}[\bar{\psi}_{B}^{a}(x),\psi_{B}^{a}(x)] = S_{B 0}[\bar{\psi}_{B}^{a}(x),\psi_{B}^{a}(x)] + S_{B \tx{dis}}[\bar{\psi}_{B}^{b}(x),\psi_{B}^{b}(x),\bar{\psi}_{B}^{c}(x),\psi_{B}^{c}(x)]$.

\subsection{Renormalization group analysis: Role of inter-valley scattering in the distance between a pair of Weyl point}

In order to perform the renormalization group analysis within the dimensional regularization \cite{QFT_Textbook}, we rewrite $S_{B}[\bar{\psi}_{B}^{a}(x),\psi_{B}^{a}(x)]$, the effective bare action of bare field variables in terms of $S_{R}[\bar{\psi}_{R}^{a},\psi_{R}^{a}]$, the effective renormalized action of renormalized field variables with $S_{CT}[\bar{\psi}_{R}^{a},\psi_{R}^{a}]$, counter terms of renormalized field variables
\begin{widetext}
\bea
S_{R}[\bar{\psi}_{R}^{a},\psi_{R}^{a}] &=&\int d^{d+1}x\bar{\psi}_{R}^{a}\big(\gamma^{0}\partial_{0}+ v_{R}\i\gamma^{k}\partial_{k}+c_{R0}\gamma^{0}\gamma^{5}+c_{Rk}\gamma^{k}\gamma^{5}\big)\psi_{R}^{a}\no\\
&-& \int d\tau
\int d\tau'\int d^{d}\x\f{\Gamma_{RV}}{2}(\bar{\psi}_{R}^{b}\gamma^{0}\psi_{R}^{b})_{\tau}(\bar{\psi}_{R}^{c}\gamma^{0}\psi_{R}^{c})_{\tau'}-\int d\tau
\int d\tau'\int d^{d}\x\f{\Gamma_{RU}}{2}(\bar{\psi}_{R}^{b}\psi_{R}^{b})_{\tau}(\bar{\psi}_{R}^{c}\psi_{R}^{c})_{\tau'},\no\\
S_{CT}[\bar{\psi}_{R}^{a},\psi_{R}^{a}] &=&\int d^{d+1}x\bar{\psi}_{R}^{a}\big(\delta_{\psi}^{\omega}\gamma^{0}\partial_{0}+\delta_{\psi}^{\k}v_{R}\i\gamma^{k}\partial_{k} +\delta_{c0}c_{R0}\gamma^{0}\gamma^{5}+\delta_{\c}c_{Rk}\gamma^{k}\gamma^{5}\big)\psi_{R}^{a}\no\\
&-& \int d\tau\int d\tau'\int d^{d}\x\f{\delta_{\Gamma V}\Gamma_{RV}}{2}(\bar{\psi}_{R}^{b}\gamma^{0}\psi_{R}^{b})_{\tau}(\bar{\psi}_{R}^{c}\gamma^{0}\psi_{R}^{c})_{\tau'}-\int d\tau\int d\tau'\int d^{d}\x\f{\delta_{\Gamma U}\Gamma_{RU}}{2}(\bar{\psi}_{R}^{b}\psi_{R}^{b})_{\tau}(\bar{\psi}_{R}^{c}\psi_{R}^{c})_{\tau'} \label{e.RGtheory} , \nn
\eea
\end{widetext}
where $S_{B}[\bar{\psi}_{B}^{a},\psi_{B}^{a}] = S_{R}[\bar{\psi}_{R}^{a},\psi_{R}^{a}] + S_{CT}[\bar{\psi}_{R}^{a},\psi_{R}^{a}]$. It is straightforward to see how bare quantities are related with renormalized ones, given by
\bqa
&& \psi_{B}^{a} = (Z_{\psi}^{\omega})^{\f{1}{2}} \psi_{R}^{a} , ~~~ v_{B} = Z_{\psi}^{\k} (Z_{\psi}^{\omega})^{-1} v_{R} , \nn && c_{B0} = Z_{c0} (Z_{\psi}^{\omega})^{-1} c_{R0} , ~~~ c_{Bk} = Z_{\c} (Z_{\psi}^{\omega})^{-1} c_{Rk} , \nn && \Gamma_{BV} = Z_{\Gamma V} (Z_{\psi}^{\omega})^{-2} \Gamma_{RV} , ~~~ \Gamma_{BU} = Z_{\Gamma U} (Z_{\psi}^{\omega})^{-2} \Gamma_{RU} , \nn \label{r.renormalization_equations}
\eqa
where $Z_{\psi}^{\omega} = 1 + \delta_{\psi}^{\omega}$, $Z_{\psi}^{\k} = 1 + \delta_{\psi}^{\k}$, $Z_{c0} = 1 + \delta_{c0}$, $Z_{\c} = 1 + \delta_{\c}$, $Z_{\Gamma V} = 1 + \delta_{\Gamma V}$, and $Z_{\Gamma U} = 1 + \delta_{\Gamma U}$.

\bf
\ing[width=0.5\tw]{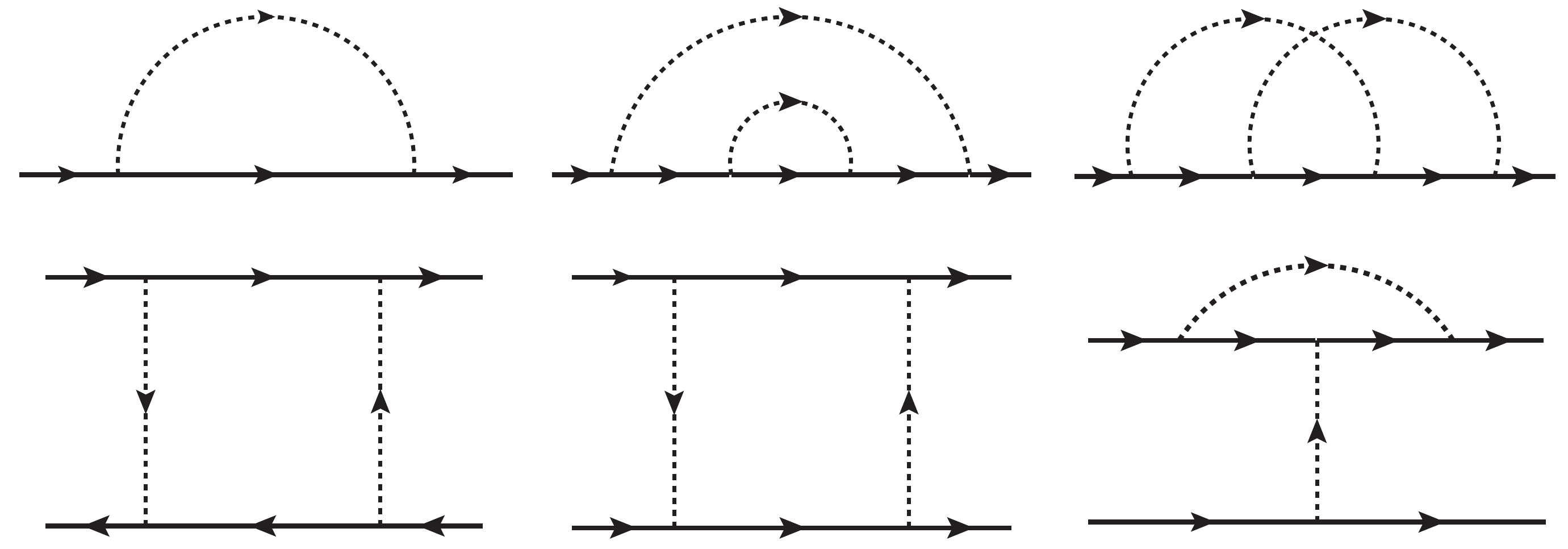}
\caption{Feynman's diagrams up to the two-loop order for self-energy corrections and the one-loop order for vertex corrections. The other diagrams disconnected to external lines or including fermion loops vanish identically in the replica limit of $R\rightarrow0$. Here, we show quantum corrections only due to intra-valley scattering, represented by single-dashed lines. In order to include inter-valley scattering, we just replace single-dashed lines with double-dashed lines one by one according to our Feynman rules. This replacement results in another Fock-type diagram and three more diagrams each for rainbow-type, crossed-type and vertex corrections.} \label{f.diagrams}
\ef

Dimensional analysis gives $\tx{dim}[\Gamma]=2-d$. In this respect we perform the dimensional regularization in $d=2+\varepsilon$ where $\varepsilon$, a ``small" parameter to control the present renormalization group analysis, will be analytically continued to $\varepsilon=1$ in the end. Performing the standard procedure for the renormalization group analysis, we find renormalization group equations, where both vertex and self-energy corrections are introduced self-consistently. See Fig. \ref{f.diagrams}, where all quantum corrections are shown as Feynman's diagrams up to the two-loop order for self-energy corrections and the one-loop order for vertex corrections. All details are shown in the supplementary material. As a result, we find counter terms with
\bea
&&\delta_{\psi}^{\omega}=\f{\Gamma_{V}+\Gamma_{U}}{2\pi\varepsilon}-\f{5\Gamma_{V}^{2}+12\Gamma_{V}\Gamma_{U}+7\Gamma_{U}^{2}}{48\pi^{2}\varepsilon},\no\\
&&\delta_{\c}=\f{\Gamma_{V}^{2}-\Gamma_{U}^{2}}{16\pi^{2}\varepsilon},~~\delta_{\Gamma V}=\f{\Gamma_{V}+\Gamma_{U}}{2\pi\varepsilon},~~\delta_{\Gamma U}=-\f{\Gamma_{V}+\Gamma_{U}}{2\pi\varepsilon} , \nn
%&&\delta_{c0}=\f{\Gamma_{V}-\Gamma_{U}}{2\pi\varepsilon}-\f{5\Gamma_{V}^{2}-14\Gamma_{V}\Gamma_{U}+5\Gamma_{U}^{2}}{48\pi^{2}\varepsilon},~~\delta_{\psi}^{\k}=-\f{(\Gamma_{V}+\Gamma_{U})^{2}}{16\pi^{2}\varepsilon}
\label{r.counterterms}
\eea

Inserting these divergent coefficients into equations (\ref{r.renormalization_equations}) and performing derivatives with respect to an energy scale for renormalization given by $\ln M$, we find renormalization group equations
\bea
\f{d\Gamma_{V}}{d\ln{M}}&=&\Gamma_{V}-\f{a_{\Gamma}}{3}\Gamma_{V}(\Gamma_{V}+\Gamma_{U}) \nn &+& b_{\Gamma}\Gamma_{V}(\Gamma_{V}+\Gamma_{U})(c_{\Gamma}\Gamma_{V}+\Gamma_{U}) , \nn
\f{d\Gamma_{U}}{d\ln{M}}&=&\Gamma_{U}-a_{\Gamma}\Gamma_{U}(\Gamma_{V}+\Gamma_{U}) \nn &+& b_{\Gamma}\Gamma_{U}(\Gamma_{V}+\Gamma_{U})(c_{\Gamma}\Gamma_{V}+\Gamma_{U}) , \nn
\f{dc_{k}}{d\ln{M}}&=&c_{k}\Big[-1-a_{\c}(\Gamma_{V}+\Gamma_{U}) \nn &+& b_{\c}(\Gamma_{V}+\Gamma_{U})(2\Gamma_{V}+\Gamma_{U})\Big] , \label{r.RGeq}
\eea
where positive numerical constants are given by
\bnn
a_{\Gamma}=\f{3}{2\pi},~~b_{\Gamma}=\f{7}{24\pi^{2}},~~c_{\Gamma}=\f{5}{7},~~a_{\c}=\f{1}{2\pi},~~b_{\c}=\f{1}{12\pi^{2}} .
\enn

Fig. \ref{f.rgflow} shows renormalization group flows for physical parameters according to Eq. (\ref{r.RGeq}). In the plane of $(\Gamma_{V},\Gamma_{U})$, we find two stable fixed points corresponding to two phases of a disordered Weyl metal state, and one unstable fixed point corresponding to the phase transition point between two phases: (1) The stable fixed point of $(0, \Gamma_{0})$ with $\Gamma_{0} = 0$ represents a clean Weyl metal phase, protected for the case of weak disorder by the pseudogap density of states of the Weyl metal state. (2) The stable fixed point of $(0, \Gamma_{2})$ with $\Gamma_{2}=\f{a_{\Gamma}+\sqrt{a_{\Gamma}^{2}-4b_{\Gamma}}}{2b_{\Gamma}} \simeq 13.68$ is identified with a diffusive Weyl metal phase, analogous to the diffusive Fermi-liquid fixed point of a conventional metallic phase \cite{CMP_QFT_Textbook}. (3) The unstable fixed point of $(0, \Gamma_{1})$ with $\Gamma_{1}=\f{a_{\Gamma}-\sqrt{a_{\Gamma}^{2}-4b_{\Gamma}}}{2b_{\Gamma}} \simeq 2.09$ denotes a critical point to separate the diffusive Weyl metal phase from the clean Weyl metal state, the existence of which originates from the pseudogap density of states. Interestingly, all these fixed points lie at the line of $\Gamma_{V}=0$, which means that inter-valley scattering shows dominant effects over intra-valley scattering for the low-energy physics in the disordered Weyl metallic state. Naively, one may suspect that their roles are similar because of the similarity of their renormalization group equations. However, the magnitude of the one-loop correction for $\Gamma_{U}$ turns out to be three times larger than that for $\Gamma_{V}$, and thus, the renormalization group flow of $(\Gamma_{V},\Gamma_{U})$ is overwhelmed by $\Gamma_{U}$. As a result, there is no chance by which $\Gamma_{V}$ has a non-trivial fixed point value. Detailed analysis on this issue is given in the supplementary material (Fig. \ref{f.VU_flow}).

In order to figure out how the distance between the pair of Weyl points renormalizes as a function of an energy scale, we focus on renormalization group equations for $\Gamma_{U}$ and $c_{k}$ at $\Gamma_{V} = 0$
\bqa
\f{d\Gamma_{U}}{d\ln{M}}&=&\Gamma_{U}-a_{\Gamma}\Gamma_{U}^{2}+b_{\Gamma}\Gamma_{U}^{3}\\
\f{dc_{k}}{d\ln{M}}&=&c_{k}\big[-1-a_{\c}\Gamma_{U}+b_{\c}\Gamma_{U}^{2}\big] .
\eqa
It is straightforward to solve the first equation and find an approximate solution for $\Gamma_{U}$ near each fixed point at $\Gamma_{V} = 0$. Inserting such fixed-point solutions into the second equation, we uncover how the distance between the pair of Weyl points evolves as a function of temperature
\be
c_{k}(T) = c_{k}(T_{0})\bigg(\f{T_{0}}{T}\bigg)^{\lambda_{\c,fn}} , \label{r.cfield}
\ee
where the energy scale $M$ has been replaced with temperature $T$. Critical exponents of $\lambda_{\c,fn}$ are found to be
\bqa
\lambda_{\c,f0}&=&1+a_{\c}\Gamma_{0}-b_{\c}\Gamma_{0}^{2}=1\\
\lambda_{\c,f1}&=&1+a_{\c}\Gamma_{1}-b_{\c}\Gamma_{1}^{2}\simeq1.34\\
\lambda_{\c,f2}&=&1+a_{\c}\Gamma_{2}-b_{\c}\Gamma_{2}^{2}\simeq1.60.
\eqa
It turns out that the distance between a pair of Weyl points increases to reach infinity, regarded to be beyond the perturbative renormalization group analysis. However, the infinity should be considered as an artifact of the continuum approximation. If the Brillouin zone is taken into account in the effective field theory, there must be a maximum of the distance within the Brillouin zone. In this respect it is natural to modify the above scaling solution as follows
\be
c_{k}(T) = c_{k}(T_{0})\bigg(\f{T_{0}}{T + T_{M}}\bigg)^{\lambda_{\c,fn}} , \label{r.cfield_modification}
\ee
where $T_{M}$ is a cutoff scale in the low-energy limit. It is interesting to notice that disorder scattering changes the temperature-dependent exponent of $c_{k}$. Inter-valley scattering gives rise to fast enhancement of the distance between a pair of Weyl points at low temperatures. This looks counter-intuitive, where anti-screening instead of screening arises from inter-valley scattering.

\bf
\ing[width=0.5\tw]{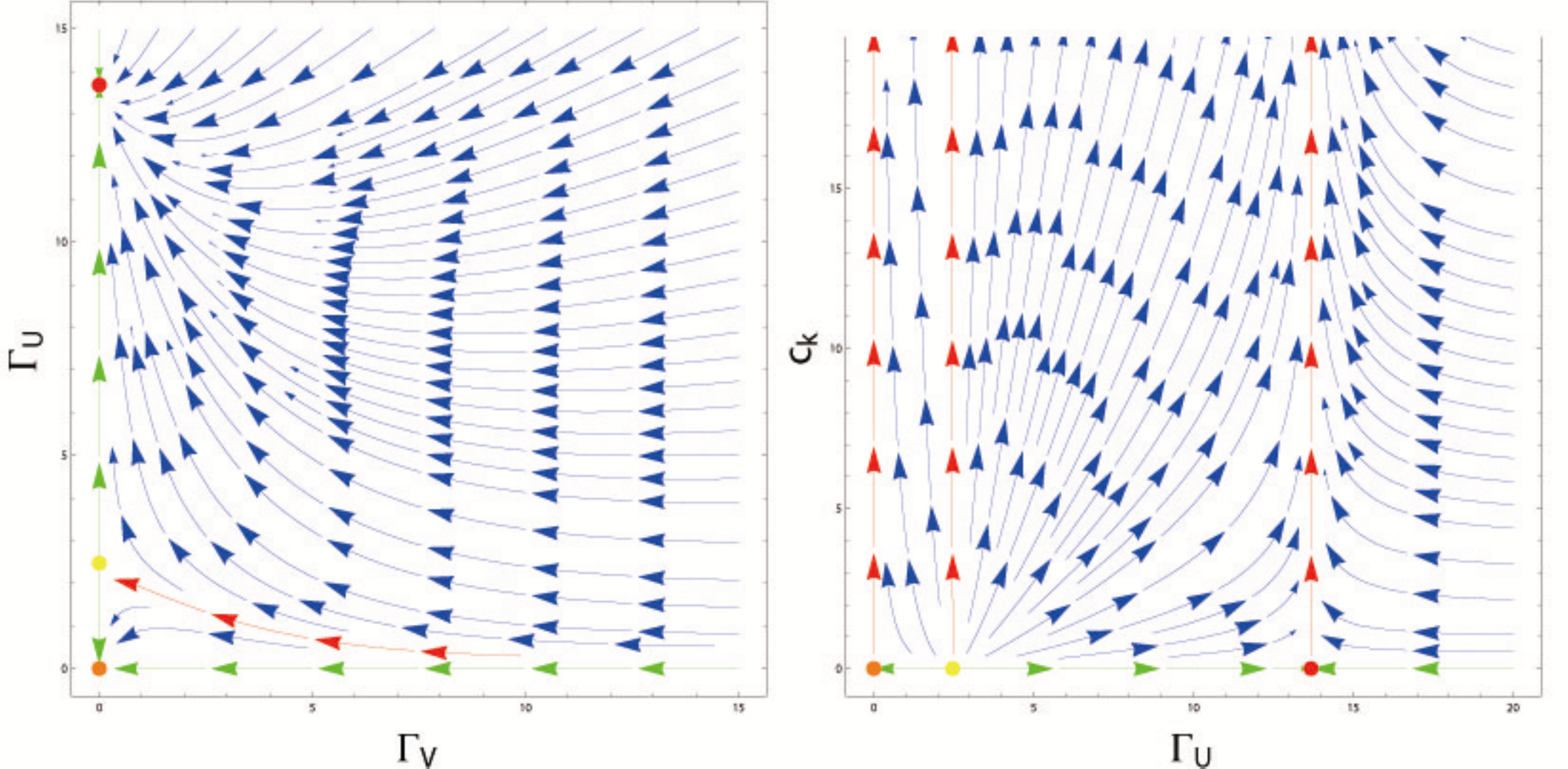}
\caption{Renormalization group flows for physical parameters. In the plane of $(\Gamma_{V},\Gamma_{U})$ (Left), there are two stable fixed points and one unstable fixed point: (1) The stable fixed point of $(0, \Gamma_{0})$ with $\Gamma_{0} = 0$ represents a clean Weyl metal phase, protected for the case of weak disorder by the pseudogap density of states of the Weyl metal state. (2) The stable fixed point of $(0, \Gamma_{2})$ with $\Gamma_{2}=\f{a_{\Gamma}+\sqrt{a_{\Gamma}^{2}-4b_{\Gamma}}}{2b_{\Gamma}} \simeq 13.68$ is identified with a diffusive Weyl metal phase, analogous to the diffusive Fermi-liquid fixed point of a conventional metallic phase. (3) The unstable fixed point of $(0, \Gamma_{1})$ with $\Gamma_{1}=\f{a_{\Gamma}-\sqrt{a_{\Gamma}^{2}-4b_{\Gamma}}}{2b_{\Gamma}} \simeq 2.09$ denotes a critical point to separate the diffusive Weyl metal phase from the clean Weyl metal state, the existence of which originates from the pseudogap density of states. In the plane of $(\Gamma_{U},c_{k})$ (Right), the renormalization group flow shows a run-away behavior for $c_{k}$, implying that the Weyl metallic state is stabilized even in the presence of disorder scattering. This run-away flow should stop at a certain energy scale if the Brillouin zone is taken into account in the effective field theory. } \label{f.rgflow}
\ef

\section{Two-parameter scaling theory for the longitudinal magnetoconductivity of a disordered Weyl metal phase within Boltzmann transport theory}

The question to address in this study is to find a scaling theory for the longitudinal magnetoconductivity. As discussed in the introduction, not only the Drude conductivity but also the distance between a pair of Weyl points or the spatial gradient of the inhomogeneous $\theta(\bm{r})$ coefficient in the topological-in-origin $\bm{E} \cdot \bm{B}$ term should be taken into account for the longitudinal magnetoconductivity in the Weyl metal phase. This situation is analogous to that of a plateau-plateau transition in the integer quantum Hall effect: Not only the Drude conductivity but also the Hall conductivity, a topological $\theta-$term, should be considered on equal footing in order to describe such a quantum phase transition involved with Anderson localization. In this respect we call the scaling theory for the longitudinal magnetoconductivity of a disordered Weyl metal phase two-parameter scaling theory as the Anderson localization transition in the case of the quantum Hall effect.

Previously, we found $\Gamma_{U}(T)$ and $c(T)$, based on the perturbative renormalization group analysis, where $\Gamma_{U}(T)$ gives the Drude conductivity and $c(T)$ describe the enhancement of the longitudinal magnetoconductivity. More precisely, we can address renormalization effects of the longitudinal magnetoconductivity based on the Boltzmann transport theory for a Weyl metal phase \cite{Boltzmann_Kim_I,Boltzmann_Kim_II}
\bqa &&
\frac{\partial n_{\chi}(\bm{p};\bm{r},t)}{\partial t} + \bm{\dot{r}}_{\chi} \cdot \bm{\nabla}_{\bm{r}} n_{\chi}(\bm{p};\bm{r},t) + \bm{\dot{p}}_{\chi} \cdot \bm{\nabla}_{\bm{p}} n_{\chi}(\bm{p};\bm{r},t) \nn && = I_{coll} [ n_{\chi}(\bm{p};\bm{r},t)] .
\eqa
Here, $n_{\chi}(\bm{p};\bm{r},t)$ is the distribution function at a chiral Fermi surface denoted by $\chi = \pm$, where $\bm{p}$ is the relative momentum of a particle-hole pair near the chiral Fermi surface, and $\bm{r}$ and $t$ are the center of mass position and time of the particle-hole pair.

$\bm{\dot{r}}_{\chi}$ and $\bm{\dot{p}}_{\chi}$ represent the change of position and momentum with respect to time, classically described and given by the so called modified Drude model \cite{AHE_Review_I,AHE_Review_II}
\bqa
&& \dot {\bm x}_F^\chi = {\bm v}_F^\chi + \dot {\bm p}_F^\chi \times \boldsymbol{\mathcal{B}}_F^\chi , \nn && \dot {\bm p}_F^\chi = {\bm E} + \dot {\bm x}_F^\chi \times {\bm B} . \eqa
$\boldsymbol{\mathcal{B}}_F^\chi$ represents a momentum-space magnetic field on the chiral Fermi surface, resulting from a momentum-space magnetic charge $\chi$ enclosed by the chiral Fermi surface. We would like to recall that the Berry curvature does not appear on the ``normal" Fermi surface that does not enclose a band-touching point. As a result, we reproduce the Drude model with $\boldsymbol{\mathcal{B}}_F^\chi = 0$. It is essential to realize the following relation between the applied magnetic field and the distance between the pair of Weyl points \bqa && \bm{B} \rightarrow g^{-1} \bm{c}(T) . \eqa It is straightforward to solve these coupled equations, the solution of which is
\bqa && \dot {\bm x}_F^\chi \approx G^\chi_3(T) \Big[ {\bm v}_F^\chi + {\bm E} \times \boldsymbol{\mathcal{B}}_F^\chi + g^{-1} \big( \boldsymbol{\mathcal{B}}_F^\chi \cdot {\bm v}_F^\chi \big) {\bm c}(T) \Big] , \nn && \dot {\bm p}_F^\chi \approx G^\chi_3(T) \Big[ {\bm E} + g^{-1} {\bm v}_F^\chi \times {\bm c}(T) + g^{-1} \big( {\bm E} \cdot {\bm c}(T) \big) \boldsymbol{\mathcal{B}}_F^\chi \Big] , \nn \eqa where $G_3^{\chi} = \big(1 + g^{-1} \boldsymbol{\mathcal{B}}_F^\chi \cdot {\bm c}(T) \big)^{-1}$ is a volume factor of the modified phase space with a pair of momentum-space magnetic charges $\chi = \pm$. They are well known the role of anomalous electromagnetic-field-dependent terms in anomalous transport phenomena: (1) The second term of ${\bm E} \times \boldsymbol{\mathcal{B}}_F^\chi$ in the first equation is responsible for the anomalous Hall effect, the Hall effect without an applied magnetic field due to an emergent magnetic field referred to as Berry curvature in the momentum space \cite{AHE_I,AHE_II,AHE_III}. (2) The third term of $g^{-1} \big( \boldsymbol{\mathcal{B}}_F^\chi \cdot {\bm v}_F^\chi \big) {\bm c}(T)$ in the first equation gives rise to the so called chiral magnetic effect that dissipationless electric currents are driven by applied magnetic fields in the limit of vanishing applied electric fields, proportional to the distance between the pair of Weyl points or applied magnetic fields \cite{CME_I,CME_II,CME_III,CME_IV,CME_V,CME_VI}. (3) The third term of $g^{-1} \big( {\bm E} \cdot {\bm c}(T) \big) \boldsymbol{\mathcal{B}}_F^\chi$ in the second equation causes the gauge anomaly for electrons on each chiral Fermi surface that gauge or electric currents on each chiral Fermi surface are not conserved \cite{NLMR_Theory_Original,NLMR_Experiment_Original,Boltzmann_Spivak_Tong,Boltzmann_Kim_I,Boltzmann_Kim_II,CME_II,CME_III,ABJ_Anomaly_I,ABJ_Anomaly_II,ABJ_Anomaly_III,ABJ_Anomaly_IV,ABJ_Anomaly_V,ABJ_Anomaly_VI,ABJ_Anomaly_VII,ABJ_Anomaly_VIII}. Of course, the breakdown of the gauge symmetry should be cured when total electric currents are considered, but chiral ``electric" currents are still not conserved, referred to as chiral anomaly.

The collision part is given by
\bqa I_{coll} [\delta n_{\chi}(\bm{p};\bm{r},t)] &=& - \frac{n_{\chi}(\bm{p};\bm{r},t) - n_{\chi}^{eq}(\bm{p})}{\tau_{intra}(T)} \nn &-& \frac{n_{\chi}(\bm{p};\bm{r},t) - n_{- \chi}(\bm{p};\bm{r},t)}{\tau_{inter}(T)} . \eqa
The first term describes the intra-valley scattering, and the second represents the inter-valley scattering. In this respect both scattering rates of $1/\tau_{intra}(T)$ and $1/\tau_{intra}(T)$ correspond to $\Gamma_{V}(T)$ and $\Gamma_{U}(T)$, respectively.

Considering homogeneity of the Weyl metal phase under constant electric fields in the dc-limit, we are allowed to solve $\bm{\dot{p}}_{\chi} \cdot \bm{\nabla}_{\bm{p}} n_{\chi}(\bm{p}) = I_{coll} [ n_{\chi}(\bm{p}) ]$. As a result, we find a two-parameter scaling theory for the longitudinal magnetoconductivity in a disordered Weyl metal phase
\bqa && \sigma_{L}(B,T) = \sigma_{D}(B,T) \big( 1 + const. [c(B,T)]^{2} \big) , \eqa
where $\sigma_{D}(B,T)$ is the Drude conductivity inversely proportional to $\Gamma_{U}(T)$ and $c(B,T)$ is the distance between a pair of Weyl points.

Rewriting the distance between the pair of Weyl points as $c(B,T) \equiv C_{W}^{1/2}(T) B$, we consider
\bqa \Delta\sigma_{L}(B,T) &\equiv& \f{\sigma_{L}(B,T)-\sigma_{D}(B,T)}{\sigma_{D}(B,T)} \nn &=& C_{W}(T) B^{2}   \eqa
for the universal scaling relation. More explicitly, inserting $C_{W}(T) = C_{W}(T_{0}) \big(T_{0}/[T+T_{M}]\big)^{- 2\lambda_{\c,fn}}$ into the above, we find
\be
\f{T_{0}^{-2(1+\eta_{n})}\Delta\sigma_{L}(B,T)}{C_{W}(T_{0})}=\bigg(\f{B}{[T + T_{M}]^{1+\eta_{n}}}\bigg)^{2} , \label{r.scaling_corr}
\ee
where ``anomalous dimensions" are given by $\eta_{0}=0, \eta_{1}=0.34$ and $\eta_{2}=0.60$, respectively, for each fixed point.

\bf[h]
\ing[width=0.5\tw]{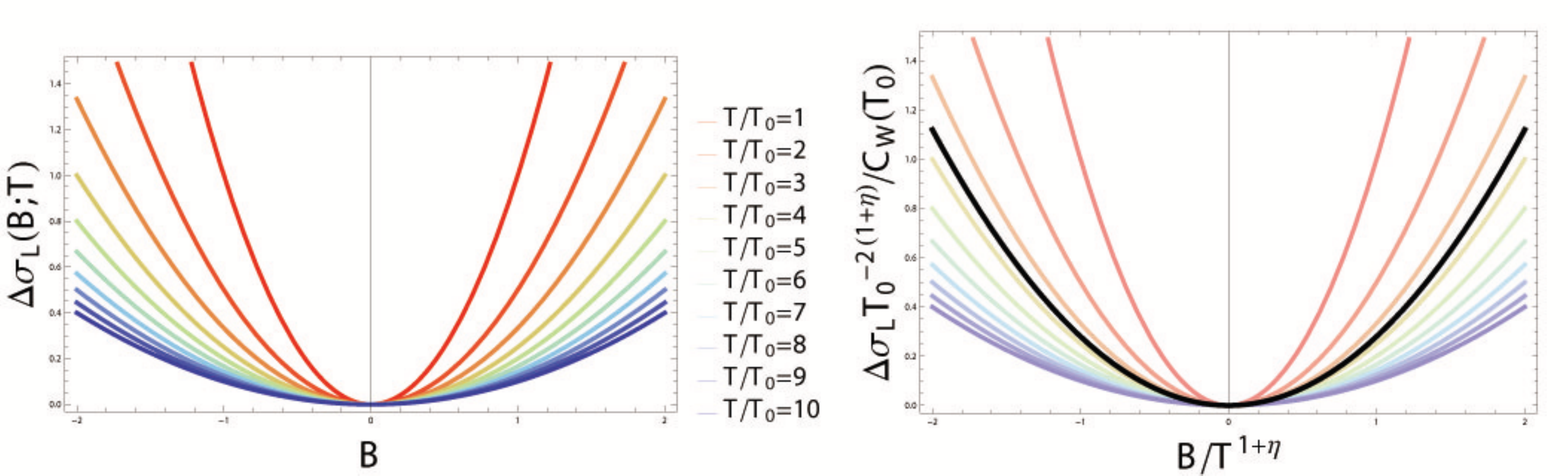} \caption{Scaling theory for the longitudinal magnetoconductivity. The longitudinal magnetoconductivity is enhanced to be proportional to $B^{2}$, the distance between the pair of Weyl point as a result of the chiral anomaly. The distance between the pair of Weyl points is renormalized to increase as temperature is reduced, which makes the degree of enhancement become larger (Left). These longitudinal transport coefficients are collapsed into a single universal curve, described by Eq. (\ref{r.scaling_corr}) (Right).} \label{f.scaling}
\ef

Fig. \ref{f.scaling} shows the longitudinal magnetoconductivity, enhanced to be proportional to $B^{2}$, the square of the distance between the pair of Weyl points, at each temperature. Our renormalization group analysis confirms that the distance between the pair of Weyl points is renormalized to increase, lowering temperature, i.e., $C_{W}(T_{H}) < C_{W}(T_L)$ with $T_{H} > T_{L}$. As a result, the degree of enhancement becomes larger as temperature is reduced (Left). Interestingly, these longitudinal transport coefficients turn out to be collapsed into a single universal curve, described by Eq. (\ref{r.scaling_corr}) (Right).

Fig. \ref{f.Cw_Exp} shows the comparison between $C_{W}(T)$ from an experimental data of $Bi_{1-x}Sb_{x}$ with $x = 3 \sim 4 \%$ and that from our renormalization group analysis \cite{WM_Review_Kim,Exp_in_preparation}. Experimentally, the enhancement coefficient $C_{W}(T)$ can be found from fitting the experimental data with Eq. (5) at a given temperature, where the Drude part is replaced with a transport coefficient of weak anti-localization corrections and additional contributions, which have nothing to do with Weyl points, are also introduced \cite{NLMR_Experiment_Original}. Repeating this fitting procedure for various temperatures, we obtain the temperature dependence of $C_{W}(T)$. The comparison between the experimental $C_{W}(T)$ and the renormalization group analysis Eq. (\ref{r.scaling_corr}) looks appealing.

\bf
\ing[width=0.5\tw]{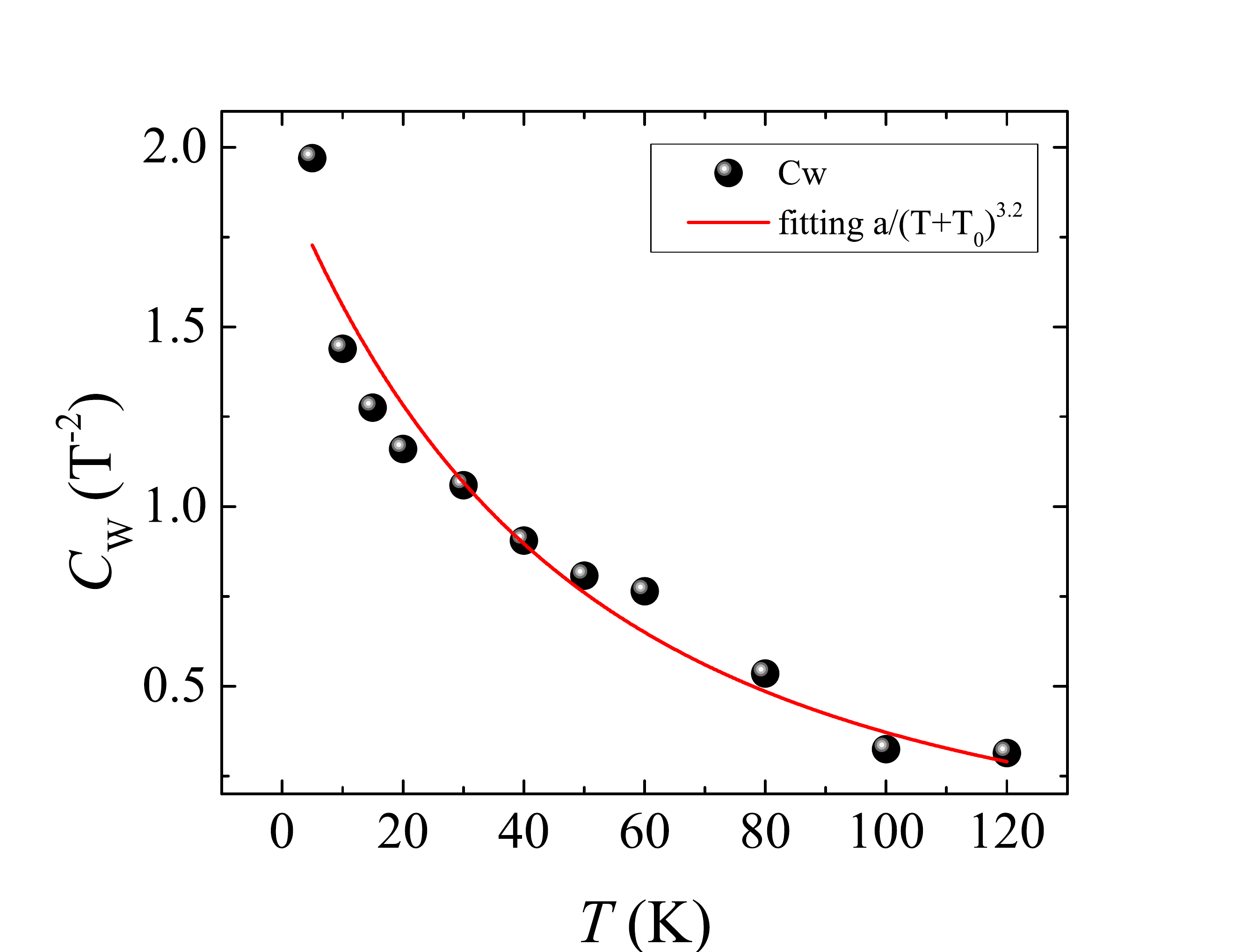}
\caption{Comparison between the theoretical prediction and an experimental data of $Bi_{1-x}Sb_{x}$ with $x = 3 \sim 4 \%$. Blackballs represent experimental data \cite{WM_Review_Kim,Exp_in_preparation} and red line denotes the theoretical prediction, given by $\frac{C_{W}(T)}{C_{W}(T_{0}) T_{0}^{2\lambda_{\c,f2}}} = \frac{1}{(T+T_{M})^{2\lambda_{\c,f2}}}$. Here, we obtain $\frac{C_{W}(T)}{C_{W}(T_{0}) T_{0}^{2\lambda_{\c,f2}}} = 1.7 * 10^{7} K^{2\lambda_{\c,f2}}$ and $T_{M} = 149 K$ with $\lambda_{\c,f2} = 1.6$ at the diffusive fixed point.} \label{f.Cw_Exp}
\ef

\section{Discussion and conclusion}

The original motivation of the present study is to reveal the existence of a topological phase transition from a Weyl metal phase to a normal metal state as a function of the strength of disorder and temperature. Our physical picture for this phase transition is as follows. Disorder scattering, in particular, inter-valley scattering is expected to kill the nature of the Weyl metallic phase since it induces mixing of chirality. We recall that the inter-valley scattering appears as an effective random-mass term. If $\bar{\psi}(x) \psi(x)$ has a nontrivial vacuum expectation value, i.e., $\langle \bar{\psi}(x) \psi(x) \rangle \not= 0$, expected to realize in the case of sufficiently strong disorder, the chiral symmetry breaks down even at the classical level and the chiral anomaly loses the physical meaning. As a result, we speculate that the distance between the pair of Weyl points renormalizes to vanish. A diffusive normal metallic state would be realized in the case of sufficiently strong disorder. Since this phase transition is not involved with symmetry breaking, it is identified with a topological phase transition.

This topological phase transition may be translated into Peccei-Quinn symmetry breaking in the context of dynamical generation of axions \cite{PQSB,PQSB_Review}. In order to realize the Peccei-Quinn symmetry breaking, there must be a scalar field. When the scalar field does not have its vacuum expectation value, any value of the $\theta-$angle can be canceled by the Peccei-Quinn transformation. On the other hand, the Peccei-Quinn symmetry breaking occurs when the scalar field has its vacuum expectation value. As a result of the continuous symmetry breaking, there exists a Goldstone boson field, referred to as an axion field. When the Peccei-Quinn symmetry is exact and thus, the axion field is massless, any value of the $\theta-$angle can be still canceled by the Peccei-Quinn transformation. However, there are instanton excitations, which do not allow the Peccei-Quinn symmetry not to be exact, giving rise to a mass term in the axion dynamics. Then, the vacuum angle is fixed to be $\theta = 0$, minimizing the energy of the system. In the present situation the corresponding scalar field results from the Hubbard-Stratonovich transformation of the random-mass term in the replica effective field theory, conventionally referred to as $Q_{ab}$, where $a$ and $b$ denote the replica index. However, there are two different aspects between the possible topological phase transition and the Peccei-Quinn symmetry breaking in high energy physics: (1) The vacuum angle is given by an inhomogeneous function of position while its gradient identified with a chiral gauge field is a constant. (2) There are no instanton-type excitations in the Weyl metal phase. This direction of research would be an interesting future task.

Unfortunately, the perturbative renormalization group analysis fails to access such an unstable fixed point, identified with the quantum critical point of the topological phase transition. In this respect the naming of the two-parameter scaling theory is not satisfactory in our opinion, basically motivated from the analogy with the plateau-plateau transition in the integer quantum Hall effect. However, it turns out that the longitudinal magnetoconductivity is governed by both parameters of the Drude conductivity and the distance between the pair of Weyl points, renormalized by inter-valley scattering, essentially analogous to $\sigma_{xx}$ and $\sigma_{xy}$ in the quantum Hall effect, respectively. In this respect we may call what we performed two-parameter scaling theory for the longitudinal magnetoconductivity in a disordered Weyl metal phase.

An unexpected result is breakdown of the $B/T$ scaling behavior near the diffusive fixed point although it is fulfilled near the clean fixed point. Actually, we could verify this prediction, comparing the proposed formula Eq. (\ref{r.scaling_corr}) of the two-parameter scaling theory with $C_{W}(T)$ in the experimental data of $Bi_{1-x}Sb_{x}$ with $x = 3 \sim 4 \%$. Here, we took into account modifying the original renormalization group analysis, introducing a cutoff scale into the equation for the distance between the pair of Weyl points as Eq. (\ref{r.cfield_modification}), in order to prohibit the divergence of the length scale within the Brillouin zone. This breakdown may be regarded to be a fingerprint of the interplay between disorder scattering and topological structure in a Weyl metal phase.

%\bf
%\ing[width=0.5\tw]{scaling_exp}
%\caption{ } \label{Phase_Diagram}
%\ef

\section*{Acknowledgement}

This study was supported by the Ministry of Education, Science, and Technology (No. NRF-2015R1C1A1A01051629 and No. 2011-0030046) of the National Research Foundation of Korea (NRF) and by TJ Park Science Fellowship of the POSCO TJ Park Foundation. This work was also supported by the POSTECH Basic Science Research Institute Grant (2015). We would like to appreciate fruitful discussions in the APCTP workshop on Delocalisation Transitions in Disordered Systems in 2015.

\begin{widetext}

\appendix

\section{model hamiltonian}

A minimal model for a Weyl metal state is given by
\bnn
\mathcal{H}=\psi^{\dag}(\x)\big(v\bm{\alpha}\cdot(-\i\gr)-\mu+g\bm{B}\cdot\bm{\sigma}\otimes I_{2\times2}\big)\psi(\x) .
\enn
$\psi=(\psi_{R},\psi_{L})^{T}$ is a four-component Dirac-spinor field in a two-component Weyl-spinor field with right(R)-left(L) chirality, and $v$ is the velocity of such fermions. $\mu$ is an electron chemical potential. $\bm{B}$ is an externally applied magnetic field with a Lande$-g$ factor $g$. $\bm{\alpha}$ is a four-by-four matrix, given by $\bm{\alpha} = \bm{\sigma} \otimes \sigma_{z}$, where $\bm{\sigma}$ is a Pauli matrix.

First, we look into a band structure. This block-diagonal matrix can be diagonalized as
\bnn
\mathcal{H}_{\k}=\phi^{\dag}_{\k}\big(\al v\k+g\bm{B}\ar\sigma_{z}\otimes P_{+}-\al v\k-g\bm{B}\ar\sigma_{z}\otimes P_{-}-\mu\big)\phi_{\k}
\enn
where $P_{+}=\bpm1&0\\0&0\epm$ and$P_{-}=\bpm0&0\\0&1\epm$ are projection matrices, and $\phi_{\k}=U_{\k}\psi_{\k}$ is an eigenstate. The unitary matrix varying with $\k$ is given by
\bnn
U_{\k}=\bpm\cos{\f{\zeta_{+}}{2}}&\sin{\f{\zeta_{+}}{2}}e^{-\i\eta_{+}}\\-\sin{\f{\zeta_{+}}{2}}e^{\i\eta_{+}}&\cos{\f{\zeta_{+}}{2}}\epm\otimes P_{+}+\bpm\cos{\f{\zeta_{-}}{2}}&\sin{\f{\zeta_{-}}{2}}e^{-\i\eta_{-}}\\-\sin{\f{\zeta_{-}}{2}}e^{\i\eta_{-}}&\cos{\f{\zeta_{-}}{2}}\epm\otimes P_{-} ,
\enn
where $\zeta_{\pm}$ $(\eta_{\pm})$ is the polar (azimuthal) angle of $\k\pm\f{g}{v}\bm{B}$, respectively. If we draw a band structure along some momentum-line, for example, $\k=(0,0,k_{z})$, then we obtain a pair of Weyl cones as shown in Fig. \ref{f.bandstructure}.

Second, we consider two types of random potentials, say, ``intra-valley scattering" and ``inter-valley scattering", given by
\bnn
&&\psi^{\dag}_{R}(\x)V(\x)\psi_{R}(\x)+\psi^{\dag}_{L}(\x)V(\x)\psi_{L}(\x)=\psi^{\dag}(\x)V(\x)\psi(\x),\\
&&\psi^{\dag}_{R}(\x)U(\x)\psi_{L}(\x)+\psi^{\dag}_{R}(\x)U(\x)\psi_{L}(\x)=\psi^{\dag}(\x)U(\x)\big(I_{2\times2}\otimes\sigma_{x}\big)\psi(\x) ,
\enn
where $V(\x)$ and $U(\x)$ are disorder potentials for intra-valley scattering and inter-valley scattering, respectively.

Now, the effective action is
\bnn
S[\psi^{\dag},\psi;V,U]=\intt\intx\psi^{\dag}(\tau,\x)\big\{\partial_{\tau}+v\bm{\alpha}\cdot(-\i\gr)-\mu+g\bm{B}\cdot\bm{\sigma}\otimes I_{2\times2}+V(\x)+U(\x)I_{2\times2}\otimes\sigma_{x}\big\}\psi(\tau,\x) ,
\enn
where the corresponding free energy is given by $F[V,U]=-T\ln{\int\D\psi^{\dag}\D\psi e^{-S[\psi^{\dag},\psi;V,U]}}$ in a given configuration of random potentials. We represent this effective action in terms of gamma matrices in the Weyl representation
\bnn
\gamma^{0}=I_{2\times2}\otimes\sigma_{x},~\gamma^{k}=\gamma^{0}(-\alpha_{k})=\sigma_{k}\otimes\i\sigma_{y}~(k=1,2,3),~\gamma^{5}=\i\gamma^{0}\gamma^{1}\gamma^{2}\gamma^{3}=-I_{2\times2}\otimes\sigma_{z} .
\enn
Then, we reach the following expression
\bnn
S[\bar{\psi},\psi;V,U]=\int d^{4}x\big\{\bar{\psi}(x)\big(\gamma^{0}\partial_{0}+\i v\gamma^{k}\partial_{k}-\mu\gamma^{0}+c_{\mu}\gamma^{\mu}\gamma^{5}\big)\psi(x)+\bar{\psi}(x)\gamma^{0}V(\x)\psi(x)+\bar{\psi}(x)U(\x)\psi(x)\big\}
\enn
with an adjoint spinor-field $\bar{\psi}\equiv\psi^{\dag}\gamma^{0}$, where we introduced $c_{k}\equiv gB_{k}$ $(k=1,2,3)$ with ``time-component" $c_{0}$. ``Space-time" of $x$ is $x^{\mu}=(\tau,\x)$ and other four-vectors are defined, similarly. For example, four-momentum is $p^{\mu}=(p^{0},\p)$ with $p^{0}=-\i\omega_{n}$. Since the action has been formulated in the imaginary time, it is defined on the Euclidean geometry as shown by $p^{\mu}p_{\mu}=-{\omega_{n}}^{2}-\p^{2}$.

\section{effective field theory for renormalization group analysis}

\subsection{Disorder Average}

We define the free part of the effective action as
\bnn
S_{0}[\bar{\psi},\psi]=\int d^{4}x\bar{\psi}(x)\big(\gamma^{0}\partial_{0}+\i v\gamma^{k}\partial_{k}-\mu\gamma^{0}+c_{\mu}\gamma^{\mu}\gamma^{5}\big)\psi(x) .
\enn
Then, a physical observable is measured as follows
\bnn
\left\langle\O(\bar{\psi},\psi)\right\rangle=\int\D V\D UP[V,U]\f{\int\D\bar{\psi}\D\psi\O(\bar{\psi},\psi)e^{-S_{0}[\bar{\psi},\psi]}e^{-\int d^{4}x\bar{\psi}(x)(\gamma^{0}V(\x)+U(\x))\psi(x)}}{\int\D\bar{\psi}\D\psi e^{-S_{0}[\bar{\psi},\psi]}e^{-\int d^{4}x\bar{\psi}(x)(\gamma^{0}V(\x)+U(\x))\psi(x)}} .
\enn
This can be reformulated as
\bnn
&&\left\langle\O(\bar{\psi},\psi)\right\rangle=\int\D V\D UP[V,U]\f{\delta}{\delta J}\br|_{J=0}\ln{Z[V,U,J]},\\
&&Z[V,U,J]=\int\D\bar{\psi}\D\psi e^{-S_{0}[\bar{\psi},\psi]}e^{-\int d^{4}x\bar{\psi}(x)(\gamma^{0}V(\x)+U(\x))\psi(x)+\int d^{4}xJ(x)\O(\bar{\psi}(x),\psi(x))},
\enn
where $J(x)$ is a source field coupled to an operator $O(\bar{\psi},\psi)$, locally.

In order to perform the averaging procedure for disorders, we resort to the replica trick of $\ln{Z}=\limR\f{Z^{R}-1}{R}$
\bnn
\left\langle\O(\bar{\psi},\psi)\right\rangle=\limR\int\D V\D UP[V,U]\f{\delta}{\delta J}\br|_{J=0}\f{Z^{R}[V,U,J]-1}{R},
\enn
where the replicated partition function is
\bnn
Z^{R}[V,U,J]=\int\D\bar{\psi}^{a}\D\psi^{a}\exp{\bigg[-\sum_{a=1}^{R}S_{0}[\bar{\psi}^{a},\psi^{a}]-\sum_{a=1}^{R}\int d^{4}x\bar{\psi}^{a}(x)\big(\gamma^{0}V(\x)+U(\x)\big)\psi^{a}(x)+\int d^{4}xJ(x)\sum_{a=1}^{R}\O(\bar{\psi}^{a},\psi^{a})\bigg]}
\enn
with a replica index ``a". In this technique a physical observable is given by
\bnn
\left\langle\O(\bar{\psi},\psi)\right\rangle&=&\limR\f{1}{R}\int\D V\D UP[V,U]\int\D\bar{\psi}^{a}\D\psi^{a}\O(\bar{\psi}^{a},\psi^{a})\\
&&\times\exp{\bigg[-\sum_{a=1}^{R}S_{0}[\bar{\psi}^{a},\psi^{a}]-\sum_{a=1}^{R}\int d^{4}x\bar{\psi}^{a}(x)\big(\gamma^{0}V(\x)+U(\x)\big)\psi^{a}(x)\bigg]}.
\enn

In this study we take into account static-and Gaussian-distributed disorders, given by
\bnn
P[V,U]=N\exp{\bigg[-\f{\intx V^{2}(\x)}{2\Gamma_{V}}-\f{\intx U^{2}(\x)}{2\Gamma_{U}}\bigg]},
\enn
where $N$ is a normalization factor. It is straightforward to perform the Gaussian integral for disorders, resulting in
\bnn
\left\langle\O(\bar{\psi},\psi)\right\rangle=\limR\f{1}{R}\sum_{a=1}^{R}\int\D\bar{\psi}\D\psi\O(\bar{\psi}^{a},\psi^{a})\exp{\bigg[-\sum_{a=1}^{R}S_{0}[\bar{\psi}^{a},\psi^{a}]-\sum_{b,c=1}^{R}S_{\tx{dis}}[\bar{\psi}^{b},\psi^{b},\bar{\psi}^{c},\psi^{c}]\bigg]},
\enn
where disorder-driven effective interactions are [Eq. (\ref{e.effectivetheory})]
\bnn
S_{\tx{dis}}[\bar{\psi}^{b},\psi^{b},\bar{\psi}^{c},\psi^{c}]&=&-\intt\intt'\int d^{3}\x\f{\Gamma_{V}}{2}\bar{\psi}^{b}(\tau,\x)\gamma^{0}\psi^{b}(\tau,\x)\bar{\psi}^{c}(\tau',\x)\gamma^{0}\psi^{c}(\tau',\x)\\
&&-\intt\intt'\int d^{3}\x\f{\Gamma_{U}}{2}\bar{\psi}^{b}(\tau,\x)\psi^{b}(\tau,\x)\bar{\psi}^{c}(\tau',\x)\psi^{c}(\tau',\x) .
\enn

\subsection{Renormalized perturbation theory}

From now on, we focus on the case of a zero-chemical potential. We start from the following effective action
\bnn
S_{B}&=&\int d^{d+1}x\bar{\psi}_{B}^{a}(x)(\gamma^{0}\partial_{0}+v_{B}\i\gamma^{k}\partial_{k}+c_{B\mu}\gamma^{\mu}\gamma^{5})\psi_{B}^{a}(x)\\
&-& \intt\intt'\int d^{d}\x\f{\Gamma_{BV}}{2}\bar{\psi}_{B}^{b}(\tau,\x)\gamma^{0}\psi_{B}^{b}(\tau,\x)\bar{\psi}_{B}^{c}(\tau',\x)\gamma^{0}\psi_{B}^{c}(\tau',\x)\\
&-& \intt\intt'\int d^{d}\x\f{\Gamma_{BU}}{2}\bar{\psi}_{B}^{b}(\tau,\x)\psi_{B}^{b}(\tau,\x)\bar{\psi}_{B}^{c}(\tau',\x)\psi_{B}^{c}(\tau',\x) ,
\enn
where summations on the replica indices are implied. The subscript $B$ denotes ``bare", meaning that this effective action is defined at an ultraviolet (UV) scale. Note that we have generalized dimensions to ``d(space)+1(time)" for dimensional regularization.

Performing the dimensional analysis, where space and time coordinates have $-1$ in mass dimension, we observe
\bnn
\tx{dim}[\psi]=\f{d}{2},~~\tx{dim}[v]=0,~~\tx{dim}[c_{\mu}]=1,~~\tx{dim}[\Gamma_{V}]=\tx{dim}[\Gamma_{U}]=2-d.
\enn
In this respect we perform the renormalization group analysis in $d+1=3+\varepsilon$ dimensions, where $\varepsilon$ is a ``small" parameter. In the end of the calculation the dimensions are analytically continued to the physical dimensions ($d+1=4$) by setting $\varepsilon=1$.

Taking into account quantum corrections, divergences would be generated. They can be absorbed into renormalization constants by redefining fields and parameters. Rewriting the bare action in terms of renormalized fields and couplings, we obtain
\bnn
S_{B}&=&\int d^{d+1}x\bar{\psi}_{R}^{a}(x)\big(Z_{\psi}^{\omega}\gamma^{0}\partial_{0}+Z_{\psi}^{\k}v_{R}\i\gamma^{k}\partial_{k}+Z_{c0}c_{R0}\gamma^{0}\gamma^{5}+Z_{\c}c_{Rk} \gamma^{k}\gamma^{5}\big)\psi_{R}^{a}(x)\\
&-& \intt\intt'\int d^{d}\x\f{Z_{\Gamma V}\Gamma_{RV}}{2}\bar{\psi}_{R}^{b}(\tau,\x)\gamma^{0}\psi_{R}^{b}(\tau,\x)\bar{\psi}_{R}^{c}(\tau',\x)\gamma^{0}\psi_{R}^{c}(\tau',\x),\\
&-& \intt\intt'\int d^{d}\x\f{Z_{\Gamma U}\Gamma_{RU}}{2}\bar{\psi}_{R}^{b}(\tau,\x)\psi_{R}^{b}(\tau,\x)\bar{\psi}_{R}^{c}(\tau',\x)\psi_{R}^{c}(\tau',\x),
\enn
where such renormalized fields and parameters are given by
\bnn
&&\psi_{B}^{a}=(Z_{\psi}^{\omega})^{\f{1}{2}}\psi_{R}^{a},~~~v_{B}=Z_{\psi}^{\k}(Z_{\psi}^{\omega})^{-1}v_{R},~~~c_{B0}=Z_{c0}(Z_{\psi}^{\omega})^{-1}c_{R0},\\
&&c_{Bk}=Z_{\c}(Z_{\psi}^{\omega})^{-1}c_{Rk},~~~\Gamma_{BV}=Z_{\Gamma V}(Z_{\psi}^{\omega})^{-2}\Gamma_{RV},~~~\Gamma_{BU}=Z_{\Gamma U}(Z_{\psi}^{\omega})^{-2}\Gamma_{RU}.
\enn

It is more elaborate to represent this theory by separating the renormalized part from counter terms that are to absorb divergences in the following way [Eq. (\ref{e.RGtheory})],
\bnn
S_{B}&=&S_{R}+S_{CT},\\
S_{R}&=&\int d^{d+1}x\bar{\psi}_{R}^{a}\big(\gamma^{0}\partial_{0}+ v_{R}\i\gamma^{k}\partial_{k}+c_{R0}\gamma^{0}\gamma^{5}+c_{Rk}\gamma^{k}\gamma^{5}\big)\psi_{R}^{a}\\
&-& \int d\tau
\int d\tau'\int d^{d}\x\f{\Gamma_{RV}}{2}(\bar{\psi}_{R}^{b}\gamma^{0}\psi_{R}^{b})_{\tau}(\bar{\psi}_{R}^{c}\gamma^{0}\psi_{R}^{c})_{\tau'}-\int d\tau
\int d\tau'\int d^{d}\x\f{\Gamma_{RU}}{2}(\bar{\psi}_{R}^{b}\psi_{R}^{b})_{\tau}(\bar{\psi}_{R}^{c}\psi_{R}^{c})_{\tau'},\\
S_{CT}&=&\int d^{d+1}x\bar{\psi}_{R}^{a}\big(\delta_{\psi}^{\omega}\gamma^{0}\partial_{0}+\delta_{\psi}^{\k}v_{R}\i\gamma^{k}\partial_{k}+\delta_{c0}c_{R0} \gamma^{0}\gamma^{5}+\delta_{\c}c_{Rk}\gamma^{k}\gamma^{5}\big)\psi_{R}^{a}\\
&-& \int d\tau\int d\tau'\int d^{d}\x\f{\delta_{\Gamma V}\Gamma_{RV}}{2}(\bar{\psi}_{R}^{b}\gamma^{0}\psi_{R}^{b})_{\tau}(\bar{\psi}_{R}^{c}\gamma^{0}\psi_{R}^{c})_{\tau'}-\int d\tau\int d\tau'\int d^{d}\x\f{\delta_{\Gamma U}\Gamma_{RU}}{2}(\bar{\psi}_{R}^{b}\psi_{R}^{b})_{\tau}(\bar{\psi}_{R}^{c}\psi_{R}^{c})_{\tau'},
\enn
where $Z_{\psi}^{\omega}=1+\delta_{\psi}^{\omega}$,~~$Z_{\psi}^{\k}=1+\delta_{\psi}^{\k}$,~~$Z_{c0}=1+\delta_ {c0}$,~~$Z_{\c}=1+\delta_{\c}$,~~$Z_{\Gamma V}=1+\delta_{\Gamma V}$ and $Z_{\Gamma U}=1+\delta_{\Gamma U}$.

\subsection{Feynman Rules}

In the momentum and frequency space the effective action is written as
\small\bnn
S[\bar{\psi}^{a},\psi^{a}]=\sum_{p}\bar{\psi}^{a}_{p}\big(\sp+\sc\gamma^{5}\big)\psi^{a}_{p} - \f{1}{L^{3}}\sum_{p_{j}}\bigg[\f{\Gamma_{V}}{2}(\bar{\psi}^{b}_{p_{1}}\gamma^{0}\psi^{b}_{p_{2}})(\bar{\psi}^{c}_{p_{3}}\gamma^{0}\psi^{c}_{p_{4}}) + \f{\Gamma_{U}}{2}(\bar{\psi}^{b}_{p_{1}}\psi^{b}_{p_{2}})(\bar{\psi}^{c}_{p_{3}}\psi^{c}_{p_{4}})\bigg]\delta^{(3)}_{\p_{1}-\p_{2},\p_{3}-\p_{4}}\delta_{p_{1}^{0}p_{2}^{0}}\delta_{p_{3}^{0}p_{4}^{0}} ,
\enn\ns
where Feynman rules are given in Fig. \ref{f.Feynmanrules}.
\bf
\ing[width=\tw]{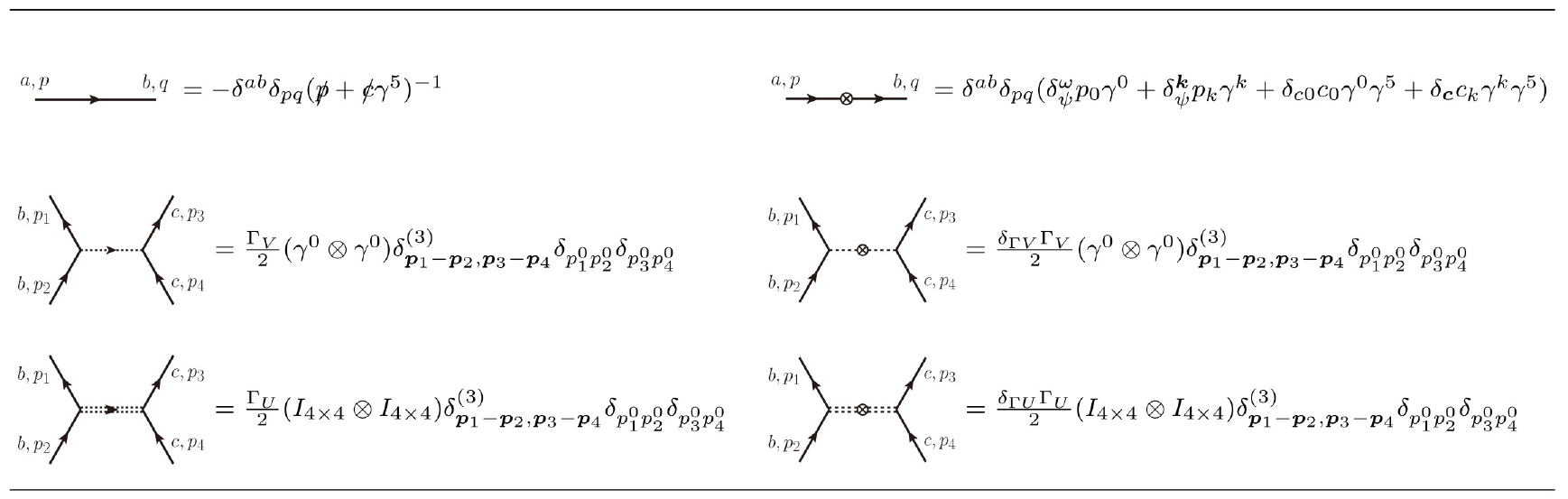} \caption{Feynman rules in the momentum and frequency space. A single-dashed line represents an intra-velly scattering while a double-dashed line, an inter-valley scattering.}\label{f.Feynmanrules}
\ef

Since there is a chiral gauge field in the kinetic-energy part, the free propagator becomes a little bit complex. Considering the following identity
\bnn
(\sp+\sc\gamma^{5})(\sp-\sc\gamma^{5})(p^{2}+c^{2}+2p\cdot c\gamma^{5})=(p^{2}+c^{2}-2p\cdot c\gamma^{5})(p^{2}+c^{2}+2p\cdot c\gamma^{5})=(p+c)^{2}(p-c)^{2} ,
\enn
we obtain an electron Green function
\bnn
G(p)=-(\sp+\sc\gamma^{5})^{-1}=-\f{(\sp-\sc\gamma^{5})(p^{2}+c^{2}+2p\cdot c\gamma^{5})}{(p+c)^{2}(p-c)^{2}}.
\enn
We introduces the following expression with a Feynman parameter for the renormalization group analysis
\bnn
G(p)=-\int^{1}_{0}dx\f{(\sp-\sc\gamma^{5})(p^{2}+c^{2}+2p\cdot c\gamma^{5})}{\big[\big(p+(1-2x)c\big)^{2}+4x(1-x)c^{2}\big]^{2}}.
\enn
For a future use, we rearrange it in terms of $\p$ as
\bea
G(p)&=&\int^{1}_{0}dx\f{\p^{2}p_{i}\gamma^{i}+\p^{2}(p_{0}\gamma^{0}-\sc\gamma^{5})+p_{i}p_{j}(-2c^{i}\gamma^{j}\gamma^{5})+p_{i}f_{1}^{i}(p_{0})+f_{0}(p_{0})}{\big[\big(\p+(1-2x)\c\big)^{2}+\Delta_{0}(p_{0};x)\big]^{2}},\label{e.Green1}\\
\Delta_{0}(p_{0};x)&=&4x(1-x)\c^{2}-(p_{0}+c_{0})^{2}+4xp_{0}c_{0},\no\\
f_{1}^{i}(p_{0})&=&-\gamma^{i}(p_{0}\gamma^{0}-\sc\gamma^{5})(p_{0}\gamma^{0}+\sc\gamma^{5})+2c^{i}\gamma^{5}(p_{0}\gamma^{0}-\sc\gamma^{5}),\no\\
f_{0}(p_{0})&=&-(p_{0}\gamma^{0}-\sc\gamma^{5})^{2}(p_{0}\gamma^{0}+\sc\gamma^{5}).\no
\eea
Alternatively, we obtain in terms of $\p'=\p+(1-2x)\c$
\bea
G(p)&=&\int^{1}_{0}dx\f{C_{3}^{i}{\p'}^{2}p'_{i}+C_{2a}{\p'}^{2}+C_{2b}^{ij}p'_{i}p'_{j}+C_{1}^{i}p'_{i}+C_{0}}{\big[{\p'}^{2}+\Delta_{0}(p_{0};x)\big]^{2}},\label{e.Green2}\\
C_{3}^{i}&=&\gamma^{i},\no\\
C_{2a}&=&\su-\sc\gamma^{5},\no\\
C_{2b}^{ij}&=&-2\gamma^{i}(u^{j}+c^{j}\gamma^{5}),\no\\
C_{1}^{i}&=&-\gamma^{i}(\su-\sc\gamma^{5})(\su+\sc\gamma^{5})-2(u^{i}-c^{i}\gamma^{5})(\su-\sc\gamma^{5}),\no\\
C_{0}&=&-(\su-\sc\gamma^{5})^{2}(\su+\sc\gamma^{5}) , \no
\eea
where $u\equiv(p_{0},\tilde{\c})$ and $\tilde{\c}_{x}\equiv(2x-1)\c$. We may use either of these expressions for convenience. Despite their complicated form, they will not be involved much in actual integration procedures.

\section{self-energy corrections}

\subsection{Relevant Feynman's diagrams}

Within the replica trick, we are allowed to perform the perturbative analysis. The full Green function of $\mathbf{G}(p,q)=\left\langle\psi_{p}\bar{\psi}_{q}\right\rangle$ is evaluated up to the $\Gamma^{2}$ order as follows
\bnn
&&\mathbf{G}(p,q)\\
&=&\limR\f{1}{R}\int\D\bar{\psi}\D\psi\big(\psi^{a}_{p}\bar{\psi}^{a}_{q}\big)e^{-S_{0}[\bar{\psi}^{\alpha},\psi^{\alpha}]}e^{\f{1}{L^{3}}\sum_{p_{j}}\big[\f{\Gamma_{V}}{2}(\bar{\psi}^{b}_{p_{1}}\gamma^{0}\psi^{b}_{p_{2}})(\bar{\psi}^{c}_{p_{3}}\gamma^{0}\psi^{c}_{p_{4}})+\f{\Gamma_{U}}{2}(\bar{\psi}^{b}_{p_{1}}\psi^{b}_{p_{2}})(\bar{\psi}^{c}_{p_{3}}\psi^{c}_{p_{4}})\big]\delta^{(3)}_{\p_{1}-\p_{2},\p_{3}-\p_{4}}\delta_{p_{1}^{0}p_{2}^{0}}\delta_{p_{3}^{0}p_{4}^{0}}}\\
&\simeq&\limR\f{1}{R}\int\D\bar{\psi}\D\psi e^{-S_{0}[\bar{\psi}^{\alpha},\psi^{\alpha}]}\bl[\psi^{a}_{p}\bar{\psi}^{a}_{q}+\f{\Gamma_{V}}{2L^{3}}\sum_{p_{j}}\big(\psi^{a}_{p}\bar{\psi}^{a}_{q}\big)\big(\bar{\psi}^{b}_{p_{1}}\gamma^{0}\psi^{b}_{p_{2}}\bar{\psi}^{c}_{p_{3}} \gamma^{0} \psi^{c}_{p_{4}}\big)\delta^{(3)}_{\p_{1}-\p_{2},\p_{3}-\p_{4}}\delta_{p_{1}^{0}p_{2}^{0}}\delta_{p_{3}^{0}p_{4}^{0}}\\
&&+\f{\Gamma_{U}}{2L^{3}}\sum_{p_{j}}\big(\psi^{a}_{p}\bar{\psi}^{a}_{q}\big)\big(\bar{\psi}^{b}_{p_{1}}\psi^{b}_{p_{2}} \bar{\psi}^{c}_{p_{3}}\psi^{c}_{p_{4}}\big)\delta^{(3)}_{\p_{1}-\p_{2},\p_{3}-\p_{4}}\delta_{p_{1}^{0}p_{2}^{0}}\delta_{p_{3}^{0}p_{4}^{0}}\\
&&+\f{\Gamma_{V}^{2}}{8(L^{3})^{2}}\sum_{p_{j}p'_{j}}\big(\psi^{a}_{p}\bar{\psi}^{a}_{q}\big)\big(\bar{\psi}^{b}_{p_{1}}\gamma^{0} \psi^{b}_{p_{2}}\bar{\psi}^{c}_{p_{3}} \gamma^{0}\psi^{c}_{p_{4}}\big)\big(\bar{\psi}^{b'}_{p'_{1}}\gamma^{0}\psi^{b'}_{p'_{2}} \bar{\psi}^{c'}_{p'_{3}}\gamma^{0}\psi^{c'}_{p'_{4}}\big) \delta^{(3)} _{\p_{1}-\p_{2},\p_{3}-\p_{4}}\delta_{p_{1}^{0}p_{2}^{0}} \delta_{p_{3}^{0}p_{4}^{0}}\delta^{(3)}_{\p'_{1}-\p'_{2},\p'_{3}-\p'_{4}}\delta_{p_{1}^{'0}p_{2}^{'0}}\delta_{p_{3}^{'0}p_{4}^{'0}}\\
&&+\f{\Gamma_{U}^{2}}{8(L^{3})^{2}}\sum_{p_{j}p'_{j}}\big(\psi^{a}_{p}\bar{\psi}^{a}_{q}\big)\big(\bar{\psi}^{b}_{p_{1}}\psi^{b}_{p_{2}}\bar{\psi}^{c}_{p_{3}} \psi^{c}_{p_{4}}\big)\big(\bar{\psi}^{b'}_{p'_{1}}\psi^{b'}_{p'_{2}}\bar{\psi}^{c'}_{p'_{3}}\psi^{c'}_{p'_{4}}\big)\delta^{(3)} _{\p_{1}-\p_{2},\p_{3}-\p_{4}}\delta_{p_{1}^{0}p_{2}^{0}}\delta_{p_{3}^{0}p_{4}^{0}}\delta^{(3)}_{\p'_{1}-\p'_{2},\p'_{3}-\p'_{4}}\delta_{p_{1}^{'0}p_{2}^{'0}}\delta_{p_{3}^{'0}p_{4}^{'0}}\\
&&+\f{\Gamma_{V}\Gamma_{U}}{4(L^{3})^{2}}\sum_{p_{j}p'_{j}}\big(\psi^{a}_{p}\bar{\psi}^{a}_{q}\big)\big(\bar{\psi}^{b}_{p_{1}}\gamma^{0}\psi^{b}_{p_{2}}\bar{\psi}^{c}_{p_{3}} \gamma^{0}\psi^{c}_{p_{4}}\big)\big(\bar{\psi}^{b'}_{p'_{1}}\psi^{b'}_{p'_{2}}\bar{\psi}^{c'}_{p'_{3}}\psi^{c'}_{p'_{4}}\big)\delta^{(3)} _{\p_{1}-\p_{2},\p_{3}-\p_{4}}\delta_{p_{1}^{0}p_{2}^{0}}\delta_{p_{3}^{0}p_{4}^{0}}\delta^{(3)}_{\p'_{1}-\p'_{2},\p'_{3}-\p'_{4}}\delta_{p_{1}^{'0}p_{2}^{'0}}\delta_{p_{3}^{'0}p_{4}^{'0}}\br] ,
\enn
where summations on the replica indices are implied. Feynman diagrams whose internal lines are not connected to external lines always vanish due to the replica symmetry (all Green's functions with different replica indices are identical) and the replica limit ($\limR\f{1}{R}$). For details, we refer to Ref. \cite{Kyoung}.

We find self-energy corrections in the first-order (Fig. \ref{f.Fock}),
\bea
\Sigma^{(1)}(p)=\f{\Gamma_{V}}{L^{3}}\sum_{q}\gamma^{0}G(p-q)\gamma^{0}\delta_{q^{0}0}+\f{\Gamma_{U}}{L^{3}}\sum_{q}G(p-q)\delta_{q^{0}0}\equiv\Sigma^{(1)}_{V}(p)+\Sigma^{(1)}_{U}(p).\label{e.Fock}
\eea
\bf[b]
\ing[width=0.4\tw]{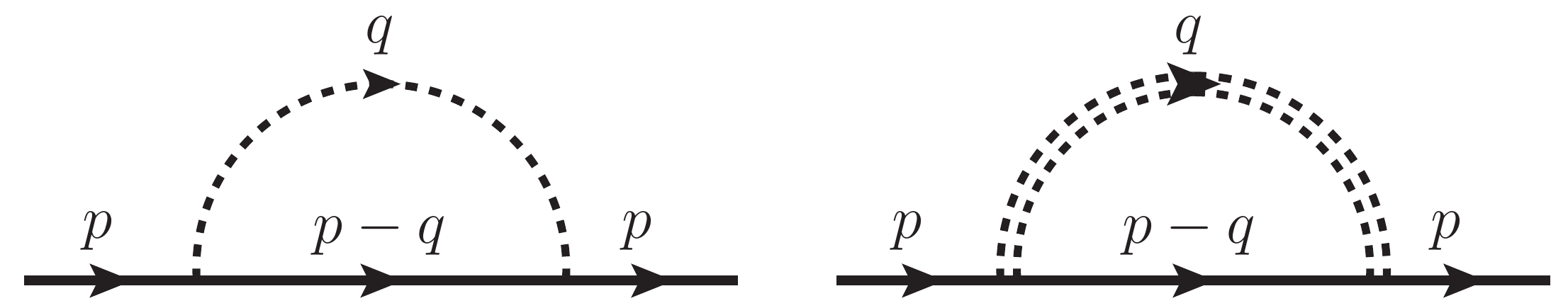}
\caption{Self-energy corrections in the first-order. There are two Fock diagrams for intra- and inter-valley scattering, respectively.} \label{f.Fock}
\ef

Likewise, we find self-energy corrections in the second order (Fig. \ref{f.self-2nd}).
\small\bea
\Sigma^{(2),r}(p)&=&\f{\Gamma_{V}^{2}}{(L^{3})^{2}}\sum_{q,l}\gamma^{0}G(p-q)\gamma^{0}G(p-q-l)\gamma^{0}G(p-q)\gamma^{0}\delta_{q^{0}0}\delta_{l^{0}0}+\f{\Gamma_{V}\Gamma_{U}}{(L^{3})^{2}}\sum_{q,l}\gamma^{0}G(p-q)G(p-q-l)G(p-q)\gamma^{0}\delta_{q^{0}0}\delta_{l^{0}0}\no\\
&&+\f{\Gamma_{U}\Gamma_{V}}{(L^{3})^{2}}\sum_{q,l}G(p-q)\gamma^{0}G(p-q-l)\gamma^{0}G(p-q)\delta_{q^{0}0}\delta_{l^{0}0}+\f{\Gamma_{U}^{2}}{(L^{3})^{2}}\sum_{q,l}G(p-q)G(p-q-l)G(p-q)\delta_{q^{0}0}\delta_{l^{0}0},\label{e.rainbow}\\
&\equiv&\Sigma^{(2),r}_{VV}(p)+\Sigma^{(2),r}_{VU}(p)+\Sigma^{(2),r}_{UV}(p)+\Sigma^{(2),r}_{UU}(p)\no\\
\Sigma^{(2),c}(p)&=&\f{\Gamma_{V}^{2}}{(L^{3})^{2}}\sum_{q,l}\gamma^{0}G(p-q)\gamma^{0}G(p-q-l)\gamma^{0}G(p-l)\gamma^{0}\delta_{q^{0}0}\delta_{l^{0}0}+\f{\Gamma_{V}\Gamma_{U}}{(L^{3})^{2}}\sum_{q,l}\gamma^{0}G(p-q)G(p-q-l)\gamma^{0}G(p-l)\delta_{q^{0}0}\delta_{l^{0}0}\no\\
&&+\f{\Gamma_{U}\Gamma_{V}}{(L^{3})^{2}}\sum_{q,l}\gamma^{0}G(p-q)G(p-q-l)\gamma^{0}G(p-l)\delta_{q^{0}0}\delta_{l^{0}0}+\f{\Gamma_{U}^{2}}{(L^{3})^{2}}\sum_{q,l}G(p-q)G(p-q-l)G(p-l)\delta_{q^{0}0}\delta_{l^{0}0}\label{e.cross}\\
&\equiv&\Sigma^{(2),c}_{VV}(p)+\Sigma^{(2),c}_{VU}(p)+\Sigma^{(2),c}_{UV}(p)+\Sigma^{(2),c}_{UU}(p)\no.
\eea\ns

\bf[b]
\ing[width=.95\tw]{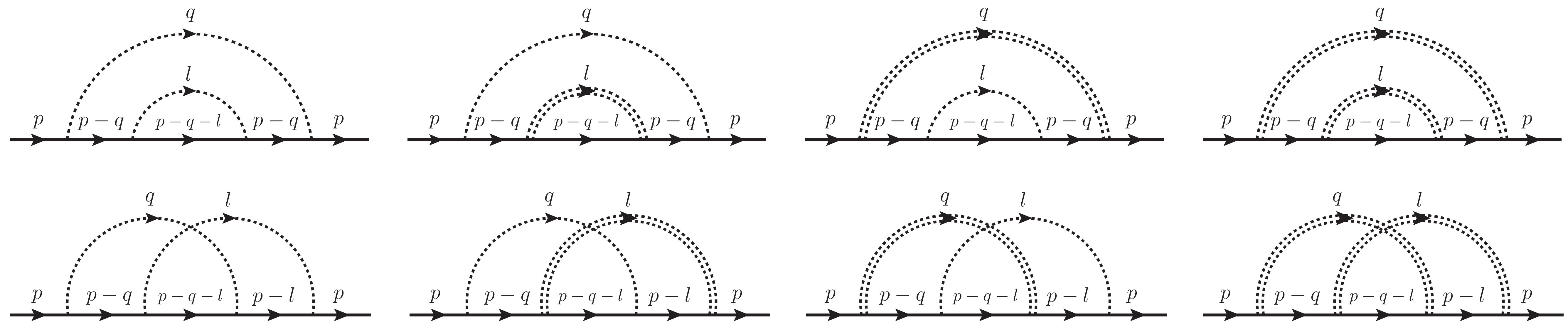}
\caption{Self-energy corrections in the second-order. There are two distinct types of diagrams, say, rainbow diagrams and crossed diagrams. Diagrams in each type are distinguished by interaction vertices (two intra-valley scattering, one intra-valley and one inter-valley scattering, and etc.). So, totally there are eight diagrams for the second-order self-energy corrections.} \label{f.self-2nd}
\ef

\subsection{Evaluation of relevant diagrams}

From now on, we evaluate self-energy diagrams one by one. Since there are two types of interactions, we have many diagrams to evaluate, especially, in the two loop-order. Instead of struggling to evaluate them one by one, we're going to find integration formulae for products of Green functions and make a use of them for the same types of diagrams.

\subsubsection{One-loop order: Fock diagrams}

First, we evaluate the first-order Fock diagram
\bnn
\Sigma^{(1)}(p)=\Gamma_{V}\gamma^{0}I_{1}(p)\gamma^{0}+\Gamma_{U}I_{1}(p),
\enn
where $I_{1}(p)$ is given by
\bnn
I_{1}(p)=\int\f{d^{d+1}q}{(2\pi)^{d+1}}2\pi\delta(q_{0})G(p-q)=\int\f{d^{d}\q}{(2\pi)^{d}}G(p_{0},\p-\q)=\int\f{d^{d}\q}{(2\pi)^{d}}G(p_{0},-\q).
\enn
With Eq. (\ref{e.Green2}), the Green function is given by
\bnn
G(p_{0},-\q)=\int^{1}_{0}dx\f{-C_{3}^{i}{\q'}^{2}q'_{i}+C_{2a}{\q'}^{2}+C_{2b}^{ij}q'_{i}q'_{j}-C_{1}^{i}q'_{i}+C_{0}}{\big[{\q'}^{2}+\Delta_{0}(p_{0};x)\big]^{2}},
\enn
where $\q'=\q+\tilde{\c}_{x}$.

Dropping $\q'$-odd terms, we have
\bea
I_{1} = \int^{1}_{0}dx\int\f{d^{d}\q'}{(2\pi)^{d}}\f{C_{2a}\q'^{2}+C_{2b}^{ij}q'_{i}q'_{j}+C_{0}}{\big[{\q'}^{2}+\Delta_{0}(p_{0};x)\big]^{2}} = \int^{1}_{0}dx\Bigg[\bigg(\f{d}{2}C_{2a}+\f{1}{2}C_{2b}^{ii}\bigg)\f{\Gamma(\f{2-d}{2})}{(4\pi)^{\f{d}{2}}\Delta_{0}^{\f{2-d}{2}}}+\f{C_{0}\Gamma(\f{4-d}{2})}{(4\pi)^{\f{d}{2}}\Delta_{0}^{\f{4-d}{2}}}\Bigg] \label{r.int1}.
\eea
Then, the self-energy is given by
\bnn
\Sigma^{(1)}(p) = \Gamma_{V}\int^{1}_{0}dx\Bigg[\f{\big(\f{d}{2}\bar{C}_{2a}+\f{1}{2}\bar{C}_{2b}^{ii}\big)\Gamma(\f{2-d}{2})}{(4\pi)^{\f{d}{2}}\Delta_{0}^{\f{2-d}{2}}} + \f{\bar{C}_{0}\Gamma(\f{4-d}{2})}{(4\pi)^{\f{d}{2}}\Delta_{0}^{\f{4-d}{2}}}\Bigg]+\Gamma_{U}\int^{1}_{0}dx\Bigg[\f{\big(\f{d}{2}C_{2a}+\f{1}{2}C_{2b}^{ii}\big) \Gamma(\f{2-d}{2})}{(4\pi)^{\f{d}{2}}\Delta_{0}^{\f{2-d}{2}}}+\f{C_{0}\Gamma(\f{4-d}{2})}{(4\pi)^{\f{d}{2}}\Delta_{0}^{\f{4-d}{2}}}\Bigg] ,
\enn
where we have introduced a bar-notation: $\bar{A}\equiv\gamma^{0}A\gamma^{0}$. Since we perform dimensional regularization in $d=2+\varepsilon$, the term containing $C_{0}$ gives only a finite value. A relevant part for renormalization is
\bea
\Sigma^{(1)}(p)
&\simeq&\f{\Gamma_{V}}{4\pi}\int^{1}_{0}dx\Bigg[\bigg(\f{d}{2}\big(p_{0}\gamma^{0}-\tilde{c}_{xk}\gamma^{k}+c_{0}\gamma^{0}\gamma^{5}-c_{k}\gamma^{k}\gamma^{5}\big)+\f{1}{2}\big(-2\tilde{c}_{xk}\gamma^{k}+2c_{k}\gamma^{k}\gamma^{5}\big)\bigg)\Gamma\bigg(\f{2-d}{2}\bigg)\bigg(\f{\Delta_{0}}{4\pi}\bigg)^{\f{d-2}{2}}\Bigg]\no\\
&&+\f{\Gamma_{U}}{4\pi}\int^{1}_{0}dx\Bigg[\bigg(\f{d}{2}\big(p_{0}\gamma^{0}+\tilde{c}_{xk}\gamma^{k}-c_{0}\gamma^{0}\gamma^{5}-c_{k}\gamma^{k}\gamma^{5}\big)+\f{1}{2}\big(2\tilde{c}_{xk}\gamma^{k}+2c_{k}\gamma^{k}\gamma^{5}\big)\bigg)\Gamma\bigg(\f{2-d}{2}\bigg)\bigg(\f{\Delta_{0}}{4\pi}\bigg)^{\f{d-2}{2}}\Bigg]\no\\
&=&-\f{\Gamma_{V}}{2\pi\varepsilon}\big(p_{0}\gamma^{0}+c_{0}\gamma^{0}\gamma^{5}\big)-\f{\Gamma_{U}}{2\pi\varepsilon}\big(p_{0}\gamma^{0}-c_{0}\gamma^{0}\gamma^{5}\big)+\mathcal{O}(1), \label{r.Fock}
\eea
where $\tilde{c}_{xk}$-terms vanish after the integration over $x$.

Based on this result, we find propagator counter terms in the following way
\bnn
-\f{\Gamma_{V}}{2\pi\varepsilon}\big(p_{0}\gamma^{0}+c_{0}\gamma^{0}\gamma^{5}\big)-\f{\Gamma_{U}}{2\pi\varepsilon}\big(p_{0}\gamma^{0}-c_{0}\gamma^{0}\gamma^{5}\big)+\O(1)+(\delta_{\psi}^{\omega}p_{0}\gamma^{0}+\delta_{\psi}^{\k}p_{k}\gamma^{k} +\delta_{c0}c_{0}\gamma^{0}\gamma^{5}+\delta_{\c}c_{k}\gamma^{k}\gamma^{5})=\tx{finite}.
\enn
As a result, propagator counter terms up to the one-loop level are obtained as
\be
\delta_{\psi}^{\omega}=\f{\Gamma_{V}}{2\pi\varepsilon}+\f{\Gamma_{U}}{2\pi\varepsilon},~~~~~\delta_{\psi}^{\k}=0,~~~~~\delta_{c0}=\f{\Gamma_{V}}{2\pi\varepsilon}-\f{\Gamma_{U}}{2\pi\varepsilon},~~~~~\delta_{\c}=0.\label{r.counter-propa1}
\ee

\subsubsection{Two-loop order I: Rainbow diagrams}

Next, we evaluate the rainbow diagrams
\bnn
\Sigma^{(2),r}(p)&=&\Gamma_{V}^{2}I_{3r}(p)[M_{1}=M_{2}=\gamma^{0}]+\Gamma_{V}\Gamma_{U}I_{3r}(p)[M_{1}=\gamma^{0},M_{2}=I_{4\times4}]\\
&&+\Gamma_{U}\Gamma_{V}I_{3r}(p)[M_{1}=I_{4\times4},M_{2}=\gamma^{0}]+\Gamma_{U}^{2}I_{3r}(p)[M_{1}=M_{2}=I_{4\times4}],
\enn
where $I_{3r}$ is given by
\bnn
I_{3r}(p)=\int\f{d^{d+1}q}{(2\pi)^{d+1}}2\pi\delta(q_{0})\int\f{d^{d+1}l}{(2\pi)^{d+1}}2\pi\delta(l_{0})M_{1}G(p-q)M_{2}G(p-q-l)M_{2}G(p-q)M_{1}.
\enn
We may simplify this expression with $I_{1}$ as
\bnn
I_{3r}(p)&=&\int\f{d^{d+1}q}{(2\pi)^{d+1}}2\pi\delta(q_{0})M_{1}G(p-q)M_{2}\bigg[\int\f{d^{d+1}l}{(2\pi)^{d+1}}2\pi\delta(l_{0})G(p-q-l)\bigg]M_{2}G(p-q)M_{1}\\
&=&\int\f{d^{d}\q}{(2\pi)^{d}}M_{1}G(p_{0},\p-\q)M_{2}I_{1}(p-q)M_{2}G(p_{0},\p-\q)M_{1}\\
&=&\int\f{d^{d}\q}{(2\pi)^{d}}M_{1}G(p_{0},-\q)M_{2}I_{1}(p_{0})M_{2}G(p_{0},-\q)M_{1},
\enn
where we used $I_{1}(p_{0},-\q)=I_{1}(p_{0})$.

Taking into account
\bnn
\f{1}{\big((p_{0}-c_{0})^{2}-(\q+\c)\big)^{2}\big((p_{0}+c_{0})^{2}-(\q-\c)\big)^{2}}=\int^{1}_{0}dy\f{6y(1-y)}{\big[{\q'}^{2}+\Delta_{0}(p_{0};y)\big]^{4}}
\enn
with $\q'=\q+\tilde{\c}_{y}$ and resorting to the representation of Eq. (\ref{e.Green2}), we reach the following expression
\fs\bnn
I_{3r} = \int^{1}_{0}dy6y(1-y)\int\f{d^{d}\q'}{(2\pi)^{d}}\f{M_{1}(-C_{3}^{i}\q'^{2}q'_{i}+C_{2a}\q'^{2}+C_{2b}^{ij}q'_{i}q'_{j}-C_{1}^{i}q'_{i}+C_{0}) M_{2}I_{1}M_{2}(-C_{3}^{k}\q'^{2}q'_{k}+C_{2a}\q'^{2}+C_{2b}^{kl}q'_{k}q'_{l}-C_{1}^{k}q'_{k}+C_{0})M_{1}}{\big[\q'^{2}+\Delta_{0}(p_{0};y)\big]^{4}} .
\enn\ns
There are many even terms contributing to the integration. However, it turns out that we have to consider the product of $C_{3}^{i}$s only. This is because the divergent part of $I_{1}$ is canceled by the one-loop self-energy diagrams containing the first-order counter term, so only the finite part of $I_{1}$ participates in the remaining calculation.${}^{\ddagger}$ In other words, divergences may arise only by the $q^{6}$-term in the $\q$-integration. For now, we just assume it (we will be back to this point later).

Keeping this term only, we have
\bnn
I_{3r}(p)&=&\int^{1}_{0}dy6y(1-y)\int\f{d^{d}\q}{(2\pi)^{d}}\f{(\q^{2})^{2}q_{i}q_{j}(M_{1}C_{3}^{i}M_{2}I_{1}M_{2}C_{3}^{j}M_{1})}{\big[\q^{2}+\Delta_{0}(p_{0};y)\big]^{4}}\\
&=&\f{(d+4)(d+2)\Gamma(\f{2-d}{2})}{8(4\pi)^{\f{d}{2}}}\int^{1}_{0}dy\f{y(1-y)}{\Delta_{0}^{\f{2-d}{2}}(y)}(M_{1}C_{3}^{i}M_{2}I_{1}M_{2}C_{3}^{i}M_{1}).
\enn
Then, the second-order self-energy correction for the rainbow diagrams is
\bnn
\Sigma^{(2),r}(p)=\f{(d+4)(d+2)\Gamma(\f{2-d}{2})}{8(4\pi)^{\f{d}{2}}} \int^{1}_{0}dy\f{y(1-y)}{\Delta_{0}^{\f{2-d}{2}}(y)} \bigg[\Gamma_{V}^{2}(\gamma^{i}I_{1}\gamma^{i})+2\Gamma_{V}\Gamma_{U}(\gamma^{0}\gamma^{i}I_{1}\gamma^{i}\gamma^{0})+\Gamma_{U}^{2}(\gamma^{i}I_{1}\gamma^{i})\bigg].
\enn

When performing the renormalization group analysis in the second order, we should include consistently one-loop self-energy corrections made of a tree-level vertex and a one-loop propagator counter term, given by (Fig. \ref{f.Fock-self-counter})
\bnn
\Sigma^{(1),\delta_{\psi}}(p)&=&\Gamma_{V}\int\f{d^{d+1}q}{(2\pi)^{d+1}}2\pi\delta(q_{0})\gamma^{0}G(p-q)(\delta_{\psi}^{\omega}p_{0}\gamma^{0} +\delta_{\psi}^{\k}p_{k}\gamma^{k}+\delta_{c0}c_{0}\gamma^{0}\gamma^{5}+\delta_{\c}c_{k}\gamma^{k} \gamma^{5})G(p-q)\gamma^{0}\\
&&+\Gamma_{U}\int\f{d^{d+1}q}{(2\pi)^{d+1}}2\pi\delta(q_{0})G(p-q)(\delta_{\psi}^{\omega}p_{0}\gamma^{0}+\delta_{\psi}^{\k}p_{k}\gamma^{k} +\delta_{c0}c_{0}\gamma^{0}\gamma^{5}+\delta_{\c}c_{k}\gamma^{k} \gamma^{5})G(p-q)\\
&=&\Gamma_{V}\f{(d+4)(d+2)\Gamma(\f{2-d}{2})}{8(4\pi)^{\f{d}{2}}}\int^{1}_{0}dy\f{y(1-y)}{\Delta_{0}^{\f{2-d}{2}}(y)}\gamma^{0}\gamma^{i}\big[-\Gamma_{V}\gamma^{0}\tx{div}(I_{1})\gamma^{0}-\Gamma_{U}\tx{div}(I_{1})\big]\gamma^{i}\gamma^{0}\\
&&+\Gamma_{U}\f{(d+4)(d+2)\Gamma(\f{2-d}{2})}{8(4\pi)^{\f{d}{2}}}\int^{1}_{0}dy\f{y(1-y)}{\Delta_{0}^{\f{2-d}{2}}(y)}\gamma^{i}\big[-\Gamma_{V}\gamma^{0}\tx{div}(I_{1})\gamma^{0}-\Gamma_{U}\tx{div}(I_{1})\big]\gamma^{i}\\
&=&-\f{(d+4)(d+2)\Gamma(\f{2-d}{2})}{8(4\pi)^{\f{d}{2}}}\int^{1}_{0}dy\f{y(1-y)}{\Delta_{0}^{\f{2-d}{2}}(y)}\bigg[\Gamma_{V}^{2}\gamma^{i}\tx{div}(I_{1})\gamma^{i}+2\Gamma_{V}\Gamma_{U}\gamma^{0}\gamma^{i}\tx{div}(I_{1})\gamma^{i}\gamma^{0}+\Gamma_{U}^{2}\gamma^{i}\tx{div}(I_{1})\gamma^{i}\bigg]\\
&\equiv&\Sigma^{(1),\delta_{\psi}}_{VV}(p)+\Sigma^{(1),\delta_{\psi}}_{VU}(p)+\Sigma^{(1),\delta_{\psi}}_{UV}(p)++\Sigma^{(1),\delta_{\psi}}_{UU}(p),
\enn
where $\tx{div}(\cdots)$ means the divergent part of $(\cdots)$. If we add these to the rainbow diagrams, the divergent part of $I_{1}$ in the rainbow diagrams is eliminated and only a finite part participates in the remaining computation (so the remark of $\ddagger$ is proved).

Writing it as $\tx{fin}(I_{1})\equiv I_{1}-\tx{div}(I_{1})=\int^{1}_{0}dx\f{C_{0}\Gamma(\f{4-d}{2})}{(4\pi)^{d/2}\Delta_{0}^{(4-d)/2}}$, we obtain
\bnn
\Sigma^{(2),r}+\Sigma^{(1),\delta_{\psi}} = \f{(d+4)(d+2)\Gamma(\f{2-d}{2})}{8(4\pi)^{\f{d}{2}}}\int^{1}_{0}dy\f{y(1-y)}{\Delta_{0}^{\f{2-d}{2}}(y)} \bigg[\Gamma_{V}^{2}\gamma^{i}\tx{fin}(I_{1})\gamma^{i}+2\Gamma_{V}\Gamma_{U}\gamma^{0}\gamma^{i}\tx{fin}(I_{1})\gamma^{i}\gamma^{0}+\Gamma_{U}^{2}\gamma^{i}\tx{fin}(I_{1})\gamma^{i}\bigg].
\enn

\bf[b]
\ing[width=.85\tw]{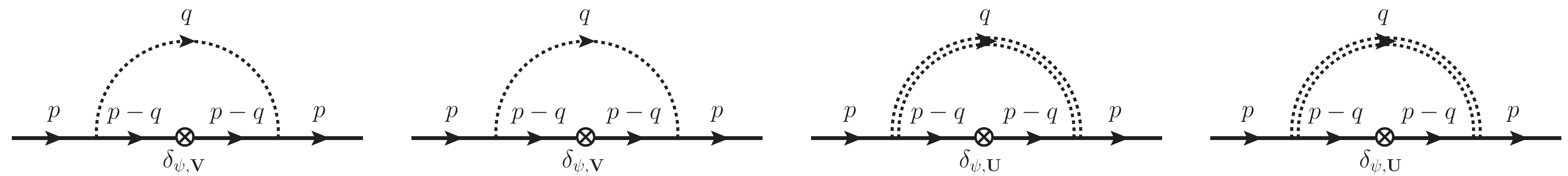}
\caption{One-loop self-energy diagrams containing the first-order propagator counter terms given in Eq. (\ref{r.counter-propa1}). Added to the rainbow diagrams, these contributions cancel the divergent part of $I_{1}$ in the rainbow diagrams, leaving only a finite part of $I_{1}$ to participate in the remaining calculation.} \label{f.Fock-self-counter}
\ef

An expansion about $d=2+\varepsilon$ gives
\bnn
\f{(d+4)(d+2)\Gamma(\f{2-d}{2})}{8(4\pi)^{\f{d}{2}}} \int^{1}_{0}dy\f{y(1-y)}{\Delta_{0}^{\f{2-d}{2}}(y)} \int^{1}_{0}dx \f{C_{0}(x)\Gamma(\f{4-d}{2})}{(4\pi)^{\f{d}{2}}\Delta_{0}^{\f{4-d}{2}}(x)}=-\f{1}{16\pi^{2}\varepsilon}\int^{1}_{0}dx\f{C_{0}(x)}{\Delta_{0}(x)}+\O(1).
\enn
As a result, the relevant part for renormalization is given as
\bnn
\Sigma^{(2),r}+\Sigma^{(1),\delta_{\psi}}=-\f{1}{16\pi^{2}\varepsilon}\bigg[\Gamma_{V}^{2}\gamma^{i}\int^{1}_{0}dx\f{C_{0}(x)}{\Delta_{0}(x)}\gamma^{i}+2\Gamma_{V}\Gamma_{U}\gamma^{0}\gamma^{i}\int^{1}_{0}dx\f{C_{0}(x)}{\Delta_{0}(x)}\gamma^{i}\gamma^{0}+\Gamma_{U}^{2}\gamma^{i}\int^{1}_{0}dx\f{C_{0}(x)}{\Delta_{0}(x)}\gamma^{i}\bigg]+\O(1).
\enn

The remaining calculation is $\int^{1}_{0}dx\f{C_{0}(x)}{\Delta_{0}(x)}$. A straightforward calculation gives
\bnn
&&p_{0}\gamma^{0}\bigg[-1-\f{\alpha}{2}\ln{\bigg(\f{\alpha-1}{\alpha+1}\bigg)}-\f{\beta}{2}\ln{\bigg(\f{\beta-1}{\beta+1}\bigg)}\bigg]+c_{0}\gamma^{0}\bigg[\f{1}{2}\ln{\bigg(\f{\alpha-1}{\alpha+1}\bigg)}+\f{1}{2}\ln{\bigg(\f{\beta-1}{\beta+1}\bigg)}\bigg]\\
&&+c_{k}\gamma^{k}\bigg[-(\alpha+\beta)+\f{1-\alpha^{2}}{2}\ln{\bigg(\f{\alpha-1}{\alpha+1}\bigg)}+\f{1-\beta^{2}}{2}\ln{\bigg(\f{\beta-1}{\beta+1}\bigg)}\bigg]+c_{0}\gamma^{0}\gamma^{5}\bigg[1+\f{\alpha}{2}\ln{\bigg(\f{\alpha-1}{\alpha+1}\bigg)}+\f{\beta}{2}\ln{\bigg(\f{\beta-1}{\beta+1}\bigg)}\bigg]\\
&&+p_{0}\gamma^{0}\gamma^{5}\bigg[-\f{1}{2}\ln{\bigg(\f{\alpha-1}{\alpha+1}\bigg)}-\f{1}{2}\ln{\bigg(\f{\beta-1}{\beta+1}\bigg)}\bigg]+c_{k}\gamma^{k}\gamma^{5}(-1) ,
\enn\ns
where $(\alpha,\beta)\equiv ab\pm\sqrt{(a^{2}-1)(b^{2}-1)}$ and $a\equiv\f{p_{0}}{\al\c\ar}$, $b\equiv\f{c_{0}}{\al\c\ar}$.

Dropping the complex logarithm terms, we have
\bnn
\gamma^{i}\int^{1}_{0}dx\f{C_{0}(x)}{\Delta_{0}(x)}\gamma^{i}&\simeq&\gamma^{i}\Big(-p_{0}\gamma^{0}-\f{2p_{0}c_{0}}{\al\c\ar^{2}}c_{k}\gamma^{k}+c_{0}\gamma^{0}\gamma^{5}-c_{k}\gamma^{k}\gamma^{5}\Big)\gamma^{i}\\
&=&-dp_{0}\gamma^{0}+(2-d)\f{2p_{0}c_{0}}{\al\c\ar^{2}}c_{k}\gamma^{k}-dc_{0}\gamma^{0}\gamma^{5}+(d-2)c_{k}\gamma^{k}\gamma^{5}.
\enn
As a result, the self-energy correction from rainbow diagrams is
\bea
\Sigma^{(2),r}(p)+\Sigma^{(1),\delta_{\psi}}(p) = \f{1}{8\pi^{2}\varepsilon}\bigg[\Gamma_{V}^{2}(p_{0}\gamma^{0}+c_{0}\gamma^{0}\gamma^{5}) +\Gamma_{U}^{2}(p_{0}\gamma^{0}+c_{0}\gamma^{0}\gamma^{5})+2\Gamma_{V}\Gamma_{U}(p_{0}\gamma^{0}-c_{0}\gamma^{0}\gamma^{5})\bigg]+\O(1) , \label{r.rainbow}
\eea
where the result is depicted pictorially in Fig. \ref{sum-rainbow}.

\bf[b]
\ing[width=.9\tw]{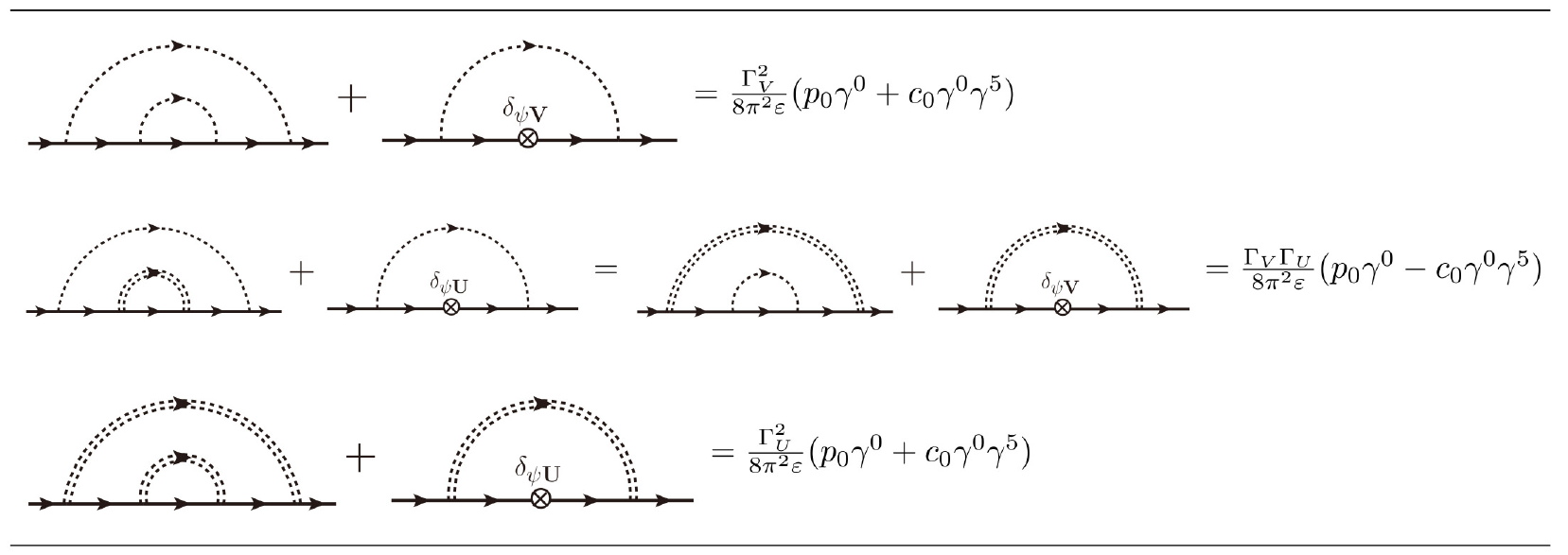} \caption{The result for the rainbow diagrams. Each rainbow diagram is added consistently by one-loop self-energy diagrams made of a tree-level vertex and a one-loop propagator counter term.} \label{sum-rainbow}
\ef

\subsubsection{Two-loop order II: Crossed diagrams}

Last, we evaluate the crossed diagrams
\bnn
\Sigma^{(2),c}(p)&=&\Gamma_{V}^{2}I_{3c}(p)[M_{1}=M_{2}=\gamma^{0}]+\Gamma_{V}\Gamma_{U}I_{3c}(p)[M_{1}=\gamma^{0},M_{2}=I_{4\times4}]\\
&&+\Gamma_{U}\Gamma_{V}I_{3c}(p)[M_{1}=I_{4\times4},M_{2}=\gamma^{0}]+\Gamma_{U}^{2}I_{3c}(p)[M_{1}=M_{2}=I_{4\times4}],
\enn
where $I_{3c}$ is given by
\bnn
I_{3c}(p)&=&\int\f{d^{d+1}q}{(2\pi)^{d+1}}2\pi\delta(q_{0})\int\f{d^{d+1}l}{(2\pi)^{d+1}}2\pi\delta(l_{0})M_{1}G(p-q)M_{2}G(p-q-l)M_{1}G(p-l)M_{2}\\
&=&\int\f{d^{d}\q}{(2\pi)^{d}}\int\f{d^{d}\l}{(2\pi)^{d}}M_{1}G(p_{0},\p-\q)M_{2}G(p_{0},\p-\q-\l)M_{1}G(p_{0},\p-\l)M_{2}\\
&=&\int\f{d^{d}\q}{(2\pi)^{d}}\int\f{d^{d}\l}{(2\pi)^{d}}M_{1}G(p_{0},-\q)M_{2}G(p_{0},-\q-\l)M_{1}G(p_{0},-\l+\p)M_{2}.
\enn
In this case the loop momenta of $\l$ and $\q$ are interwoven and this makes the analysis more complicated.

First, we perform the integration on $\q$. Using Eq. (\ref{e.Green2}), we have
\bnn
&&\int^{1}_{0}dx\int^{1}_{0}dy\int\f{d^{d}\q}{(2\pi)^{d}}\f{-C_{3}^{i}\q'^{2}q_{i}+C_{2a}\q'^{2}+C_{2b}^{ij}q'_{i}q'_{j}-C_{1}^{i}q'_{i}+C_{0}}{\big[\q'^{2}+\Delta_{0}(p_{0};x)\big]^{2}}\\
&&\times M_{2}\f{-C_{3}^{k}(\q'+\l)^{2}(q'_{k}+l_{k})+C_{2a}(\q'+\l)^{2}+C_{2b}^{kl}(q'_{k}+l_{k})(q'_{l}+l_{l})-C_{1}^{l}(q'_{l}+l_{l})+C_{0}}{\big[(\q'+\l)^{2}+\Delta_{0}(p_{0};y)\big]^{2}}.
\enn
Denominators are combined as
\bnn
\f{1}{\big[\q'^{2}+\Delta_{0}(p_{0};x)\big]^{2}\big[(\q'+\l)^{2}+\Delta_{0}(p_{0};y)\big]^{2}}=\int^{1}_{0}dz\f{6z(1-z)}{\big[(\q'+z\l)^{2}+\Delta_{1}(p_{0},\l;x,y,z)\big]^{4}},
\enn
where $\Delta_{1}=z(1-z)\l^{2}+(1-z)\Delta_{0}(p_{0};x)+z\Delta_{0}(p_{0};y)$. Shifting $\q'\rightarrow\q'-z\l$ and renaming $\q'$ as $\q$, we have
\fs\bnn
&&\int^{1}_{0}dx\int^{1}_{0}dy\int^{1}_{0}dz\int\f{d^{d}\q}{(2\pi)^{d}}\f{6z(1-z)}{\big[\q^{2}+\Delta_{1}\big]^{4}}\Big(-C_{3}^{i}(\q-z\l)^{2}(q_{i}-zl_{i})+C_{2a}(\q-z\l)^{2}+C_{2b}^{ij}(q_{i}-zl_{i})(q_{j}-zl_{j})-C_{1}^{i}(q_{i}-zl_{i})+C_{0}\Big)\\
&&\times M_{2}\Big(-C_{3}^{k}\big(\q+(1-z)\l\big)^{2}\big(q_{k}+(1-z)l_{k}\big)+C_{2a}\big(\q+(1-z)\l\big)^{2}+C_{2b}^{kl}\big(q_{k}+(1-z)l_{k}\big)\big(q_{l}+(1-z)l_{l}\big)-C_{1}^{k}\big(q_{k}+(1-z)l_{k}\big)+C_{0}\Big).
\enn\ns

Despite this complex expression, we need to consider only a few terms for renormalization. This can be understood, considering a simple integral
\bea
\int\f{d^{d}\q}{(2\pi)^{d}}\f{(\q^{2})^{m}}{[\q^{2}+\Delta]^{4}}=\f{\Gamma(\f{8-d-2m}{2})\Gamma(\f{d}{2}+m)}{(4\pi)^{\f{d}{2}}\Gamma(\f{d}{2})\Gamma(4)\Delta^{\f{8-d-2m}{2}}}\label{e.simplified integral}.
\eea
Since we resort to the dimensional regularization in $d=2+\varepsilon$, an integral for $m$ smaller than $3$ gives a finite value, and it doesn't participate in renormalization. The product of the $q^{3}$-terms (i.e. $q^{6}$-term) certainly gives renormalization effects. Other than $q^{6}$-term, even terms of $q^{4}l^{2}$, $q^{2}l^{4}$ and $l^{6}$ possibly contribute to renormalization after the $\l$-integral because there will be an equal number of momentum $l$ in the denominator and numerator (considering the dimension of an integrand, this fact may be easily estimated, because any dimensionful constant in numerator lowers the superficial degree of divergence of the integral). All of those come from the product of the $C_{3}$-terms, so the relevant part is
\bnn
\int^{1}_{0}dx\int^{1}_{0}dy\int^{1}_{0}dz6z(1-z)\int\f{d^{d}\q}{(2\pi)^{d}}\f{(\q-z\l)^{2}(\slashed{\q}-z\slashed{\l})M_{2}\big(\q+(1-z)\l\big)^{2}\big(\slashed{\q}+(1-z)\slashed{\l}\big)}{\big[\q^{2}+\Delta_{1}(p_{0},\l;x,y,z)\big]^{4}},
\enn
where $\slashed{\q}\equiv q_{i}\gamma^{i}(i=1,2,3)$.

The numerator is arranged as
\bnn
N&=&(-1)^{1+\f{1}{4}\tr[M_{2}]}M_{2}(\q-z\l)^{2}\big(\q+(1-z)\l\big)^{2}\big(\slashed{\q}-z\slashed{\l}\big)\big(\slashed{\q}+(1-z)\slashed{\l}\big)\\
&=&(-1)^{\f{1}{4}\tr[M_{2}]}M_{2}\Big[D_{6}(\q^{2})^{3}+D_{4a}(\q^{2})^{2}+D_{4b}^{ij}\q^{2}q_{i}q_{j}+D_{2a}\q^{2}+D_{2b}^{ij}q_{i}q_{j}+D_{0}\Big]+(\tx{odd terms}) ,
\enn
where the coefficients are given by
\bnn
D_{6}&=&1,\\
D_{4a}&=&(3z^{2}-3z+1)\l^{2},\\
D_{4b}^{ij}&=&(12z^{2}-8z)l^{i}l^{j}+(2-4z)l^{i}\gamma^{j}\slashed{\l},\\
D_{2a}&=&(3z^{4}-6z^{3}+4z^{2}-z)(\l^{2})^{2},\\
D_{2b}^{ij}&=&(12z^{4}-20z^{3}+8z^{2})\l^{2}l^{i}l^{j}+(-4z^{3}+6z^{2}-2z)\l^{2}l^{i}\gamma^{j}\slashed{\l},\\
D_{0}&=&z^{3}(z-1)^{3}(\l^{2})^{3}.
\enn

Now, the integral is easily performed to be
\small\bnn
&&\int\f{d^{d}\q}{(2\pi)^{d}}\f{(-1)^{\f{1}{4}\tr[M_{2}]}M_{2}\big[D_{6}(\q^{2})^{3}+D_{4a}(\q^{2})^{2}+D_{4b}^{ij}\q^{2}q_{i}q_{j}+D_{2a}\q^{2}+D_{2b}^{ij}q_{i}q_{j}+D_{0}\big]}{\big[\q^{2}+\Delta_{1}(\l;x,y,z)\big]^{4}}\\
&=&\f{(-1)^{\f{1}{4}\tr[M_{2}]}M_{2}}{(4\pi)^{\f{d}{2}}}\bigg[\f{d(d+4)(d+2)\Gamma\big(\f{2-d}{2}\big)}{8\Gamma(4)}\f{D_{6}}{\Delta_{1}^{\f{2-d}{2}}}+\f{d(d+2)\Gamma\big(\f{4-d}{2}\big)}{4\Gamma(4)}\f{D_{4a}+\f{D_{4b}^{ii}}{d}}{\Delta_{1}^{\f{4-d}{2}}}+\f{d\Gamma\big(\f{6-d}{2}\big)}{2\Gamma(4)}\f{D_{2a}+\f{D_{2b}^{ii}}{d}}{\Delta_{1}^{\f{6-d}{2}}}+\f{\Gamma\big(\f{8-d}{2}\big)}{\Gamma(4)}\f{D_{0}}{\Delta_{1}^{\f{8-d}{2}}}\bigg].
\enn\ns

Next, we perform the $\l$-integral. Using $M_{1}M_{2}M_{1}=M_{2}$ (since the matrices of $M_{1}$ and $M_{2}$ are either $I_{4\times4}$ or $\gamma^{0}$), we have
\bnn
I_{3c}(p)&=&\f{(-1)^{\f{1}{4}\tr[M_{2}]}}{(4\pi)^{\f{d}{2}}}\int^{1}_{0}dx\int^{1}_{0}dy\int^{1}_{0}dzz(1-z)\int\f{d^{d}\l}{(2\pi)^{d}}\Bigg[\f{d(d+4)(d+2)\Gamma\big(\f{2-d}{2}\big)}{8}\f{D_{6}}{\Delta_{1}^{\f{2-d}{2}}}\\
&&+\f{d(d+2)\Gamma\big(\f{4-d}{2}\big)}{4}\f{D_{4a}+\f{D_{4b}^{ii}}{d}}{\Delta_{1}^{\f{4-d}{2}}}+\f{d\Gamma\big(\f{6-d}{2}\big)}{2}\f{D_{2a}+\f{D_{2b}^{ii}}{d}}{\Delta_{1}^{\f{6-d}{2}}}+\f{\Gamma\big(\f{8-d}{2}\big)D_{0}}{\Delta_{1}^{\f{8-d}{2}}}\Bigg]M_{2}G(p_{0},-\l+\p)M_{2}+\tx{(finite parts)}.
\enn

Taking out $z(1-z)$ from $\Delta_{1}=z(1-z)\l^{2}+(1-z)\Delta_{0}(x)+z\Delta_{0}(y)$ first, we find that the remaining integrals are such a simple form:
\bnn
&&\int^{1}_{0}dv\int\f{d^{d}\l}{(2\pi)^{d}}\f{(\l^{2})^{n}}{\big[\l^{2}+\f{1}{z}\Delta_{0}(p_{0};x)+\f{1}{1-z}\Delta_{0}(p_{0};y)\big]^{n+\f{2-d}{2}}}\\
&&\times\f{-(\l-\p)^{2}(l_{i}-p_{i})\gamma^{i}+(\l-\p)^{2}(p_{0}\gamma^{0}-\sc\gamma^{5})+(l_{i}-p_{i})(l_{j}-p_{j})(-2c^{i}\gamma^{j}\gamma^{5})-(l_{i}-p_{i})f_{1}^{i}(p_{0})+f_{0}(p_{0})}{\big[\big(\l-\p+\tilde{\c}_{v}\big)^{2}+\Delta_{0}(p_{0};v)\big]^{2}},
\enn
where the cases of $n=0,1,2,3$ correspond to integrals for $D_{6},~D_{4},~D_{2}$ and $D_{0}$, respectively. Such integrations result in $\Gamma(n+\f{2-d}{2}+2-\f{d}{2}-n-m)=\Gamma(3-d-m)$, where $m=1$ stands for $l^{2}$ (the leading even-term) and $m=0$ for a constant term in the propagator. Within the dimensional regularization in $d=2+\varepsilon$, only the integral of $m=1$ possibly gives a divergent factor of $\Gamma(2-d)$. However, in the $n=0$ case we already got $\Gamma(\f{2-d}{2})$, and we need to consider the constant ($m=0$) term, which turns out to be important. We first compute this term.

The denominator is transformed as
\bnn
\int^{1}_{0}dv\f{1}{\big[\l^{2}+\f{\Delta_{0}(x)}{z}+\f{\Delta_{0}(y)}{1-z}\big]^{\f{2-d}{2}}\big[\big(\l-\p+\tilde{\c}_{v}\big)^{2}+\Delta_{0}(v)\big]^{2}}=\int^{1}_{0}dv\int^{1}_{0}dw\f{w(1-w)^{-\f{d}{2}}\Gamma\big(\f{6-d}{2}\big)/\Gamma\big(\f{2-d}{2}\big)}{\big[\big(\l-w(\p-\tilde{\c}_{v})\big)^{2}+\Delta_{2}(\p;x,y,z,v,w)\big]^{\f{6-d}{2}}},
\enn
where $\Delta_{2}=w(1-w)(\p-\tilde{\c}_{v})^{2}+\f{1-w}{z}\Delta_{0}(x)+\f{1-w}{1-z}\Delta_{0}(y)+w\Delta_{0}(v)$. This suggests that we may use Eq. (\ref{e.Green2}) with a slight change. Then, the integral for $m=0$ is
\bnn
\f{\Gamma\big(\f{6-d}{2}\big)}{\Gamma\big(\f{2-d}{2}\big)}\int^{1}_{0}dv\int^{1}_{0}dww(1-w)^{-\f{d}{2}}\int\f{d^{d}\l'}{(2\pi)^{d}}\f{C_{0}(w,v)}{\big[\l'^{2}+\Delta_{2}(w,v)\big]^{\f{6-d}{2}}}=\f{\Gamma(3-d)}{(4\pi)^{\f{d}{2}}\Gamma\big(\f{2-d}{2}\big)}\int^{1}_{0}dv\int^{1}_{0}dww(1-w)^{-\f{d}{2}}\f{C_{0}(w,v)}{\Delta_{2}(w,v)},
\enn
where $C_{0}(w,v)$ is same with that of  Eq. (\ref{e.Green2}) except for $u=(p_{0},(1-w)\p+w\tilde{\c}_{v})$. Note $C_{0}(w=1,v)=C_{0}(v)$ (the original definition of $C_{0}$) and $\Delta_{2}(w=1,v)=\Delta_{0}(v)$.

This implies that we may take out a relevant part in the following way
\bnn
&&\f{\Gamma(3-d)}{(4\pi)^{\f{d}{2}}\Gamma\big(\f{2-d}{2}\big)}\int^{1}_{0}dv\int^{1}_{0}dww(1-w)^{-\f{d}{2}}\Bigg[\f{C_{0}(v)}{\Delta_{0}(v)}+\f{d}{dw}\left.\f{C_{0}(w,v)}{\Delta_{2}(w,v)}\right|_{w=1}(w-1)+\O(w-1)^{2}\Bigg]\\
&=&\f{\Gamma(3-d)}{(4\pi)^{\f{d}{2}}\Gamma\big(\f{6-d}{2}\big)}\int^{1}_{0}dv\f{C_{0}(v)}{\Delta_{0}(v)}-\f{\f{4\Gamma(3-d)}{(6-d)(4-d)}}{(4\pi)^{\f{d}{2}}\Gamma\big(\f{2-d}{2}\big)}\int^{1}_{0}dv\f{d}{dw}\left.\f{C_{0}(w,v)}{\Delta_{2}(w,v)}\right|_{w=1}+\cdots.
\enn
Note that $\Gamma(\f{2-d}{2})$ in the first term is canceled after the $w$-integral, but $\Gamma\big(\f{2-d}{2}\big)$ in the second term is not. Together with $\Gamma\big(\f{2-d}{2}\big)$ originating from the $\q$-integral, the first term contributes to a divergent part while the other higher-order terms give only finite values. In short, the above analysis suggests that we should include $\int^{1}_{0}dv\f{C_{0}(v)}{\Delta_{0}(v)}$.

Now, we focus on the $m=1$ case. Since $l^{2}$ may arise from $l^{2}$ (surely) and $l^{3}$ (after momentum shift), we're keeping them. After the similar analysis as the above, we obtain
\fs\bnn
&&\int^{1}_{0}dv\int\f{d^{d}\l}{(2\pi)^{d}}\f{(\l^{2})^{n}}{\big[\l^{2}+\f{1}{z}\Delta_{0}(x)+\f{1}{1-z}\Delta_{0}(y)\big]^{n+\f{2-d}{2}}}\f{\l^{2}(\sp-\sc\gamma^{5})-\l^{2}l_{i}\gamma^{i}+l_{i}l_{j}(-2p^{i}\gamma^{j}-2c^{i}\gamma^{j}\gamma^{5})}{\big[\big(\l-\p+\tilde{\c}_{v}\big)^{2}+\Delta_{0}(p_{0};v)\big]^{2}}\\
&=&\int^{1}_{0}dv\int^{1}_{0}dw\f{w(1-w)^{n-\f{d}{2}}\Gamma\big(n+\f{6-d}{2}\big)}{\Gamma(2)\Gamma\big(n+\f{2-d}{2}\big)}\int\f{d^{d}\l}{(2\pi)^{d}}\f{(\l^{2})^{n}\big[\l^{2}(\sp-\sc\gamma^{5})-\l^{2}l_{i}\gamma^{i}+l_{i}l_{j}(-2p^{i}\gamma^{j}-2c^{i}\gamma^{j}\gamma^{5})\big]}{\big[\big(\l-w(\p-\tilde{\c}_{v})\big)^{2}+\Delta_{2}(\p;x,y,z,v,w)\big]^{n+\f{6-d}{2}}}\\
&\simeq&\f{\Gamma\big(n+\f{6-d}{2}\big)}{\Gamma\big(n+\f{2-d}{2}\big)}\int^{1}_{0}dv\int^{1}_{0}dww(1-w)^{n-\f{d}{2}}\int\f{d^{d}\l}{(2\pi)^{d}}\f{(\l^{2})^{n}\big[\l^{2}(\sp-\sc\gamma^{5}-w(p_{i}-\tilde{c}_{vi})\gamma^{i})+l_{i}l_{j}\big(2(n+1)w(p^{i}-\tilde{c}_{v}^{i})\gamma^{j}-2p^{i}\gamma^{j}-2c^{i}\gamma^{j}\gamma^{5}\big)\big]}{\big[\l^{2}+\Delta_{2}(\p;x,y,z,v,w)\big]^{n+\f{6-d}{2}}},
\enn\ns
where we have shifted $\l\rightarrow\l+w(\p-\tilde{\c}_{v})$ and kept only the leading even terms including shifted contributions from $(\l^{2})^{n}$ and $\l^{2}l_{i}$.

After the $\bm{l}$-integration, we reach the following expression
\fs\bnn
\f{\Gamma(2-d)\Gamma\big(\f{d}{2}+n+1\big)}{(4\pi)^{\f{d}{2}}\Gamma\big(\f{d}{2}\big)\Gamma\big(n+\f{2-d}{2}\big)}\int^{1}_{0}dv\int^{1}_{0}dww(1-w)^{n-\f{d}{2}}\Delta_{2}^{d-2}\bigg(\sp-\sc\gamma^{5}-w(p_{i}-\tilde{c}_{vi})\gamma^{i}+\f{1}{d}\big(-2(n+1)w(p_{i}-\tilde{c}_{vi})\gamma^{i}+2p_{i}\gamma^{i}+2c_{i}\gamma^{i}\gamma^{5}\big)\bigg).
\enn\ns

Considering $d=2+\varepsilon$, $\Delta_{2}^{d-2}$ is not involved in the $w$- and the $v$-integral. The effect of the $v$-integral is just to remove $\tilde{c}_{vi}$. The $w$-integral gives
\bnn
\f{\Gamma(2-d)}{(4\pi)^{\f{d}{2}}\Gamma\big(\f{d}{2}\big)}\f{\Gamma\big(\f{d}{2}+n+1\big)}{\Gamma\big(n+\f{6-d}{2}\big)}\bigg(\sp-\sc\gamma^{5}+\f{2}{d}\big(p_{i}\gamma^{i}+c_{i}\gamma^{i}\gamma^{5}\big)-\f{2}{n+\f{6-d}{2}}\f{2(n+1)+d}{d}p_{i}\gamma^{i}\bigg),
\enn
where $2/(n+\f{6-d}{2})$ makes up for the difference due to additional $w$. Among the remaining Feynman parameters of $x,y$ and $z$, only $z$ is effective since there are polynomials of $z$ in the $D$s.

The $z$-integrals for each $n$ are performed as (from the first line, $n=0,1,2,3$)
\bnn
&&\int^{1}_{0}dz\f{z(1-z)}{z^{\f{2-d}{2}}(1-z)^{\f{2-d}{2}}}=\f{\big[\Gamma\big(\f{d+2}{2}\big)\big]^{2}}{\Gamma(d+2)},\\
&&\int^{1}_{0}dzz(1-z)\f{1-3z(1-z)+\f{2-12z(1-z)}{d}}{z^{\f{4-d}{2}}(1-z)^{\f{4-d}{2}}}=\f{d^{2}+8}{d^{2}}\f{\big[\Gamma\big(\f{d+2}{2}\big)\big]^{2}}{\Gamma(d+2)},\\
&&\int^{1}_{0}dzz(1-z)\f{z(1-z)\Big(-1+3z(1-z)+\f{-2+12z(1-z)}{d}\Big)}{z^{\f{6-d}{2}}(1-z)^{\f{6-d}{2}}}=-\f{d^{2}+8}{d^{2}}\f{\big[\Gamma\big(\f{d+2}{2}\big)\big]^{2}}{\Gamma(d+2)},\\
&&\int^{1}_{0}dzz(1-z)\f{-z^{3}(1-z)^{3}}{z^{\f{8-d}{2}}(1-z)^{\f{8-d}{2}}}=-\f{\big[\Gamma\big(\f{d+2}{2}\big)\big]^{2}}{\Gamma(d+2)}.
\enn
As a result, we obtain
\bnn
I_{3c}(p)&=&\f{\Gamma\big(\f{2-d}{2}\big)\Gamma(2-d)}{(4\pi)^{d}\Gamma\big(\f{d}{2}\big)}\f{\big[\Gamma\big(\f{d+2}{2}\big)\big]^{2}}{\Gamma(d+2)}\Bigg[(-1)^{\f{1}{4}\tr[M_{2}]}M_{2}\bigg(\sp-\sc\gamma^{5}+\f{2}{d}\big(p_{k}\gamma^{k}+c_{k}\gamma^{k}\gamma^{5}\big)\bigg)M_{2}\\
&&\times\Bigg(\f{(d+4)(d+2)d}{8}\f{\Gamma\big(\f{d+2}{2}\big)}{\Gamma\big(\f{6-d}{2}\big)}+\f{(d+2)d(2-d)}{8}\f{\Gamma\big(\f{d+2}{2}+1\big)}{\Gamma\big(\f{6-d}{2}+1\big)}\f{d^{2}+8}{d^{2}}\\
&&-\f{d(4-d)(2-d)}{8}\f{\Gamma\big(\f{d+2}{2}+2\big)}{\Gamma\big(\f{6-d}{2}+2\big)}\f{d^{2}+8}{d^{2}}-\f{(6-d)(4-d)(2-d)}{8}\f{\Gamma\big(\f{d+2}{2}+3\big)}{\Gamma\big(\f{6-d}{2}+3\big)}\Bigg)\\
&&-\f{2}{d}(-1)^{\f{1}{4}\tr[M_{2}]}M_{2}(p_{k}\gamma^{k})M_{2}\Bigg(\f{(d+4)(d+2)d}{8}\f{2+d}{\f{6-d}{2}}\f{\Gamma\big(\f{d+2}{2}\big)}{\Gamma\big(\f{6-d}{2}\big)}+\f{(d+2)d(2-d)}{8}\f{4+d}{\f{6-d}{2}+1}\f{\Gamma\big(\f{d+2}{2}+1\big)}{\Gamma\big(\f{6-d}{2}+1\big)}\f{d^{2}+8}{d^{2}}\\
&&-\f{d(4-d)(2-d)}{8}\f{6+d}{\f{6-d}{2}+2}\f{\Gamma\big(\f{d+2}{2}+2\big)}{\Gamma\big(\f{6-d}{2}+2\big)}\f{d^{2}+8}{d^{2}}-\f{(6-d)(4-d)(2-d)}{8}\f{8+d}{\f{6-d}{2}+3}\f{\Gamma\big(\f{d+2}{2}+3\big)}{\Gamma\big(\f{6-d}{2}+3\big)}\Bigg)\Bigg]\\
&&+\f{\Gamma\big(\f{2-d}{2}\big)\Gamma(3-d)}{(4\pi)^{d}\Gamma\big(\f{6-d}{2}\big)}\f{\big[\Gamma\big(\f{d+2}{2}\big)\big]^{2}}{\Gamma(d+2)}\f{(d+4)(d+2)d}{8}(-1)^{\f{1}{4}\tr[M_{2}]}M_{2}\bigg[\int^{1}_{0}dv\f{C_{0}(v)}{\Delta_{0}(v)}\bigg]M_{2}\\
&=&\bigg(\f{1}{8\pi^{2}\varepsilon^{2}}+\f{5+6\gamma-6\ln{4\pi}}{8\pi^{2}\varepsilon}\bigg)(-1)^{\f{1}{4}\tr[M_{2}]}(p_{0}\gamma^{0})+\f{1}{16\pi^{2}\varepsilon}(p_{k}\gamma^{k})+\bigg(\f{1}{8\pi^{2}\varepsilon^{2}}+\f{5+6\gamma-6\ln{4\pi}}{8\pi^{2}\varepsilon}\bigg)(c_{0}\gamma^{0}\gamma^{5})\\
&&-\f{1}{16\pi^{2}\varepsilon}(-1)^{\f{1}{4}\tr[M_{2}]}(c_{k}\gamma^{k}\gamma^{5})-\f{1}{8\pi^{2}\varepsilon}(-1)^{\f{1}{4}\tr[M_{2}]}M_{2}\bigg[\int^{1}_{0}dv\f{C_{0}(v)}{\Delta_{0}(v)}\bigg]M_{2}+\O(1),
\enn
where we have used the matrix identities:
\bnn
&&(-1)^{\f{1}{4}\tr[M_{2}]}M_{2}\gamma^{0}M_{2}=(-1)^{\f{1}{4}\tr[M_{2}]}\gamma^{0},~~(-1)^{\f{1}{4}\tr[M_{2}]}M_{2}\gamma^{0}\gamma^{5}M_{2}=-\gamma^{0}\gamma^{5},\\
&&(-1)^{\f{1}{4}\tr[M_{2}]}M_{2}\gamma^{k}M_{2}=-\gamma^{k},~~(-1)^{\f{1}{4}\tr[M_{2}]}M_{2}\gamma^{k}\gamma^{5}M_{2}=(-1)^{\f{1}{4}\tr[M_{2}]}\gamma^{k}\gamma^{5}.
\enn

Finally, the self-energy correction from the crossed diagrams is
\bnn
&&\Sigma^{(2),c}(p)\\
&=&\Gamma_{V}^{2}\bigg[\bigg(\f{1}{8\pi^{2}\varepsilon^{2}}+\f{5+6\gamma-6\ln{4\pi}}{48\pi^{2}\varepsilon}\bigg)(p_{0}\gamma^{0}+c_{0}\gamma^{0}\gamma^{5})+\f{1}{16\pi^{2}\varepsilon}p_{k}\gamma^{k}-\f{1}{16\pi^{2}\varepsilon}c_{k}\gamma^{k}\gamma^{5}-\f{1}{8\pi^{2}\varepsilon}\int^{1}_{0}dv\f{\bar{C}_{0}(v)}{\Delta_{0}(v)}\bigg]\\
&&+\Gamma_{V}\Gamma_{U}\bigg[\bigg(\f{1}{8\pi^{2}\varepsilon^{2}}+\f{5+6\gamma-6\ln{4\pi}}{48\pi^{2}\varepsilon}\bigg)(-p_{0}\gamma^{0}+c_{0}\gamma^{0}\gamma^{5})+\f{1}{16\pi^{2}\varepsilon}p_{k}\gamma^{k}+\f{1}{16\pi^{2}\varepsilon}c_{k}\gamma^{k}\gamma^{5}+\f{1}{8\pi^{2}\varepsilon}\int^{1}_{0}dv\f{C_{0}(v)}{\Delta_{0}(v)}\bigg]\\
&&+\Gamma_{U}\Gamma_{V}\bigg[\bigg(\f{1}{8\pi^{2}\varepsilon^{2}}+\f{5+6\gamma-6\ln{4\pi}}{48\pi^{2}\varepsilon}\bigg)(p_{0}\gamma^{0}+c_{0}\gamma^{0}\gamma^{5})+\f{1}{16\pi^{2}\varepsilon}p_{k}\gamma^{k}-\f{1}{16\pi^{2}\varepsilon}c_{k}\gamma^{k}\gamma^{5}-\f{1}{8\pi^{2}\varepsilon}\int^{1}_{0}dv\f{\bar{C}_{0}(v)}{\Delta_{0}(v)}\bigg]\\
&&+\Gamma_{U}^{2}\bigg[\bigg(\f{1}{8\pi^{2}\varepsilon^{2}}+\f{5+6\gamma-6\ln{4\pi}}{48\pi^{2}\varepsilon}\bigg)(-p_{0}\gamma^{0}+c_{0}\gamma^{0}\gamma^{5})+\f{1}{16\pi^{2}\varepsilon}p_{k}\gamma^{k}+\f{1}{16\pi^{2}\varepsilon}c_{k}\gamma^{k}\gamma^{5}+\f{1}{8\pi^{2}\varepsilon}\int^{1}_{0}dv\f{C_{0}(v)}{\Delta_{0}(v)}\bigg]+\O(1).
\enn\ns

\bf[h]
\ing[width=.85\tw]{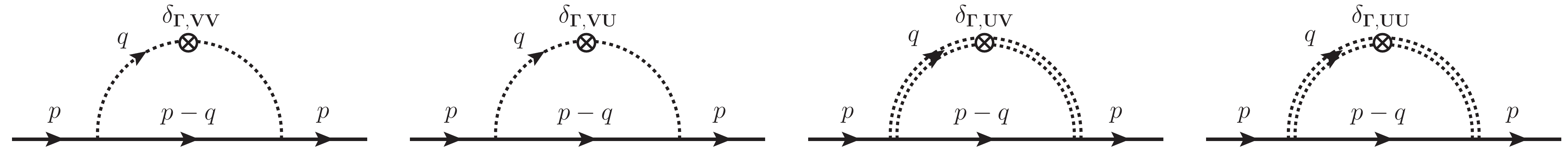}
\caption{One-loop self-energy diagrams containing a vertex counter term. Added to the crossed diagrams, these contributions cancel non-local divergences, arising in the crossed diagrams and leaving local divergences only.} \label{f.Fock-vertex-counter}
\ef

When we take into account the vertex renormalization, we should introduce consistently self-energy corrections made of a vertex counter term, given by (Fig. \ref{f.Fock-vertex-counter})
\bnn
\Sigma^{(1),\delta_{\Gamma}}(p)=\delta_{\Gamma V}\Gamma_{V}\gamma^{0}I_{1}(p)\gamma^{0}+\delta_{\Gamma U}\Gamma_{U}I_{1}(p)\equiv\Sigma^{(1),\delta_{\Gamma}}_{VV}(p)+\Sigma^{(1),\delta_{\Gamma}}_{VU}(p)+\Sigma^{(1),\delta_{\Gamma}}_{UV}(p)+\Sigma^{(1),\delta_{\Gamma}}_{UU}(p).
\enn
Recall $I_{1}=\int^{1}_{0}dx\Big(\f{d}{2}(p_{0}\gamma^{0}-\sc\gamma^{5})+\f{1}{2}(2\tilde{c}_{xk}\gamma^{k}+2c_{k}\gamma^{k}\gamma^{5})\Big)\f{\Gamma(\f{2-d}{2})}{(4\pi)^{d/2}\Delta_{0}^{(2-d)/2}}+\int^{1}_{0}dx\f{C_{0}\Gamma(\f{4-d}{2})}{(4\pi)^{d/2}\Delta_{0}^{(4-d)/2}}$ in Eq. (\ref{r.int1}).

Expanding $I_{1}$ about $d=2+\varepsilon$ and inserting $\delta_{\Gamma V}=\f{\Gamma_{V}}{2\pi\varepsilon}+\f{\Gamma_{U}}{2\pi\varepsilon}$ and $\delta_{\Gamma U}=-\f{\Gamma_{U}}{2\pi\varepsilon}-\f{\Gamma_{V}}{2\pi\varepsilon}$ into the above expression, which will be computed in the next section, we obtain
\bnn
\Sigma^{(1),\delta_{\Gamma}}(p)&=&\Gamma_{V}^{2}\bigg[-\bigg(\f{1}{4\pi^{2}\varepsilon^{2}}+\f{1+\gamma-\ln{4\pi}}{8\pi^{2}\varepsilon}\bigg)\big(p_{0}\gamma^{0}+c_{0}\gamma^{0}\gamma^{5}\big)+\f{1}{8\pi^{2}\varepsilon}\int^{1}_{0}dx\f{\bar{C}_{0}(x)}{\Delta_{0}(x)}\bigg]\\
&&+\Gamma_{V}\Gamma_{U}\bigg[-\bigg(\f{1}{4\pi^{2}\varepsilon^{2}}+\f{1+\gamma-\ln{4\pi}}{8\pi^{2}\varepsilon}\bigg)\big(p_{0}\gamma^{0}+c_{0}\gamma^{0}\gamma^{5}\big)+\f{1}{8\pi^{2}\varepsilon}\int^{1}_{0}dx\f{\bar{C}_{0}(x)}{\Delta_{0}(x)}\bigg]\\
&&+\Gamma_{U}\Gamma_{V}\bigg[-\bigg(\f{1}{4\pi^{2}\varepsilon^{2}}+\f{1+\gamma-\ln{4\pi}}{8\pi^{2}\varepsilon}\bigg)\big(-p_{0}\gamma^{0}+c_{0}\gamma^{0}\gamma^{5}\big)-\f{1}{8\pi^{2}\varepsilon}\int^{1}_{0}dx\f{C_{0}(x)}{\Delta_{0}(x)}\bigg]\\
&&+\Gamma_{U}^{2}\bigg[-\bigg(\f{1}{4\pi^{2}\varepsilon^{2}}+\f{1+\gamma-\ln{4\pi}}{8\pi^{2}\varepsilon}\bigg)\big(-p_{0}\gamma^{0}+c_{0}\gamma^{0}\gamma^{5}\big)-\f{1}{8\pi^{2}\varepsilon}\int^{1}_{0}dx\f{C_{0}(x)}{\Delta_{0}(x)}\bigg]+\O(1).
\enn\ns

Adding these contributions to the crossed diagrams, we finally obtain
\bea
\Sigma^{(2),c}(p)+\Sigma^{(1),\delta_{\Gamma}}(p)&=&\Gamma_{V}^{2}\bigg[\bigg(-\f{1}{8\pi^{2}\varepsilon^{2}}-\f{1}{48\pi^{2}\varepsilon}\bigg)(p_{0}\gamma^{0}+c_{0}\gamma^{0}\gamma^{5})+\f{1}{16\pi^{2}\varepsilon}p_{k}\gamma^{k}-\f{1}{16\pi^{2}\varepsilon}c_{k}\gamma^{k}\gamma^{5}\bigg]\no\\
&&+\Gamma_{U}^{2}\bigg[\bigg(-\f{1}{8\pi^{2}\varepsilon^{2}}-\f{1}{48\pi^{2}\varepsilon}\bigg)(-p_{0}\gamma^{0}+c_{0}\gamma^{0}\gamma^{5})+\f{1}{16\pi^{2}\varepsilon}p_{k}\gamma^{k}+\f{1}{16\pi^{2}\varepsilon}c_{k}\gamma^{k}\gamma^{5}\bigg]\no\\
&&+2\Gamma_{V}\Gamma_{U}\bigg[\bigg(-\f{1}{8\pi^{2}\varepsilon^{2}}-\f{1}{48\pi^{2}\varepsilon}\bigg)c_{0}\gamma^{0}\gamma^{5}+\f{1}{16\pi^{2}\varepsilon}p_{k}\gamma^{k}\bigg]+\O(1)\label{r.cross}.
\eea
This result is depicted pictorially in Fig. \ref{sum-cross}.

\bf[b]
\ing[width=.9\tw]{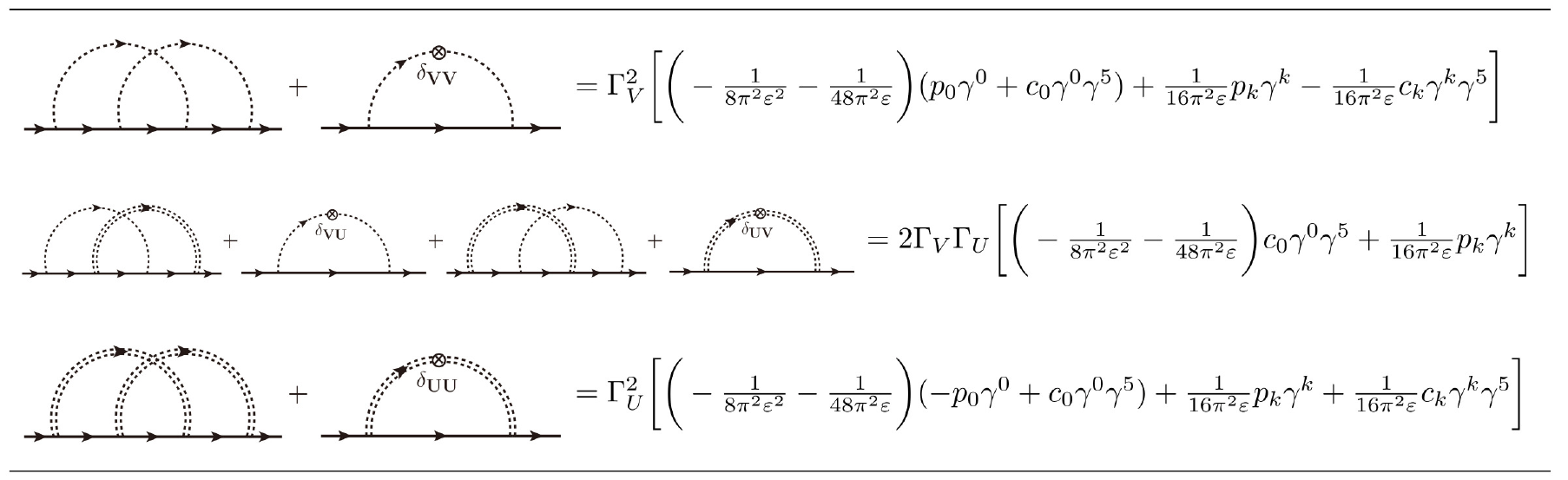}\caption{The result for the crossed diagrams. Each crossed diagram is added consistently by one-loop self-energy diagram made of a tree-level vertex and a vertex-correction counter term. Note that there are simple poles for $c_{k}$, resulting in renormalization of $c_{k}$, while the sign difference between $\Gamma_{V}$ and $\Gamma_{U}$ channels implies that their roles ($\Gamma_{V}$ and $\Gamma_{U}$) are different. Also, one may notice that non-local divergences of $(\gamma-\ln{4\pi})$ are canceled, which is the result of BPHZ theorem \cite{QFT_Textbook}, and so the problematic term of $\int^{1}_{0}dx\f{C_{0}}{\Delta_{0}}$ is.}\label{sum-cross}
\ef

\section{vertex correction}

\subsection{Relevant Feynman diagrams}

The vertex renormalization can be found from a four-point function of $\mathbf{D}(p,p',q,q')=\left\langle\psi_{p}\bar{\psi}_{p'}\psi_{q}\bar{\psi}_{q'} \right\rangle$. Performing the perturbative analysis up to the $\Gamma^{2}$ order, we obtain
\bnn
&&\mathbf{D}(p,p',q,q')\\
&=&\limR\f{1}{R}\int\D\bar{\psi}\D\psi\big(\psi^{a}_{p}\bar{\psi}^{a}_{p'}\psi^{a}_{q}\bar{\psi}^{a}_{q'}\big)e^{-S_{0}[\bar{\psi}^{\alpha},\psi^{\alpha}]}e^{\f{1}{L^{3}}\sum_{p_{j}}\big[\f{\Gamma_{V}}{2}(\bar{\psi}^{b}_{p_{1}}\gamma^{0}\psi^{b}_{p_{2}})(\bar{\psi}^{c}_{p_{3}}\gamma^{0}\psi^{c}_{p_{4}})+\f{\Gamma_{U}}{2}(\bar{\psi}^{b}_{p_{1}}\psi^{b}_{p_{2}})(\bar{\psi}^{c}_{p_{3}}\psi^{c}_{p_{4}})\big]\delta^{(3)}_{\p_{1}-\p_{2},\p_{3}-\p_{4}}\delta_{p_{1}^{0}p_{2}^{0}}\delta_{p_{3}^{0}p_{4}^{0}}}\\
&\simeq&\limR\f{1}{R}\int\D\bar{\psi}\D\psi e^{-S_{0}[\bar{\psi}^{\alpha},\psi^{\alpha}]}\br[\psi^{a}_{p}\bar{\psi}^{a}_{p'}\psi^{a}_{q}\bar{\psi}^{a}_{q'}+\f{\Gamma_{V}}{2L^{3}}\sum_{p_{j}}\big(\psi^{a}_{p}\bar{\psi}^{a}_{p'}\psi^{a}_{q}\bar{\psi}^{a}_{q'}\big)\big(\bar{\psi}^{b}_{p_{1}}\gamma^{0}\psi^{b}_{p_{2}} \bar{\psi}^{c}_{p_{3}}\gamma^{0}\psi^{c}_{p_{4}}\big)\delta^{(3)}_{\p_{1}-\p_{2},\p_{3}-\p_{4}}\delta_{p^{0}_{1}p^{0}_{2}}\delta_{p^{0}_{3}p^{0}_{4}}\\
&&+\f{\Gamma_{U}}{2L^{3}}\sum_{p_{j}}\big(\psi^{a}_{p}\bar{\psi}^{a}_{p'}\psi^{a}_{q}\bar{\psi}^{a}_{q'}\big)\big(\bar{\psi}^{b}_{p_{1}}\psi^{b}_{p_{2}} \bar{\psi}^{c}_{p_{3}}\psi^{c}_{p_{4}}\big)\delta^{(3)}_{\p_{1}-\p_{2},\p_{3}-\p_{4}}\delta_{p^{0}_{1}p^{0}_{2}}\delta_{p^{0}_{3}p^{0}_{4}}\\
&&+\f{\Gamma_{V}^{2}}{8(L^{3})^{2}}\sum_{p_{j}p_{j}'}\big(\psi^{a}_{p}\bar{\psi}^{a}_{p'}\psi^{a}_{q}\bar{\psi}^{a}_{q'}\big) \big(\bar{\psi}^{b}_{p_{1}}\gamma^{0}\psi^{b}_{p_{2}}\bar{\psi}^{c}_{p_{3}}\gamma^{0}\psi^{c}_{p_{4}}\big)\big(\bar{\psi}^{b'}_{p'_{1}} \gamma^{0}\psi^{b'}_{p'_{2}}\bar{\psi}^{c'}_{p'_{3}}\gamma^{0} \psi^{c'}_{p'_{4}}\big)\delta^{(3)}_{\p_{1}-\p_{2},\p_{3}-\p_{4}} \delta_{p^{0}_{1}p^{0}_{2}}\delta_{p^{0}_{3}p^{0}_{4}}\delta^{(3)}_{\p'_{1}-\p'_{2},\p'_{3}-\p'_{4}}\delta_{p^{'0}_{1}p^{'0}_{2}} \delta_{p^{'0}_{3}p^{'0}_{4}}\\
&&+\f{\Gamma_{U}^{2}}{8(L^{3})^{2}}\sum_{p_{j}p_{j}'}\big(\psi^{a}_{p}\bar{\psi}^{a}_{p'}\psi^{a}_{q}\bar{\psi}^{a}_{q'}\big) \big(\bar{\psi}^{b}_{p_{1}}\psi^{b}_{p_{2}} \bar{\psi}^{c}_{p_{3}}\psi^{c}_{p_{4}}\big)\big(\bar{\psi}^{b'}_{p'_{1}}\psi^{b'}_{p'_{2}} \bar{\psi}^{c'}_{p'_{3}}\psi^{c'}_{p'_{4}}\big)\delta^{(3)}_{\p_{1}-\p_{2},\p_{3}-\p_{4}}\delta_{p^{0}_{1}p^{0}_{2}}\delta_{p^{0}_{3}p^{0}_{4}} \delta^{(3)}_{\p'_{1}-\p'_{2},\p'_{3}-\p'_{4}}\delta_{p^{'0}_{1}p^{'0}_{2}}\delta_{p^{'0}_{3}p^{'0}_{4}}\\
&&+\f{\Gamma_{V}\Gamma_{U}}{4(L^{3})^{2}}\sum_{p_{j}p_{j}'}\big(\psi^{a}_{p}\bar{\psi}^{a}_{p'}\psi^{a}_{q}\bar{\psi}^{a}_{q'}\big) \big(\bar{\psi}^{b}_{p_{1}}\gamma^{0}\psi^{b}_{p_{2}} \bar{\psi}^{c}_{p_{3}}\gamma^{0}\psi^{c}_{p_{4}}\big)\big(\bar{\psi}^{b'}_{p'_{1}} \psi^{b'}_{p'_{2}}\bar{\psi}^{c'}_{p'_{3}}\psi^{c'}_{p'_{4}}\big)\delta^{(3)}_{\p_{1}-\p_{2},\p_{3}-\p_{4}}\delta_{p^{0}_{1}p^{0}_{2}} \delta_{p^{0}_{3}p^{0}_{4}}\delta^{(3)}_{\p'_{1}-\p'_{2},\p'_{3}-\p'_{4}}\delta_{p^{'0}_{1}p^{'0}_{2}}\delta_{p^{'0}_{3}p^{'0}_{4}}\bl].
\enn

Among the first-order contributions, fully-connected diagrams give scattering elements (Fig. \ref{f.vertex-tree}). The four-point function and the scattering matrix element at the tree level are
\bea
M^{(0)}(p,p;q)&\equiv&M^{(0)}_{V}(p,p;q)+M^{(0)}_{U}(p,p;q)=2\Gamma_{V}(\gamma^{0}\otimes\gamma^{0})+2\Gamma_{V}(I_{4\times4}\otimes I_{4\times4}).\label{e.ver-tree}
\eea

\bf[b]
\ing[width=0.5\tw]{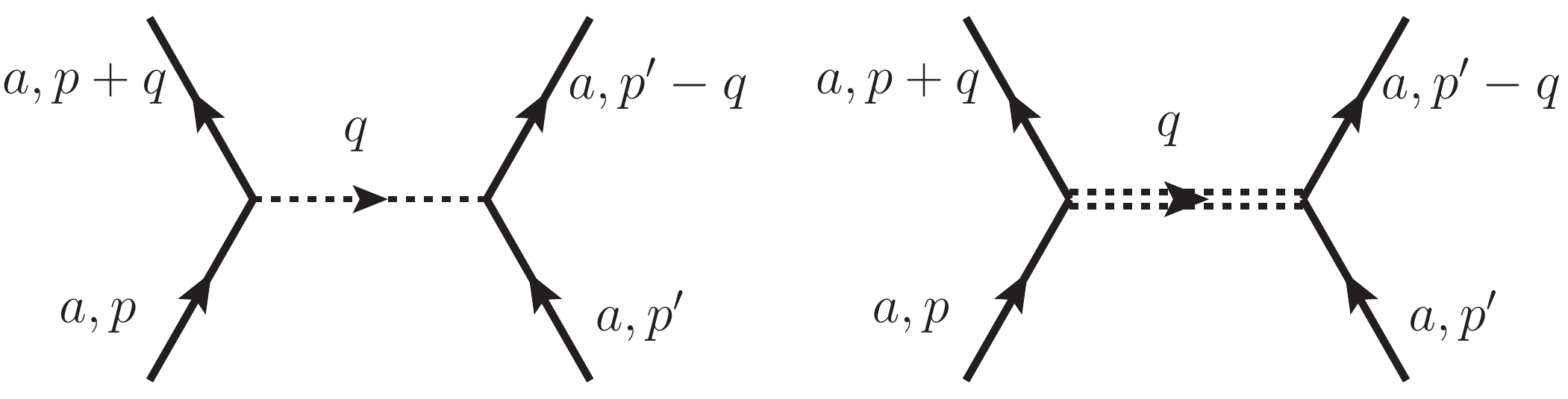}
\caption{Tree-level vertex. There are two contributions from intra-valley and inter-valley scattering.} \label{f.vertex-tree}
\ef

Among the second order contributions, only diagrams fully connected with the external lines survive in the replica limit of $R\rightarrow0$ and give scattering matrix elements. Thus, the scattering matrix elements in the second order are given by (Fig. \ref{f.vertex corrections})
\bnn
\M^{(1)}_{ph}&=&\f{2\Gamma_{V}^{2}}{L^{3}}\sum_{l}\gamma^{0}G(p-l)\gamma^{0}\otimes\gamma^{0}G(p'-l-q)\gamma^{0}\delta_{l^{0}0}+\f{2\Gamma_{V}\Gamma_{U}}{L^{3}}\sum_{l}\gamma^{0}G(p-l)\otimes G(p'-l-q)\gamma^{0}\delta_{l^{0}0}\\
&&+\f{2\Gamma_{U}\Gamma_{V}}{L^{3}}\sum_{l}G(p-l)\gamma^{0}\otimes\gamma^{0}G(p'-l-q)\delta_{l^{0}0}+\f{2\Gamma_{U}^{2}}{L^{3}}\sum_{l}G(p-l)\otimes G(p'-l-q)\delta_{l^{0}0},\\
&\equiv&\M^{ph}_{VV}+\M^{ph}_{VU}+\M^{ph}_{UV}+\M^{ph}_{UU}\\
\M^{(1)}_{pp}&=&\f{2\Gamma_{V}^{2}}{L^{3}}\sum_{l}\gamma^{0}G(p-l)\gamma^{0}\otimes\gamma^{0}G(p'+l)\gamma^{0}\delta_{l^{0}0}+\f{2\Gamma_{V}\Gamma_{U}}{L^{3}}\sum_{l}\gamma^{0}G(p-l)\otimes \gamma^{0}G(p'+l)\delta_{l^{0}0}\\
&&+\f{2\Gamma_{U}\Gamma_{V}}{L^{3}}\sum_{l}G(p-l)\gamma^{0}\otimes G(p'+l)\gamma^{0}\delta_{l^{0}0}+\f{2\Gamma_{U}^{2}}{L^{3}}\sum_{l}G(p-l)\otimes G(p'+l)\delta_{l^{0}0},\\
&\equiv&\M^{pp}_{VV}+\M^{pp}_{VU}+\M^{pp}_{UV}+\M^{pp}_{UU}\\
\M^{(1)}_{ver}&=&\f{2\Gamma_{V}^{2}}{L^{3}}\sum_{l}\gamma^{0}G(p-l)\gamma^{0}G(p+q-l)\gamma^{0}\otimes\gamma^{0}\delta_{l^{0}0}+\f{2\Gamma_{V}\Gamma_{U}}{L^{3}}\sum_{l}\gamma^{0}G(p-l)G(p+q-l)\gamma^{0}\otimes I_{4\times4}\delta_{l^{0}0}\\
&&+\f{2\Gamma_{U}\Gamma_{V}}{L^{3}}\sum_{l}G(p-l)\gamma^{0}G(p+q-l)\otimes\gamma^{0}\delta_{l^{0}0}+\f{2\Gamma_{U}^{2}}{L^{3}}\sum_{l}G(p-l)G(p+q-l)\otimes I_{4\times4}\delta_{l^{0}0},\\
&\equiv&\M^{ver}_{VV}+\M^{ver}_{VU}+\M^{ver}_{UV}+\M^{ver}_{UU},
\enn
where $``ph"$, $``pp"$ and $``ver"$ represent ``particle-hole", ``particle-particle" and ``vertex", respectively.

\bf[h]
\ing[width=0.85\tw]{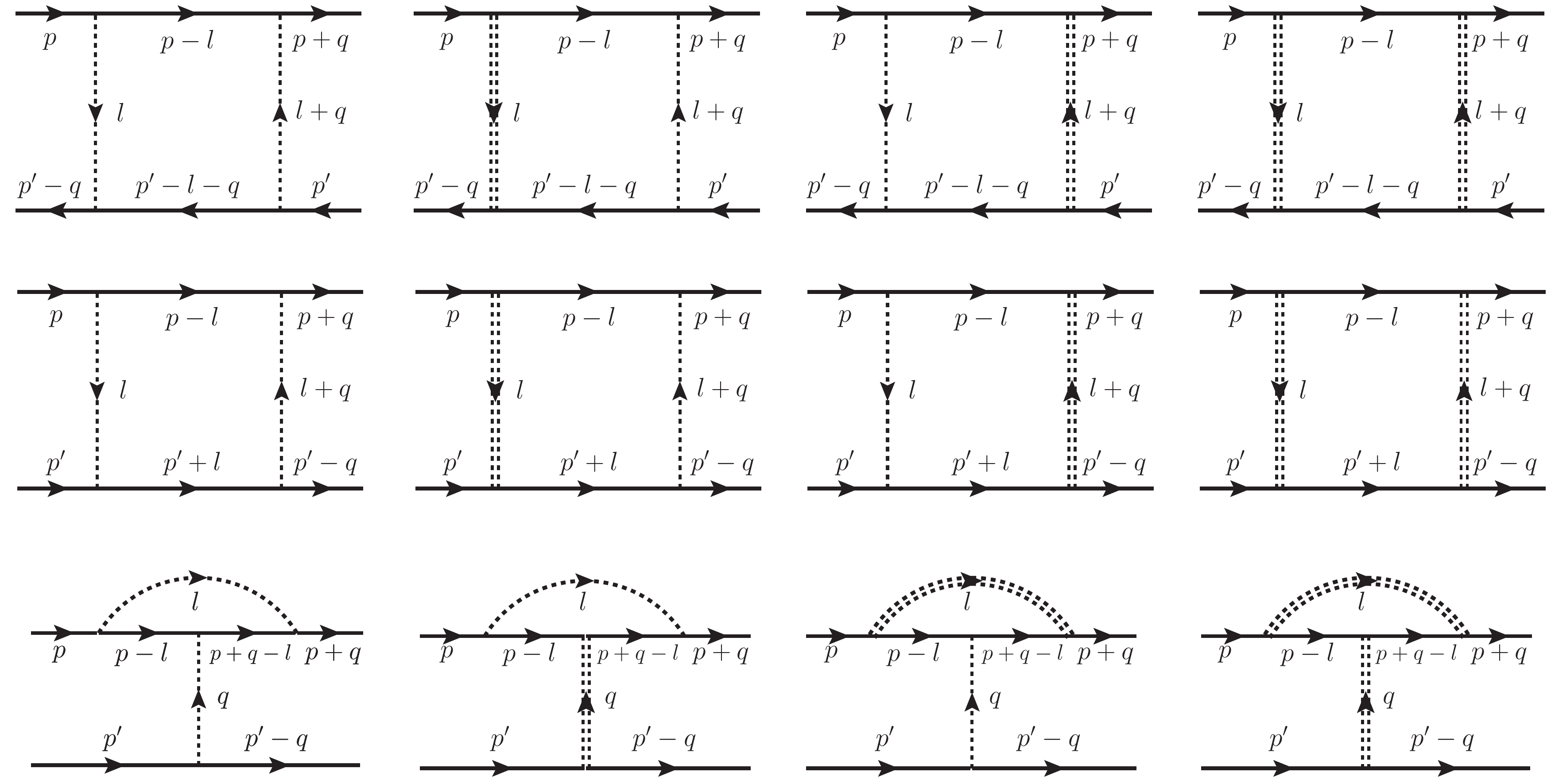}
\caption{Vertex corrections in the second order. There are three distinct types of diagrams, say, particle-hole, particle-particle and vertex diagrams. Diagrams in each type are distinguished by interaction vertices (two intra-valley scattering, one intra-valley and one inter-valley scattering, and etc.). So, totally there are twelve diagrams for the second-order vertex corrections.} \label{f.vertex corrections}
\ef

\subsection{Evaluation of relevant diagrams}

\subsubsection{Particle-hole channel}

First, we evaluate the particle-hole diagrams (the first line in Fig. \ref{f.vertex corrections})
\bnn
\M^{(1)}_{ph}&=&2\Gamma_{V}^{2}I_{2ph}[M_{1}=M_{2}=\gamma^{0}]+2\Gamma_{V}\Gamma_{U}I_{2ph}[M_{1}=\gamma^{0}, M_{2}=I_{4\times4}]\no\\
&&+2\Gamma_{U}\Gamma_{V}I_{2ph}[M_{1}=I_{4\times4}, M_{2}=\gamma^{0}]+2\Gamma_{U}^{2}I_{2ph}[M_{1}=M_{2}=I_{4\times4}],
\enn
where $I_{2ph}$ is given by ($\k\equiv\p-\p'+\q$)
\bnn
I_{2ph}&=&\int\f{d^{d+1}l}{(2\pi)^{d+1}}2\pi\delta (l_{0})M_{1}G(p-l)M_{2}\otimes M_{2}G(p'-l-q)M_{1}\\
&=&\int\f{d^{d}\l}{(2\pi)^{d}}M_{1}G(p_{0},\p-\l)M_{2}\otimes M_{2}G(p_{0}'-q_{0},\p'-\l-\q)M_{1}\\
&=&\int \f{d^{d}\l}{(2\pi)^{d}}M_{1}G(p_{0},-\l)M_{2}\otimes M_{2}G(p_{0}'-q_{0},-\l-\k)M_{1}.
\enn

Using Eq. (\ref{e.Green1}), we have
\fs\bnn
I_{2ph}&=&\int^{1}_{0}dx\int^{1}_{0}dy\int\f{d^{d}\l}{(2\pi)^{d}}M_{1}\f{-\l^{2}l_{i}\gamma^{i}+\l^{2}(p_{0}\gamma^{0}-\sc\gamma^{5})+l_{i}l_{j}(-2c^{i}\gamma^{j}\gamma^{5})-l_{i}f_{1}^{i}(p_{0})+f_{0}(p_{0})}{\big[\big(\l-(1-2x)\c\big)^{2}+\Delta_{0}(p_{0};x)\big]^{2}}M_{2}\otimes M_{2}\\
&&\f{-(\l+\k)^{2}(l_{j}+k_{j})\gamma^{j}+(\l+\k)^{2}((p'_{0}-q_{0})\gamma^{0}-\sc\gamma^{5})+(l_{i}+k_{i})(l_{j}+k_{j})(-2c^{i}\gamma^{j}\gamma^{5})-(l_{i}+k_{i})f_{1}^{i}(p'_{0}-q_{0})+f_{0}(p'_{0}-q_{0})}{\big[\big(\l+\k-(1-2y)\c\big)^{2}+\Delta_{0}(p'_{0}-q_{0};y)\big]^{2}}M_{1}.
\enn\ns

Despite this complicated expression, only the product of the $l$-cubic terms contributes to renormalization by the same reason that we considered in Eq. (\ref{e.simplified integral}). Keeping this term only, we obtain
\bnn
I_{2ph}&\simeq&\int^{1}_{0}dx\int^{1}_{0}dy\int\f{d^{d}\l}{(2\pi)^{d}}\f{\l^{2}(\l+\k)^{2}l_{i}(l_{j}+k_{j})(M_{1}\gamma^{i}M_{2}\otimes M_{2}\gamma^{j}M_{1})}{\big[\big(\l-(1-2x)\c\big)^{2}+\Delta_{0}(x)\big]^{2}\big[\big(\l+\k-(1-2y)\c\big)^{2}+\Delta_{0}(y)\big]^{2}}\\
&=&\int^{1}_{0}dx\int^{1}_{0}dy\int^{1}_{0}dz6z(1-z)\int\f{d^{d}\l}{(2\pi)^{d}}\f{\l^{2}(\l+\k)^{2}l_{i}(l_{j}+k_{j})(M_{1}\gamma^{i}M_{2}\otimes M_{2}\gamma^{j}M_{1})}{\big[{\l'}^{2}+\Delta_{1}(\k;x,y,z)\big]^{4}},
\enn
where $\Delta_{1}=z(1-z)\big(\k+2(y-x)\c\big)^{2}+(1-z)\Delta_{0}(x)+z\Delta_{0}(y)$ and $\l'=\l+z\k-z(1-2y)\c-(1-z)(1-2x)\c$.

Renaming momentum as $\l'\rightarrow\l$ and keeping only a relevant term again, we reach the following expression
\bnn
I_{2ph}&=&(M_{1}\gamma^{i}M_{2}\otimes M_{2}\gamma^{j}M_{1})\int^{1}_{0}dx\int^{1}_{0}dy\int^{1}_{0}dz6z(1-z)\int\f{d^{d}\l}{(2\pi)^{d}}\f{(\l^{2})^{2}l_{i}l_{j}}{\big[\l^{2}+\Delta_{1}\big]^{4}}\\
&=&(M_{1}\gamma^{i}M_{2}\otimes M_{2}\gamma^{i}M_{1})\int^{1}_{0}dx\int^{1}_{0}dy\int^{1}_{0}dzz(1-z)\f{(d+4)(d+2)}{32\pi}\Gamma\bigg(\f{2-d}{2}\bigg)\bigg(\f{\Delta_{1}}{4\pi}\bigg)^{\f{d-2}{2}}\\
&=&-\f{1}{4\pi\varepsilon}(M_{1}\gamma^{i}M_{2}\otimes M_{2}\gamma^{i}M_{1})+\O(1).
\enn
Thus, the scattering matrix element for the particle-hole diagrams is
\be
\M^{(1)}_{ph}=-\f{\Gamma_{V}^{2}}{2\pi\varepsilon}(\gamma^{i}\otimes\gamma^{i})+\f{\Gamma_{V}\Gamma_{U}}{\pi\varepsilon}(\gamma^{0}\gamma^{i}\otimes\gamma^{0}\gamma^{i})-\f{\Gamma_{U}^{2}}{2\pi\varepsilon}(\gamma^{i}\otimes\gamma^{i})+\O(1).\label{r.ph}
\ee

\subsubsection{Particle-particle channel}

Next, we evaluate the particle-particle diagrams (the second line in Fig. \ref{f.vertex corrections})
\bnn
\M^{(1)}_{pp}&=&2\Gamma_{V}^{2}I_{2pp}[M_{1}=M_{2}=\gamma^{0}]+2\Gamma_{V}\Gamma_{U}I_{2pp}[M_{1}=\gamma^{0}, M_{2}=I_{4\times4}]\no\\
&&+2\Gamma_{U}\Gamma_{V}I_{2pp}[M_{1}=I_{4\times4}, M_{2}=\gamma^{0}]+2\Gamma_{U}^{2}I_{2pp}[M_{1}=M_{2}=I_{4\times4}],
\enn
where $I_{2pp}$ is given by ($\k\equiv\p+\p'$)
\bnn
I_{2pp}
&=&\int\f{d^{d+1}l}{(2\pi)^{d+1}}2\pi\delta (l_{0})M_{1}G(p-l)M_{2}\otimes M_{1}G(p'+l)M_{2}\\
&=&\int\f{d^{d}\l}{(2\pi)^{d}}M_{1}G(p_{0},\p-\l)M_{2}\otimes M_{1}G(p'_{0},\p'+\l)M_{2}\\
&=&\int\f{d^{d}\l}{(2\pi)^{d}}M_{1}G(p_{0},-\l)M_{2}\otimes M_{1}G(p'_{0},\l+\k)M_{2}.
\enn

The analysis is quite similar with that of the particle-hole channel. Keeping only a relevant term, we have
\bnn
I_{2pp}&\simeq&\int^{1}_{0}dx\int^{1}_{0}dy\int\f{d^{d}\l}{(2\pi)^{d}}\f{-\l^{2}(\l+\k)^{2}l_{i}(l_{j}+k_{j})(M_{1}\gamma^{i}M_{2}\otimes M_{1} \gamma^{j}M_{2})}{\big[\big(\l-(1-2x)\c\big)^{2}+\Delta_{0}(x)\big]^{2}\big[\big(\l+\k+(1-2y)\c\big)^{2}+\Delta_{0}(y)\big]^{2}}\\
&=&-\int^{1}_{0}dx\int^{1}_{0}dy\int^{1}_{0}dz6z(1-z)\int\f{d^{d}\l}{(2\pi)^{d}}\f{\l^{2}(\l+\k)^{2}l_{i}(l_{j}+k_{j})(M_{1}\gamma^{i}M_{2}\otimes M_{1}\gamma^{j}M_{2})}{\big[{\l'}^{2}+\Delta_{1}(\k;x,y,z)\big]^{4}},
\enn
where $\Delta_{1}=z(1-z)\big(\k+2(1-x-y)\c\big)^{2}+(1-z)\Delta_{0}(x)+z\Delta_{0}(y)$ and $\l'=\l+z\k+z(1-2y)\c-(1-z)(1-2x)\c$. Note a minus sign in front of the integral that essentially originates from the opposite sign in the loop-momentum of the two propagators. Due to this sign difference, the contribution from the pp-diagram will cancel that of the ph-diagram.

The remaining calculation is the same as before. As a result, we reach the following expression
\bnn
I_{2pp}=+\f{1}{4\pi\varepsilon}(M_{1}\gamma^{i}M_{2}\otimes M_{1}\gamma^{i}M_{2})+\O(1).
\enn
Thus, the scattering matrix elements for the particle-particle diagrams is
\be
\M^{(1)}_{pp}=\f{\Gamma_{V}^{2}}{2\pi\varepsilon}(\gamma^{i}\otimes\gamma^{i})+\f{\Gamma_{V}\Gamma_{U}}{\pi\varepsilon}(\gamma^{0}\gamma^{i}\otimes\gamma^{0}\gamma^{i})+\f{\Gamma_{U}^{2}}{2\pi\varepsilon}(\gamma^{i}\otimes\gamma^{i})+\O(1).\label{r.pp}
\ee

\subsubsection{Vertex channel}

Lastly, we evaluate the vertex diagrams (the third line in Fig. \ref{f.vertex corrections})
\bnn
\M^{(1)}_{ver}&=&2\Gamma_{V}^{2}I_{2ver}[M_{1}=M_{2}=\gamma^{0}]+2\Gamma_{V}\Gamma_{U}I_{2ver}[M_{1}=\gamma^{0},M_{2}=I_{4\times4}]\\
&&+2\Gamma_{U}\Gamma_{V}I_{2ver}[M_{1}=I_{4\times4},M_{2}=\gamma^{0}]+2\Gamma_{U}^{2}I_{2ver}[M_{1}=M_{2}=I_{4\times4}],
\enn
where $I_{2ver}$ is given by
\bnn
I_{2ver}&=&\int\f{d^{d+1}l}{(2\pi)^{d+1}}M_{1}G(p-l)M_{2}G(p+q-l)M_{1}\otimes M_{2}\delta_{l^{0}0}\\
&=&\int\f{d^{d}\l}{(2\pi)^{d}}M_{1}G(p_{0},\p-\l)M_{2}G(p_{0}+q_{0},\p+\q-\l)M_{1}\otimes M_{2}\\
&=&\int\f{d^{d}\l}{(2\pi)^{d}}M_{1}G(p_{0},-\l)M_{2}G(p_{0}+q_{0},-\l+\q)M_{1}\otimes M_{2}.
\enn

The analysis is also similar with the ph case except for the fact that ``$\otimes$" are not located between propagators.

Keeping only a relevant term, we have
\bnn
I_{2ver}&\simeq&\int^{1}_{0}dx\int^{1}_{0}dy\int\f{d^{d}\l}{(2\pi)^{d}}\f{\l^{2}(\l-\q)^{2}l_{i}(l_{j}-q_{j})(M_{1}\gamma^{i}M_{2}\gamma^{j}M_{1}\otimes M_{2})}{\big[\big(\l-(1-2x)\c\big)^{2}+\Delta_{0}(x)\big]^{2}\big[\big(\l-\q-(1-2y)\c\big)^{2}+\Delta_{0}(y)\big]^{2}}\\
&=&\int^{1}_{0}dx\int^{1}_{0}dy\int^{1}_{0}dz6z(1-z)\int\f{d^{d}\l}{(2\pi)^{d}}\f{\l^{2}(\l-\q)^{2}l_{i}(l_{j}-q_{j})(M_{1}\gamma^{i}M_{2}\gamma^{j}M_{1}\otimes M_{2})}{\big[{\l'}^{2}+\Delta_{1}(\q;x,y,z)\big]^{4}},
\enn
where $\Delta_{1}=z(1-z)\big(\q+2(x-y)\c\big)^{2}+(1-z)\Delta_{0}(x)+z\Delta_{0}(y)$ and $\l'=\l-z\q-z(1-2y)\c-(1-z)(1-2x)\c$.

Renaming momentum as $\l'\rightarrow\l$ and keeping only a relevant term again, we reach the following expression
\bnn
I_{2ver}&=&(M_{1}\gamma^{i}M_{2}\gamma^{j}M_{1}\otimes M_{2})\int^{1}_{0}dx\int^{1}_{0}dy\int^{1}_{0}dz6z(1-z)\int\f{d^{d}\l}{(2\pi)^{d}}\f{(\l^{2})^{2}l_{i}l_{j}}{\big[\l^{2}+\Delta_{1}\big]^{4}}\\
&=&(M_{1}\gamma^{i}M_{2}\gamma^{i}M_{1}\otimes M_{2})\int^{1}_{0}dx\int^{1}_{0}dy\int^{1}_{0}dzz(1-z)\f{(d+4)(d+2)}{32\pi}\Gamma\bigg(\f{2-d}{2}\bigg)\bigg(\f{\Delta_{1}}{4\pi}\bigg)^{\f{d-2}{2}}\\
&=&-\f{M_{1}\gamma^{i}M_{2}\gamma^{i}M_{1}\otimes M_{2}}{4\pi\varepsilon}+\O(1).
\enn
Thus, the scattering matrix element for the vertex-diagrams is
\be
\M^{(1)}_{ver}=-\f{\Gamma_{V}^{2}}{\pi\varepsilon}(\gamma^{0}\otimes\gamma^{0})+\f{\Gamma_{V}\Gamma_{U}}{\pi\varepsilon}(I_{4\times4}\otimes I_{4\times4})-\f{\Gamma_{U}\Gamma_{V}}{\pi\varepsilon}(\gamma^{0}\otimes\gamma^{0})+\f{\Gamma_{U}^{2}}{\pi\varepsilon}(I_{4\times4}\otimes I_{4\times4})+\O(1) , \label{r.ver}
\ee
where the result is depicted pictorially in Fig. \ref{sum-vertex}.

\bf[b]
\ing[width=.9\tw]{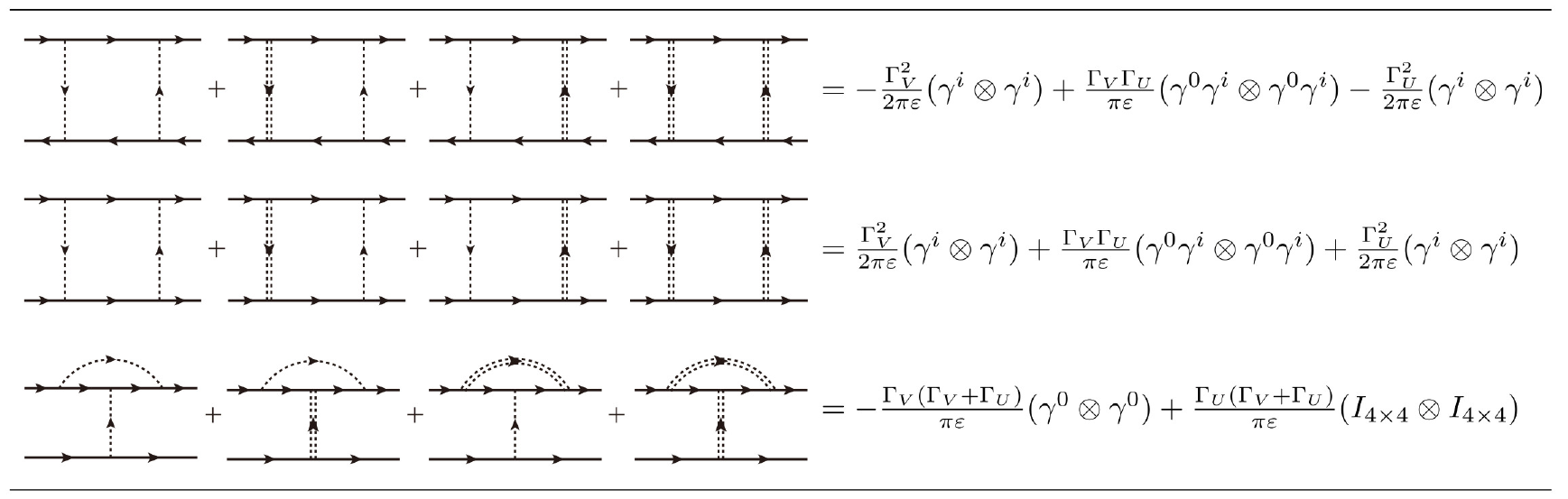}\caption{The result for the vertex corrections in the second order. Note that the contribution from the particle-hole diagrams (the first line) will be canceled to that of the particle-particle diagrams (the second line). A novel coupling term of $\gamma^{0}\gamma^{i}$ appears, but not concerned here. As a result, vertex diagrams (the third line) participate in renormalization of intra-valley scattering ($\Gamma_{V}$) and inter-valley scattering ($\Gamma_{U}$). Note the sign difference in the two factors, which results in the distinction between two types of scatterings. That is, inter-valley scattering becomes relevant while intra-valley scattering irrelevant in the low-energy physics.}\label{sum-vertex}
\ef

\section{renormalization group equations}

Combining Eq. (\ref{r.Fock}), Eq. (\ref{r.rainbow}) and Eq. (\ref{r.cross}) in the following way
\bnn
&&\Sigma^{(1)}(p)+\big(\Sigma^{(2),r}(p)+\Sigma^{(1),\delta_{\psi}}(p)\big)+\big(\Sigma^{(2),c}(p)+\Sigma^{(1),\delta_{\Gamma}}(p)\big)+\big(\tx{propagator counterterms}\big)\\
&=&-\f{\Gamma_{V}}{2\pi\varepsilon}\big(p_{0}\gamma^{0}+c_{0}\gamma^{0}\gamma^{5}\big)-\f{\Gamma_{U}}{2\pi\varepsilon}\big(p_{0}\gamma^{0}-c_{0}\gamma^{0}\gamma^{5}\big)\\
&&+\f{\Gamma_{V}^{2}}{8\pi^{2}\varepsilon}(p_{0}\gamma^{0}+c_{0}\gamma^{0}\gamma^{5})+\f{\Gamma_{U}^{2}}{8\pi^{2}\varepsilon}(p_{0}\gamma^{0}+c_{0}\gamma^{0}\gamma^{5})+\f{2\Gamma_{V}\Gamma_{U}}{8\pi^{2}\varepsilon}(p_{0}\gamma^{0}-c_{0}\gamma^{0}\gamma^{5})\\
&&+\Gamma_{V}^{2}\bigg[\bigg(-\f{1}{8\pi^{2}\varepsilon^{2}}-\f{1}{48\pi^{2}\varepsilon}\bigg)(p_{0}\gamma^{0}+c_{0}\gamma^{0}\gamma^{5})+\f{1}{16\pi^{2}\varepsilon}p_{k}\gamma^{k}-\f{1}{16\pi^{2}\varepsilon}c_{k}\gamma^{k}\gamma^{5}\bigg]\no\\
&&+\Gamma_{U}^{2}\bigg[\bigg(-\f{1}{8\pi^{2}\varepsilon^{2}}-\f{1}{48\pi^{2}\varepsilon}\bigg)(-p_{0}\gamma^{0}+c_{0}\gamma^{0}\gamma^{5})+\f{1}{16\pi^{2}\varepsilon}p_{k}\gamma^{k}+\f{1}{16\pi^{2}\varepsilon}c_{k}\gamma^{k}\gamma^{5}\bigg]\no\\
&&+2\Gamma_{V}\Gamma_{U}\bigg[\bigg(-\f{1}{8\pi^{2}\varepsilon^{2}}-\f{1}{48\pi^{2}\varepsilon}\bigg)c_{0}\gamma^{0}\gamma^{5}+\f{1}{16\pi^{2}\varepsilon}p_{k}\gamma^{k}\bigg]+\mathcal{O}(1)+(\delta_{\psi}^{\omega}p_{0}\gamma^{0}+\delta_{\psi}^{\k}p_{k}\gamma^{k}+\delta_{c0}c_{0}\gamma^{0}\gamma^{5} +\delta_{\c}c_{k}\gamma^{k}\gamma^{5}),
\enn
we find propagator counter terms in Eq. (\ref{r.counterterms})
\bnn
&&\delta_{\psi}^{\omega}=\f{\Gamma_{V}+\Gamma_{U}}{2\pi\varepsilon}-\f{(\Gamma_{V}+\Gamma_{U})^{2}}{8\pi^{2}\varepsilon}+\f{\Gamma_{V}^{2}-\Gamma_{U}^{2}}{48\pi^{2}\varepsilon},~~~~~~~~\delta_{\psi}^{\k}=-\f{(\Gamma_{V}+\Gamma_{U})^{2}}{16\pi^{2}\varepsilon},\no\\
&&\delta_{c0}=\f{\Gamma_{V}-\Gamma_{U}}{2\pi\varepsilon}-\f{(\Gamma_{V}-\Gamma_{U})^{2}}{8\pi^{2}\varepsilon}+\f{(\Gamma_{V}+\Gamma_{U})^{2}}{48\pi^{2}\varepsilon},~~~~\delta_{\c}=\f{\Gamma_{V}^{2}-\Gamma_{U}^{2}}{16\pi^{2}\varepsilon}.
\enn

Similarly, combining Eq.(\ref{r.ph}), Eq.(\ref{r.pp}) and Eq.(\ref{r.ver}) as follows
\bnn
&&\M^{(1)}_{ph}+\M^{(1)}_{pp}+\M^{(1)}_{ver}+4\times\delta_{\Gamma V}\f{\Gamma_{V}}{2}(\gamma^{0}\otimes\gamma^{0})+4\times\delta_{\Gamma U}\f{\Gamma_{U}}{2}(I_{4\times4}\otimes I_{4\times4})+4\times\f{\delta_{\Gamma T}}{2}(\gamma^{0}\gamma^{i}\otimes\gamma^{0}\gamma^{i})\\
&=&-\f{\Gamma_{V}^{2}}{2\pi\varepsilon}(\gamma^{i}\otimes\gamma^{i})+\f{\Gamma_{V}\Gamma_{U}}{\pi\varepsilon}(\gamma^{0}\gamma^{i}\otimes\gamma^{0}\gamma^{i})-\f{\Gamma_{U}^{2}}{2\pi\varepsilon}(\gamma^{i}\otimes\gamma^{i})+\f{\Gamma_{V}^{2}}{2\pi\varepsilon}(\gamma^{i}\otimes\gamma^{i})+\f{\Gamma_{V}\Gamma_{U}}{\pi\varepsilon}(\gamma^{0}\gamma^{i}\otimes\gamma^{0}\gamma^{i})+\f{\Gamma_{U}^{2}}{2\pi\varepsilon}(\gamma^{i}\otimes\gamma^{i})\\
&&-\f{\Gamma_{V}^{2}}{\pi\varepsilon}(\gamma^{0}\otimes\gamma^{0})+\f{\Gamma_{V}\Gamma_{U}}{\pi\varepsilon}(I_{4\times4}\otimes I_{4\times4})-\f{\Gamma_{U}\Gamma_{V}}{\pi\varepsilon}(\gamma^{0}\otimes\gamma^{0})+\f{\Gamma_{U}^{2}}{\pi\varepsilon}(I_{4\times4}\otimes I_{4\times4})+\O(1)\\
&&+2\delta_{\Gamma V}\Gamma_{V}(\gamma^{0}\otimes\gamma^{0})+2\delta_{\Gamma U}\Gamma_{U}(I_{4\times4}\otimes I_{4\times4})+2\delta_{\Gamma T}(\gamma^{0}\gamma^{i}\otimes\gamma^{0}\gamma^{i}),
\enn
we find vertex counter terms in Eq. (\ref{r.counterterms})
\bnn
\delta_{\Gamma V}=\f{\Gamma_{V}}{2\pi\varepsilon}+\f{\Gamma_{U}}{2\pi\varepsilon},~~~~~\delta_{\Gamma U}=-\f{\Gamma_{U}}{2\pi\varepsilon}-\f{\Gamma_{V}}{2\pi\varepsilon},~~~~~\delta_{\Gamma T}=-\f{\Gamma_{V}\Gamma_{U}}{\pi\varepsilon}.
\enn

As a result, we obtain the renormalization factors:
\bea
Z_{\psi}^{\omega}&\simeq&\exp{\Big[-\f{\Gamma_{V}+\Gamma_{U}}{2\pi}\ln{M}+\f{5\Gamma_{V}^{2}+12\Gamma_{V}\Gamma_{U}+7\Gamma_{U}^{2}}{48\pi^{2}}\ln{M}\Big]},\no\\
Z_{\psi}^{\k}&\simeq&\exp{\Big[\f{(\Gamma_{V}+\Gamma_{U})^{2}}{16\pi^{2}}\ln{M}\Big]},\no\\
Z_{c0}&\simeq&\exp{\Big[-\f{\Gamma_{V}-\Gamma_{U}}{2\pi}\ln{M}+\f{5\Gamma_{V}^{2}-14\Gamma_{V}\Gamma_{U}+5\Gamma_{U}^{2}}{48\pi^{2}}\ln{M}\Big]},\no\\
Z_{\c}&\simeq&\exp{\Big[-\f{\Gamma_{V}^{2}-\Gamma_{U}^{2}}{16\pi^{2}}\ln{M}\Big]},\no\\
Z_{\Gamma V}&\simeq&\exp{\Big[-\f{\Gamma_{V}+\Gamma_{U}}{2\pi}\ln{M}\Big]},\no\\
Z_{\Gamma U}&\simeq&\exp{\Big[\f{\Gamma_{V}+\Gamma_{U}}{2\pi}\ln{M}\Big]},\label{r.Rfactors}
\eea
where we have replaced $\f{1}{\varepsilon}$ with a cut-off scale, $\ln{\f{1}{M}}$, and approximated the renormalization factor as $Z=1+\delta\simeq\exp{(\delta)}$.

Recall the relations between the bare and renormalized quantities: $\Gamma_{V}=M^{d-2}(Z_{\psi}^{\omega})^{2}(Z_{\Gamma V})^{-1}\Gamma_{BV},~\Gamma_{U}=M^{d-2}(Z_{\psi}^{\omega})^{2}(Z_{\Gamma U})^{-1}\Gamma_{U},~v_{R}=Z^{\omega}_{\psi}(Z^{\k}_{\psi})^{-1}v_{B}~c_{R0}=M^{-1}Z_{\psi}^{\omega}(Z_{c0})^{-1}c_{B0}$, and $c_{Rk}=M^{-1}Z_{\psi}^{\omega}(Z_{\c})^{-1}c_{Bk}$. Based on these equations, it is straightforward to find the renormalization group equations
\bea
\f{d\ln{\Gamma_{V}}}{d\ln{M}}&=&d-2+2\f{d\ln{Z_{\psi}^{\omega}}}{d\ln{M}}-\f{d\ln{Z_{\Gamma V}}}{d\ln{M}},\no\\
\f{d\ln{\Gamma_{U}}}{d\ln{M}}&=&d-2+2\f{d\ln{Z_{\psi}^{\omega}}}{d\ln{M}}-\f{d\ln{Z_{\Gamma U}}}{d\ln{M}},\no\\
\f{d\ln{v}}{d\ln{M}}&=&\f{d\ln{Z_{\psi}^{\omega}}}{d\ln{M}}-\f{d\ln{Z_{\psi}^{\k}}}{d\ln{M}},\no\\
\f{d\ln{c_{0}}}{d\ln{M}}&=&-1+\f{d\ln{Z_{\psi}^{\omega}}}{d\ln{M}}-\f{d\ln{Z_{c0}}}{d\ln{M}},\no\\
\f{d\ln{c_{k}}}{d\ln{M}}&=&-1+\f{d\ln{Z_{\psi}^{\omega}}}{d\ln{M}}-\f{d\ln{Z_{\c}}}{d\ln{M}}.\label{e.RGE}
\eea

Substituting the results of (\ref{r.Rfactors}) into Eq. (\ref{e.RGE}), we obtain the renormalization group equations [Eq.(\ref{r.RGeq})]
\bnn
\f{d\Gamma_{V}}{d\ln{M}}&=&\Gamma_{V}\bigg[1-\f{\Gamma_{V}+\Gamma_{U}}{2\pi}+\f{(\Gamma_{V}+\Gamma_{U})(5\Gamma_{V}+7\Gamma_{U})}{24\pi^{2}}\bigg],\no\\
\f{d\Gamma_{U}}{d\ln{M}}&=&\Gamma_{U}\bigg[1-\f{3(\Gamma_{V}+\Gamma_{U})}{2\pi}+\f{(\Gamma_{V}+\Gamma_{U})(5\Gamma_{V}+7\Gamma_{U})}{24\pi^{2}}\bigg],\no\\
\f{dv}{d\ln{M}}&=&v\bigg[-\f{\Gamma_{V}+\Gamma_{U}}{2\pi}+\f{(\Gamma_{V}+\Gamma_{U})(\Gamma_{V}+2\Gamma_{U})}{24\pi^{2}}\bigg],\no\\
\f{dc_{0}}{d\ln{M}}&=&c_{0}\bigg[-1-\f{\Gamma_{U}}{\pi}+\f{\Gamma_{U}(\Gamma_{U}+13\Gamma_{V})}{24\pi^{2}}\bigg],\no\\
\f{dc_{k}}{d\ln{M}}&=&c_{k}\bigg[-1-\f{\Gamma_{V}+\Gamma_{U}}{2\pi}+\f{(\Gamma_{V}+\Gamma_{U})(2\Gamma_{V}+\Gamma_{U})}{12\pi^{2}}\bigg].
\enn

We notice that $\Gamma_{V}$ and $\Gamma_{U}$ affect renormalization of the other parameters, but the reverse way is not the case. In other words, $\Gamma_{V}$ and $\Gamma_{U}$ determine renormalization effects of all parameters, including themselves. In this respect we focus first on the equations for $\Gamma_{V}$ and $\Gamma_{U}$:
\bnn
\f{d\Gamma_{V}}{d\ln{M}}&=&\Gamma_{V}\bigg[1-\f{\Gamma_{V}+\Gamma_{U}}{2\pi}+\f{(\Gamma_{V}+\Gamma_{U})(5\Gamma_{V}+7\Gamma_{U})}{24\pi^{2}}\bigg],\\
\f{d\Gamma_{U}}{d\ln{M}}&=&\Gamma_{U}\bigg[1-\f{3(\Gamma_{V}+\Gamma_{U})}{2\pi}+\f{(\Gamma_{V}+\Gamma_{U})(5\Gamma_{V}+7\Gamma_{U})}{24\pi^{2}}\bigg].
\enn
It turns out that despite their structural similarity of these equations the fates of two types of disorders are very distinct as depicted in Fig. \ref{f.VU_flow}. If we include one-loop corrections only (Left), there appear two critical lines each for $\Gamma_{V}$ and $\Gamma_{U}$. Over the red line $\Gamma_{U}$ starts to increase and over the blue line $\Gamma_{V}$ does, too. However, the total gradient is overwhelmed by that of $\Gamma_{U}$, i.e. almost upward. This means that the anti-screening of $\Gamma_{V}$ is much weaker than that of $\Gamma_{U}$. If we include two-loop corrections also that give rise to screening in both disorders (Right), there appears another critical line for $\Gamma_{U}$ while the critical line for $\Gamma_{V}$ disappears, so $\Gamma_{V}$ becomes irrelevant. As a result, we have two nonzero fixed points on the line of $\Gamma_{V}=0$ as shown in this figure and the first figure in Fig. \ref{f.rgflow}.

\bf
\ing[width=0.8\tw]{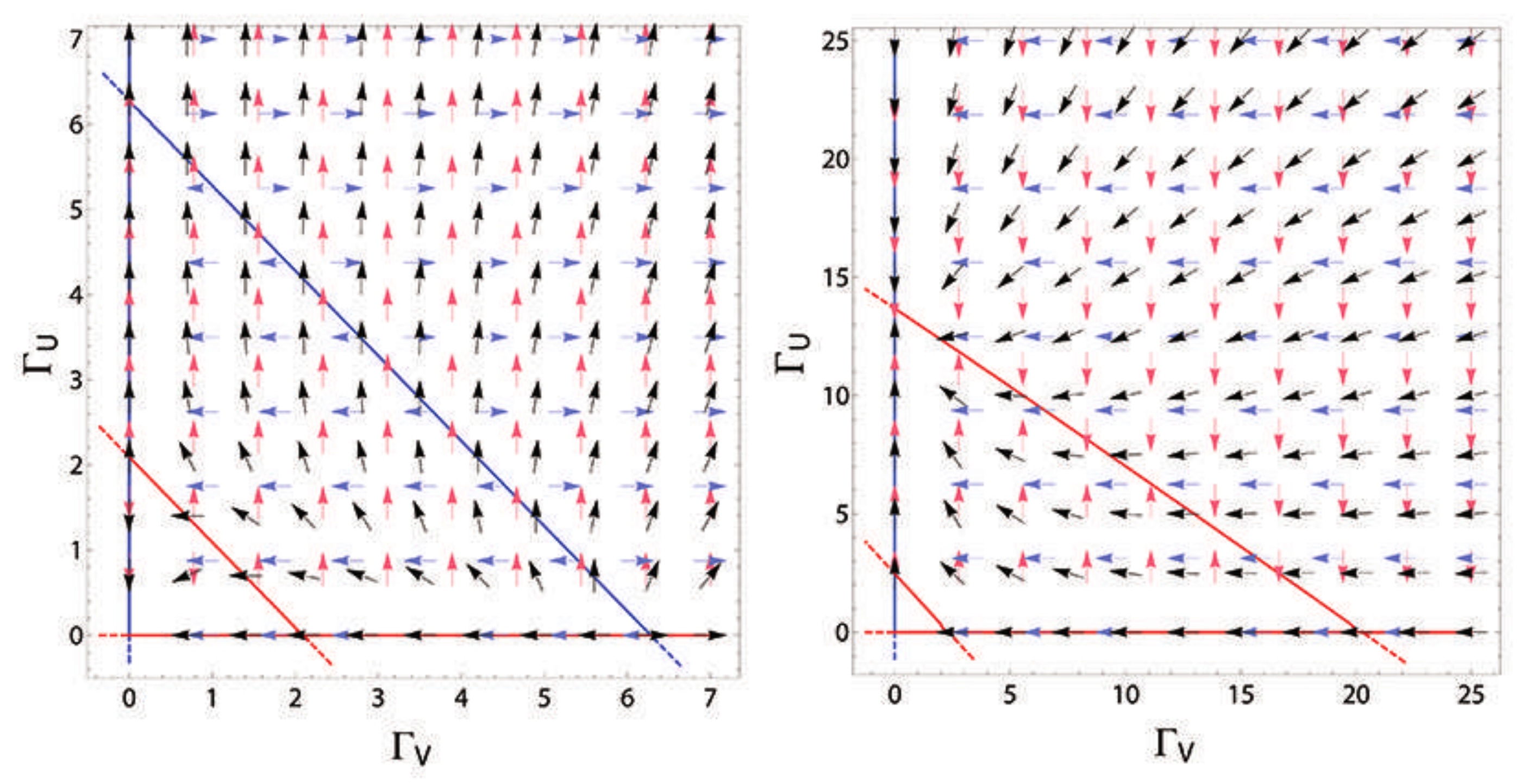}\caption{Topography of the renormalization group equations for $\Gamma_{V}$ and $\Gamma_{U}$. At each point red, blue and black arrows denote the direction in which $\Gamma_{V}$, $\Gamma_{U}$ and $(\Gamma_{V},\Gamma_{U})$ are heading as lowering the scale of the system. In the left figure, where only one-loop corrections are included, there are two critical lines each for $\Gamma_{U}$ (the red line) and $\Gamma_{V}$ (the blue line). In the right figure, where two-loop corrections are also included, there appears another critical line for $\Gamma_{U}$ while the critical line for $\Gamma_{V}$ disappears. As a result, the direction of $\Gamma_{V}$ remains negative so there are two nonzero fixed points on the line of $\Gamma_{V}=0$.} \label{f.VU_flow}
\ef

This observation suggests that $\Gamma_{U}$ has dominant effects over $\Gamma_{V}$ for the low-energy physics. Since we are interested in the renormalization of $c_{k}$, we need to consider two equations at $\Gamma_{V} = 0$:
\bnn
\f{d\Gamma_{U}}{d\ln{M}}&=&\Gamma_{U}\Big[1-a_{\Gamma}\Gamma_{U}+b_{\Gamma}\Gamma_{U}^{2}\Big],\\
\f{dc_{k}}{d\ln{M}}&=&c_{k}\Big[-1-a_{\c}\Gamma_{U}+b_{\c}\Gamma_{U}^{2}\Big],
\enn
where the positive numerical constants are given by
\bnn
a_{\Gamma}=\f{3}{2\pi},~b_{\Gamma}=\f{7}{24\pi^{2}},~a_{\c}=\f{1}{2\pi},~b_{\c}=\f{1}{12\pi^{2}}.
\enn

In the first equation for $\Gamma_{U}$, there are three fixed points: $\Gamma_{0}=0, \Gamma_{1}=\f{a_{\Gamma}-\sqrt{a_{\Gamma}^{2}-4b_{\Gamma}}}{2b_{\Gamma}}$, and $\Gamma_{2}=\f{a_{\Gamma}+\sqrt{a_{\Gamma}^{2}-4b_{\Gamma}}}{2b_{\Gamma}}$. Two stable fixed points of $\Gamma_{0}$ and $\Gamma_{2}$ are identified as a clean Weyl metal state and a diffusive Weyl metal phase, respectively. An unstable fixed point of $\Gamma_{1}$ is identified as the phase transition point from the clean Weyl metal state to the diffusive Weyl metal phase.

Let's move on the second equation for $c_{k}$. The formal solution is given by
\bnn
c_{k}(T)=c_{k}(T_{0})\exp{\bigg[-\int^{\ln{T}}_{\ln{T_{0}}}d\ln{M}-a_{\c}\int^{\ln{T}}_{\ln{T_{0}}}d\ln{M}~\Gamma_{U}(M)+b_{\c}\int^{\ln{T}}_{\ln{T_{0}}}d\ln{M}~\Gamma_{U}^{2}(M)\bigg]} ,
\enn
where $T_{0}$ is a UV cutoff. Inserting the solution of $\Gamma_{U}(M)$ into the above, we find that the distance between the pair of Weyl points shows a power-law divergent behavior
\be
c_{k}(T)=c_{k}(T_{0})\bigg(\f{T_{0}}{T}\bigg)^{\lambda_{\c,fn}} \label{r.cfield} ,
\ee
where $\lambda_{\c,fn}$ is a critical exponent around each fixed point, given by
\bnn
\lambda_{\c,f0}&=&1+a_{\c}\Gamma_{0}-b_{\c}\Gamma_{0}^{2}=1,\\
\lambda_{\c,f1}&=&1+a_{\c}\Gamma_{1}-b_{\c}\Gamma_{1}^{2}\simeq1.34,\\
\lambda_{\c,f2}&=&1+a_{\c}\Gamma_{2}-b_{\c}\Gamma_{2}^{2}\simeq1.60.
\enn
Disorder scattering changes the temperature-dependent exponent of $c_{k}$ (see Fig. \ref{f.cfield}).

\bf[t]
\ing[width=.6\tw]{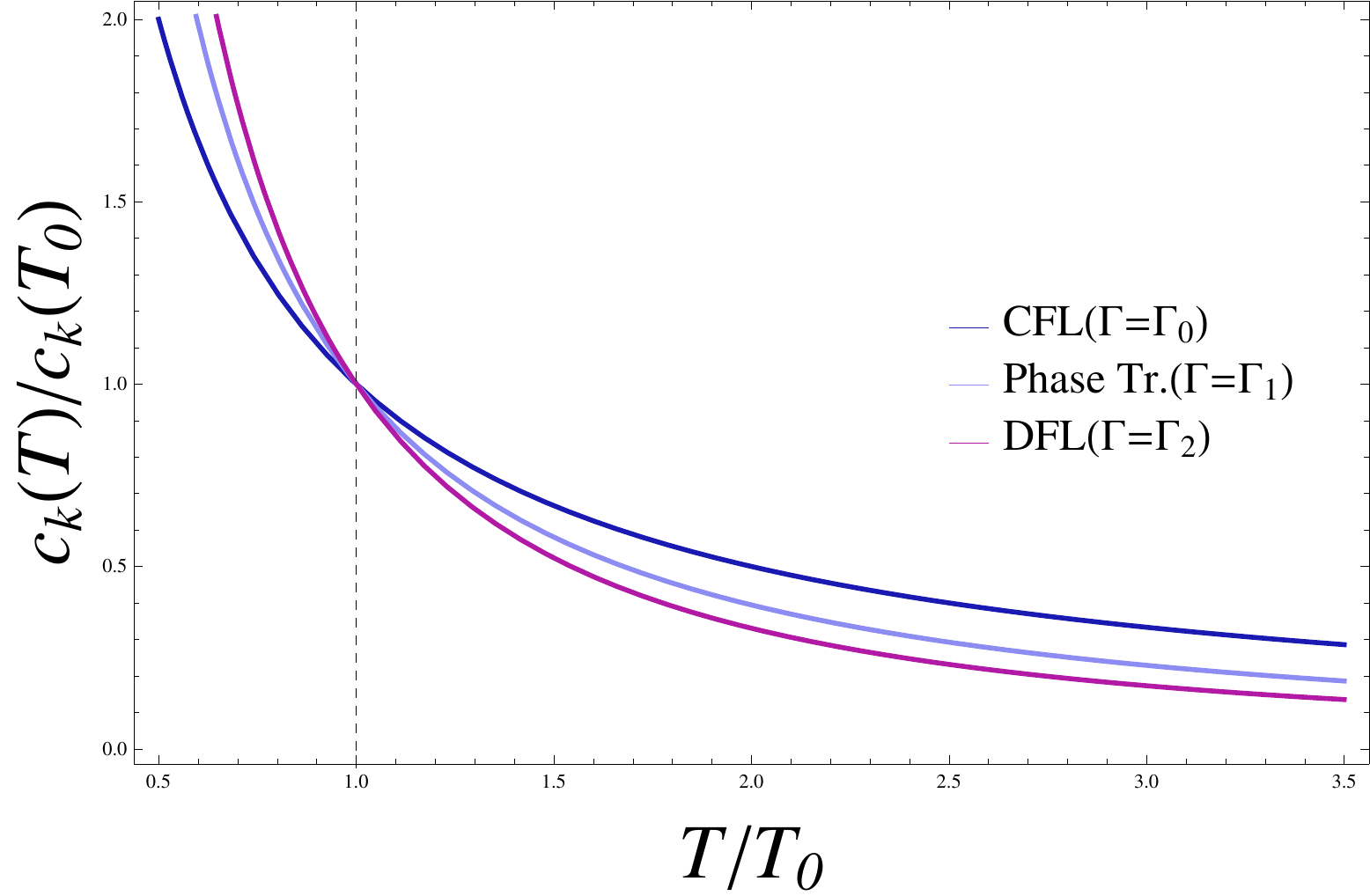} \caption{Evolution of $c_{k}$ with temperature $(T)$ near each fixed point. At the CFL $(\Gamma=\Gamma_{0})$ (clean Fermi-liquid), the exponent of $c_{k}$ is $\lambda_{\c,f0}=1$ as the dimensional analysis suggests. On the other hand, at the phase transition point $(\Gamma=\Gamma_{1})$ and DFL $(\Gamma=\Gamma_{2})$ (diffusive Fermi-liquid), the exponents of $c_{k}$ are changed to be $\lambda_{\c,f1}\simeq1.34$ and $\lambda_{\c,f2}\simeq1.60$, respectively due to additional contributions from nonzero values of $\Gamma$.}
\label{f.cfield}
\ef

\end{widetext}

\end{document}